\documentclass[fleqn,usenatbib]{mnras}

\usepackage{amsmath}
\usepackage{url}
\usepackage{amsfonts}
\usepackage{amsbsy}
\usepackage{verbatim}
\usepackage{array}
\usepackage{amssymb}
\usepackage{amsbsy}
\usepackage{graphicx,xcolor,color,subfig}

\renewcommand{\u}{\ensuremath{\mathbf{u}}}
\renewcommand{\d}{\ensuremath{\partial}}

\newcommand{\ex}{\ensuremath{\mathbf{e}_{x}}}
\newcommand{\ey}{\ensuremath{\mathbf{e}_{y}}}
\newcommand{\ez}{\ensuremath{\mathbf{e}_{z}}}

\graphicspath{{figures/}{../figures/}}

\title[Simulations of the convective overstability]{Axisymmetric simulations of
  the convective overstability in protoplanetary discs}
\author[R.J.~Teed, H.N.~Latter]{Robert J.~Teed\thanks{E-mail:
   Robert.Teed@glasgow.ac.uk}$^{1,}$$^{2}$, Henrik N.~Latter$^{2}$ \\
$^{1}$ School of Mathematics and Statistics, University of Glasgow, University Place, Glasgow G12 8SQ, UK\\
$^{2}$ DAMTP, University of Cambridge, CMS, Wilberforce Road,
Cambridge CB3 0WA, UK}

\begin{document}

\maketitle

\begin{abstract}
 Protoplanetary discs at certain radii exhibit adverse radial entropy
gradients that can drive
 oscillatory convection (`convective
overstability'; COS).
The ensuing hydrodynamical activity may reshape the radial thermal
structure of the disc while mixing solid material radially and vertically
or, alternatively, concentrating it in vortical structures.
We perform local axisymmetric simulations of the COS
using the code SNOOPY, showing first how parasites halt the 
instability's exponential growth, and second, the different saturation routes it
takes subsequently.
As the Reynolds and (pseudo-) Richardson numbers increase, the system
moves successively from (a) a weakly nonlinear state characterised by relatively ordered nonlinear waves,
to (b) wave turbulence, and finally to (c) the
formation of intermittent and then persistent zonal flows.
In three-dimensions, we expect the latter flows to spawn vortices in the orbital plane.
Given the very high Reynolds numbers in
protoplanetary discs, the third regime should be the most
prevalent.
As a consequence, we argue that the COS is an important dynamical process
in planet formation, especially near features such as dead
zone edges, ice lines, gaps, and dust rings. 

\end{abstract}


\section{Introduction}

For most of their lives, protoplanetary (PP) discs are too cold and poorly
ionised to support a form of the magnetorotational instability
unhindered by non-ideal MHD (e.g.\ Turner et
al.~2014). This state of affairs has renewed interest in the ability of purely
hydrodynamic processes to supply the turbulent activity necessary 
for disc accretion (Lesur \& Fromang 2017, Lyra \&
Umurhan 2019). Indeed, a number of hydrodynamic instabilities may attack discs at certain radii and
at certain evolutionary stages; the most commonly discussed are
the vertical shear instability (VSI), the Rossby
wave instability, the subcritical baroclinic instability, and radial
oscillatory convection (also called `convective overstability'; COS)
(e.g.\ Urpin and Brandenburg 1997, Lovelace et al. 1999, Lesur and
Papaloizou 2009, Klahr and Hubbard 2014, Lyra 2014). 
These instabilities, however, produce flows that are probably too weak to solve the problem of angular momentum
transport, though
they may be important for other processes, such as dust diffusion and aggregation. Moreover, 
observations in scattered
light and the (sub-) mm do suggest a
weak level of background turbulence: sufficient to loft small
$\mu $m dust, yet unable to stir up larger mm-sized grains (e.g.\ Perrin et
al.~2009, Pinte et al.~2016). This low-level activity is also consistent with recently
measured turbulent broadening of molecular lines (Flaherty et
al.~2015, 2017, 2018).

Ordinarily, a negative radial entropy gradient is stabilised by a
disc's strong angular momentum gradient (cf.\ the Solberg-H\o iland
criterion; Tassoul 2000). But if cooling is not too strong
nor too weak a form of oscillatory double-diffusive convection, the COS,
finds a way around this constraint and produces hydrodynamical activity (Latter
2016, hereafter L16). Actually, most PP discs possess a positive entropy gradient on average (L16), though more recent observations indicate there is a significant subset that bucks this trend (de Gregorio-Monsalvo et al.~2013, Tazzari et al.~2017). In any case,  
all discs should undergo sharp transitions at special radii, such as dead zone edges, ice lines, gaps, and
dust rings where it is likely a strongly decreasing entropy profile might develop. It is the goal of
this paper 
to assess the behaviour and vigour
of the COS
in such special regions.

Earlier work has shown that COS modes
cannot grow too large before being attacked by
parasitic instability (L16). If the parasites set the level of
hydrodynamic activity, then the COS saturates at a relatively
low level. But it is also possible that, after its initial breakdown,
the turbulent flow splits
into a sequence of zonal flows, by analogy with semiconvection, thus leading to a far
more vigorous and interesting state (Rosenblum et al.~2011, Zaussinger and Spruit 2013). 
These flows, in turn, may be subject to Kelvin-Helmholtz
instability and hence will shed vortices that could accumulate
solids (Lyra 2014, Raettig et al.~2021).

To understand the saturation properties of the COS, and in particular its propensity to form zonal flows,
we undertake axisymmetric simulations in the Boussinesq shearing
box with SNOOPY, a commonly used pseudo-spectral code (Lesur 2007).
We traverse a range of parameters, primarily the Reynolds and (pseudo-) Richardson numbers (Re and $R$ respectively), to determine the various nonlinear outcomes available to the
system.

We find that at low Re and $R$, the system exhibits laminar nonlinear wave states that can be modelled by 
a simple dynamical system based on a three-wave resonance (e.g.\ Craik 1985). 
Increasing either Re or $R$ sees the system
enter a `wave-turbulent' state, comprised of a disordered field of
inertial waves (e.g. Galtier 2003). At larger values of Re and $R$, this state supports additional vertical elevator modes and 
the intermittent formation of zonal flows via 
a mean-field anti-diffusive instability, similar in nature to the
layer formation witnessed in
thermohaline and semi convection (e.g. Rosenblum et al.~2011, Traxler et
al.~2011, Mirouh et
al.~2012, Spruit 2013, Zaussinger and Spruit 2013), and also jet
production in turbulent planetary
atmospheres and tokamak plasmas
(e.g.\ Dritschel and McIntyre 2008, Diamond et al.~2005). For higher Re and $R$ the zonal flows persist and strengthen.
We expect these flows to generate vortices in the orbital plane when the restriction
of axisymmetry is lifted. We delineate
the boundaries between these different saturation regimes in the Re-$R$
parameter space, and
argue that if the COS is to appear on reasonable timescales it is 
the persistent zonal flow regime that is most likely in PP
discs. 

The paper is organised as follows. In Section 2 we briefly outline the
background physics of the COS and discuss its prevalence in PP discs,
while Section 3 presents the tools we employ to understand it: 
the model equations and their numerical implementation. Our results
appear in Sections  4 and 5, where first we examine the onset and
initial
breakdown of the instability, making comparison with previous work, and
second describe the subsequent evolutionary paths the system might
take, and their dependence on the physical parameters. In Section 6 we outline a model that helps explain why zonal flows form. 
Conclusions are
drawn and future work pointed out in Section 7.

\section{Background}
\subsection{Instability mechanism and basic properties}

We begin by briefly describing the physical cause of instability.
In essence, the fastest growing COS mode is an epicycle accompanied by a thermodynamic
oscillation. If thermal diffusion is present, a crucial
time-lag develops between the dynamical component of the oscillation
(the epicycle) and the thermodynamic component. This means that after
half
an epicycle a fluid blob returns to its starting radius with a
different
temperature than that which it started (and hence its immediate
surroundings at that instant). In the presence of a negative radial entropy gradient,
the blob will hence suffer a buoyancy acceleration that amplifies the
initial oscillation and leads to runaway growth. For more
details see Section 3.3 in L16. 

Note that the instability
condition is the Schwarzschild criterion, rather than the
Solberg-H\o iland
criterion: the COS has found a way to use thermal diffusion to negate
the stabilising influence of rotation.
As with many hydrodynamical instabilities in accretion discs, the COS
is `double diffusive', relying on thermal diffusion, in this
case, to vastly overwhelm viscous diffusion. A final point is that the instability
mechanism is similar to the subcritical baroclinic instability (SBI;
Klahr \& Bodenheimer 2003, Petersen et al.~2007, Lesur
and Papaloizou 2010), but we emphasise the two instabilities are distinct. The
COS is linear and axisymmetric and requires a vertical wavenumber ($k_z\neq 0$),
while the SBI is nonlinear and non-axisymmetric and works in razor-thin discs (where there is no $k_z$). 
 
Other key COS properties include its maximum growth
rate and characteristic lengthscale. The former reaches
$-(1/4)(N^2/\kappa)$,
where $N^2$ is the squared radial buoyancy frequency and $\kappa$ is
the epicyclic frequency. If radiative cooling is described in the
diffusive approximation, maximum growth occurs on a distinct wavelength
$\approx \sqrt{\xi/\kappa}$ where $\xi$ is thermal
diffusivity. Typically this lengthscale is longer than
the photon mean free path in the inner radii of typical PP disc models (see Table I), and thus the
diffusive approximation is acceptable, certainly for linear analyses,
and probably for nonlinear simulations.

\subsection{Prevalence in PP discs}

Early observations in the (sub-) mm continuum permitted researchers to infer
the large-scale (smooth) radial profiles of PP disc temperatures,
densities, etc.  (e.g.\ Andrews et
al.~2009, Isella et al.~2009, Guilloteau et al.~2011).  
Generally, they indicated that most discs possess a \emph{positive} entropy
gradient, on account of the surface density's
steep fall-off with radius, thus suggesting that
discs were stable (in bulk) to the COS (see Sections 8.2 in Lin \& Youdin
2015 and 3.4.1 in L16). However,
the high angular resolution afforded by ALMA, especially,
has shown that, within 100 AU, PP discs exhibit far flatter
surface-density 
profiles than previously thought. In fact (within the modelling
errors) HD 163296 and
 several examples in Lupus (e.g.\ Sz 65, Sz 71, and Sz 98), 
possess almost no variation in surface density at 10s of AU,
while a number of other discs see their surface densities gently
increase (de Gregorio-Monsalvo et al.~2013, Tazzari et al.~2017). 
The COS's chances are much improved in such environments, though
it still very much depends on the large-scale temperature profile
and the degree of flaring (L16). We conclude that there are probably
a significant subset of PP discs that support the COS throughout
a broad range of radii between 1 and 10 AU. 

On the other hand, PP discs exhibit a great deal of complicated and
sometimes abrupt radial structure. Multiple observations reveal gaps, dust
rings, and spirals, while theory posits special radii such as
ice lines and dead/active zone boundaries (e.g.\ Muto et al.~2012, Brogan et
al.~2015, Fedele et al.~2017, Lecar et al.~2006, 
Gammie 1996, Armitage 2011). It is conceivable that around
such features the entropy gradient will flip, especially if 
these regions suffer radially inhomogeneous heating, as might be
expected at the inner dead-zone edge, at the outer edge of a
ring or gap, or in regions shielded from the central star by vertical disc
deformations (e.g. Jankovic et al.~2021, Natta et al.~ 2001, Dullemond et al.~2001, Chrenko and Nesvorny 2020); 
moreover, sublimation/condensation fronts and opacity transitions can display abrupt `thermal anomolies' and associated
entropy jumps (e.g.\ Garaud and Lin 2007). 
We might expect the COS to appear in these localised
pockets,
and perhaps to best function on account of the stronger local gradients. 
These regions will then be reshaped, thanks to the COS, via mixing of both
heat and solid particles. 

Separate to the observations, theoretical models of global disc structure have attempted to
answer the question of the prevalence and location of instabilities
such as the COS. Models based on passively irradiated discs
generally admit stable radial entropy profiles, unless the surface
density varies very slowly; work using such set-ups have investigated
cooling rates and their impact on instability, but cooling rates alone cannot
decide whether the COS grows or not, and are hence potentially
misleading (e.g.\ Malygin et al.~2017, Lyra
\& Umurhan 2019). `Active' alpha-disc models, on the other hand, can
yield radial intervals in which the entropy gradient becomes negative
and the COS unstable (Pfeil \& Klahr 2019). But then what is supplying
the `alpha', if not the COS itself? If this background heating issues
from a separate source of turbulence, will this not interfere with the
emergence of the COS? Furthermore, the vertical profile of heat deposition may not be in accord with an
alpha-type viscosity (see for example Mori et al.~2019).
Clearly, the theoretical models need further work. 
For the moment we simply posit that the COS can
prevail in a subset of PP discs, possibly on a range of radii or in
isolated pockets associated with abrupt structure.

\subsection{Saturation, zonal flows, and elevator flows}

To assess the influence of the COS we must track its nonlinear
evolution. L16 showed that the fastest growing modes are
attacked by a parametric instability involving a three-way resonance
with inertial waves. Typically these limit the initial saturation
amplitude of the COS to relatively low levels, with random velocities
some $10^{-5}$ the local sound speed, probably too low to be of
interest. However, for certain parameters, the subsequent evolution of
the turbulence may be quite different, breaking it up into radial layers of strong stratification in temperature and angular momentum accompanied
by much stronger velocities and transport. In fact, this is what is seen in some of
the simulations of axisymmetric COS by Lyra (2014) and of 
2D semi-convection, which is mathematically identical (Rosenblum et
al.~2011, Mirouh et al.~2012). In full 3D, these layers, or rather
`zonal flows', could shed vortices via Kelvin-Helmholtz instability (RWI). This is an outcome of much greater interest as it is likely to impact not only
on the disc structure but on particle accumulation and even planet
formation.

The emergence of zonal flows is an interesting potential feature of
the COS, and is generic to turbulent and rotating systems.
The production of these flows is witnessed not
only in planetary atmospheres and tokamaks, 
but in several accretion disc context: simulations of the MRI (Steinacker
et al 2002, Johansen
et al 2009, Kunz and Lesur 2013, Simon et al.~2012), the parametric
instabilities that afflict warped and eccentric discs (Wienckers \&
Ogilvie 2018, Paardekooper \& Ogilvie 2019) and the VSI (Richard et al.~2016). Moreover, the geostrophic balance,
underlying zonal flows, controls the linear instability mechanism
of several disc processes: the VSI itself
(Latter \& Papaloizou 2018), 
the streaming instability (Jacquet et al.~2011), 
the secular dust gravitational instability (Latter \& Rosca 2017),
the diffusive gravitational instability (Vanon \& Ogilvie 2017), 
and the double-diffusive resistive instability (Latter et al.~2010).

Geostrophic balance comprises a quasi-steady equilibrium between the Coriolis force and the 
pressure gradient (alongside, possibly, radial buoyancy or Lorentz forces), and is characterised by zonal flows, a radial sequence of super and sub Keplerian jets (zonal flows).  
In the absence of viscosity, these jets form by a nonlinear process
that swaps rings of material
at different radii: the outer ring moves in and orbits slower than
the material immediately surrounding it, while the inner ring moves out
and rotates faster. The two rings are held in place by pressure, 
otherwise they would fall back to whence they came (via the Coriolis
force). Note
that zonal flows need not arise from exchange of angular momentum between
fluid rings, but from exchange of radial location. This is important
because in (inviscid) axisymmetry no angular momentum exchange is possible. 

The obvious question is how these `swaps' can be arranged by the 
turbulent flow. In geophysics
it is common to invoke a mean-field `anti-diffusive' 
instability (e.g.\ the `zonostrophic' and `gamma'
instabilities; Srinivasan and Young 2012, Radko 2003); in 2D
semiconvection, Garaud and coworkers have shown that simple
turbulence closure models based on these ideas are consistent with
numerical simulations (Rosenblum et al.~2011, Traxler et al.~2011,
Stellmach et al.~2011, Mirouh et al.~2012, Wood et al.~2013). 
However, the physical insight obtained is somewhat limited and, moreover,
the mean field models are not
predictive and can only be deployed in the post processing of
numerical data. Alternatively, the development of such flows has been explored directly from the system's underlying nonlinear wave couplings: though in ideal hydrodynamics a three-wave coupling is incapable of generating geostrophic modes (Greenspan 1969), higher order resonant interactions might be able to, as could the introduction of irreversibility via an instability (such as provided by the COS, or possibly the VSI), dissipation, turbulence, or even the detuning of a wave triad (e.g.\ Smith and Waleffe 1999, Kerswell 1999, Le Reun et al.~2020).

In this paper, we take the former approach and 
build a physical mean-field model to aid our intuition of the anti-diffusive behaviour, based on the local
angular momentum fluxes generated by forced inertial waves. Because
the wave forcing (from the COS) is itself sensitive to the angular momentum gradient, the possibility of up-gradient angular momentum
transport (hence anti-diffusion) can be brought out in a relatively transparent way. 

Finally, we acknowledge that vertically local disc models (including cylindrical discs; Dewberry et al.~2020) often develop elevator flows, which usually consist of a radial sequence of updrafts and downdrafts exhibiting no vertical variation (see for example, Calzavarini et al.~2006), and some of our COS simulations are no exception. They may be interpreted as attempts by the system to manifest a larger scale circulation, and thus are not unphysical necessarily, but are certainly poorly described by local models. 

\section{Governing equations and numerical methods}

\subsection{Equations and parameters}

Being interested in small scales and subsonic flow, we
employ the Boussinesq shearing
box (Latter \& Papaloizou 2017, hereafter LP17). This model describes a small `block' of disc centred upon a
cylindrical radius $r_0$ moving on the circular orbit prescribed by $r_0$ and at
an orbital frequency of $\Omega$. We are also free to stipulate the vertical height of the block, $z_0$, above the midplane, though throughout we assume $z_0=0$. The block is represented in
Cartesian coordinates with the $x$ and $y$ directions corresponding to
the radial and azimuthal directions, respectively 
(see Goldreich \& Lynden-Bell 1965). The model can include both vertical and radial stratification, as well as vertical shear, but to keep things simple we incorporate only radial stratification - a necessary ingredient for the COS. Future work might explore the interplay between the COS and other physics --- and indeed the VSI, if present. We stress that the Boussinesq shearing box equations can be derived self-consistently and do not rely on any ad-hoc or problematic assumptions\footnote{This is in contrast to recent work using compressible local frameworks and an explicit radial pressure gradient (e.g. Lyra and Klahr 2011), which can be marred by spurious overstabilities (LP17).}.

The governing equations are
\begin{align} \label{GE1}
&\d_t \u + \u\cdot\nabla\u = -\frac{1}{\rho}\nabla P -2\Omega \ez\times
\u \notag\\ 
& \hskip1.5cm + 2q\Omega\, x\,\ex -N^2\theta\,\ex  +\nu\nabla^2\u, \\
& \d_t\theta + \u\cdot\nabla\theta = u_x + \xi\,\nabla^2\theta,\label{GE2} \\
& \nabla\cdot \u = 0, \label{GE3}
\end{align}
where $\u$ is the fluid velocity, $P$ is pressure, $\rho$ is the
(constant) background density, $\theta$ is the buoyancy variable. The dimensionless
shear parameter of the sheet is denoted by $q$, equal to $(3/2)$ in a
Keplerian disc, and the buoyancy frequency arising from the radial
stratification is denoted by $N$. 
We employ thermal diffusion rather
than an optically thin cooling law, as is done in Klahr \& Hubbard
(2014) and Lyra (2014), with $\xi$ the thermal
diffusivity\footnote{An unnerving consequence of the adoption of a single cooling time in local models of the COS is that the $k=0$ mode grows; i.e. the box itself is unstable! (See Eq.~21 in Lyra 2014.)}. Viscous diffusion is also included, with $\nu$ the
kinematic viscosity, understood to be molecular.

The (squared) buoyancy frequency can be determined from
\begin{align} \label{Nsq}
N^2 = - \frac{1}{\gamma \rho}\frac{\d P}{\d r}\frac{\d \ln\left(P\rho^{-\gamma}\right)}{\d r},
\end{align}
evaluated at $r=r_0$. In the above $\gamma$ is the adiabatic index.
Another important quantity is the (squared) epicyclic frequency
\begin{equation} 
\kappa^2 = 2(2-q)\Omega^2,
\end{equation}
which describes the angular momentum structure of the disc.

Following Lesur \& Papaloizou (2010), the stratification length has
been absorbed into $\theta$, so that
$\theta = -(\d_R S)_0^{-1} S'$, where $S'$ is the dimensionless entropy perturbation. The total entropy in the box may then be associated with $\theta_x=-x+\theta$. On the other hand, the total angular momentum is $h=2\Omega x + u_y$ (LP17).

In addition to $q$, the system can be specified by three other
dimensionless parameters.
The `$R$' number measures the relative strength of the (unstable)
radial stratification to the stabilising angular momentum gradient:
\begin{equation}
R = - \frac{N^2}{\kappa^2}.
\end{equation}
In some previous work this has been (incorrectly) identified with the
Richardson number, which instead possess the squared shear rate in the
denominator. Though the distinction is unimportant in most contexts,
we emphasise that the COS is sensitive to the angular momentum
gradient, not the shear rate per se; and when explaining layer formation (which
is caused by radial variations in $R$) this is a key point.
In thin astrophysical discs we might expect $R$ to be small, as discussed in
L16. (Note that in L16 $R$ is denoted by $n^2$.)

Finally, the relative importance of the diffusivities is measured by the
Peclet and Reynolds numbers
\begin{equation} \label{Renum}
\text{Pe}= \frac{L^2\kappa}{\xi}, \qquad \text{Re}=\frac{L^2\kappa}{\nu},
\end{equation}
where $L$ is a characteristic outer lengthscale. Our model has no
intrinsic
physical outer scale, so $L$ must be taken to be our box size. Please
be aware that our Pe and Re do not correspond to the usual
definitions, because $L$ need not be $H$ the disc scaleheight.
Lastly, we occasionally make use of the Prandtl number, Pr$=\nu/\xi$. 

\subsection{Characteristic lengthscales}

To get a feel for the physical scales in our problem we adopt
a specific disc model, a minimum-mass
solar nebula developed by Chiang \& Youdin (2010) and Lin \&
Youdin (2015). This provides scaling laws for relevant midplane 
properties. For example, in cgs units,
\begin{align*}
\Sigma = 2200\,r_\text{AU}^{-3/2}, \,
 H = 3.3\times 10^{11}\, r_\text{AU}^{9/7}, 
\,
 \ell_\text{mfp} = 1.3\times 10^8\, r_\text{AU}^{51/14}, 
\end{align*}
where $\Sigma$ is the surface density, $\ell_\text{mfp}$ is the photon mean free path, and
$r_\text{AU}$ is disc radius in AU. 

In addition, we can can estimate
the wavelength of fastest COS growth $\lambda_\text{max}\approx \sqrt{\xi/\kappa}$,
and the critical wavelength below which the linear instability switches off,
$\lambda_\text{crit}$ (cf.\ Eq.~(A2) in Latter 2016\footnote{This equation suffers from a typo:
  $\xi^2/\kappa^2$ should be replaced by $\kappa^2/\xi^2$. There are
  also a handful of other unfortunate instances in L16 where this ratio is swapped. }):
\begin{align*}
 \lambda_\text{max} = 10^{9}\, r_\text{AU}^{93/28}, \qquad
 \lambda_\text{crit} = 10^8\,\left(\frac{R}{0.01}\right)^{-1/4} r_\text{AU}^{75/28}. 
\end{align*}
Notice the very weak scaling with $R$ in our expression for 
$\lambda_\text{crit}$. 
In the following discussion
we omit this factor: values of $R$ any less than
$10^{-3}$ produce linear COS growth times too small to be important ($\gtrsim 10^4 \Omega^{-1}$).   

On lengthscales shorter than $\lambda_\text{crit}$ we expect the COS to instigate a direct turbulent
cascade, which ultimately approaches the isotropic Kolmogorov regime (Nazarenko \& Schekochihin 2011,
LP17). Kinetic energy will be thermalised on the Kolmogorov dissipation
length $\ell_\text{visc} = (\nu^3/\varepsilon)^{1/4}$, where $\varepsilon$ is the (specific) 
energy injection rate, equal to $N^2\langle \theta u_x\rangle$ in our problem, with the angle brackets
denoting a suitable length and time average. Assuming a quasilinear approach, by setting the relevant
injection length and time scales to be those of the fastest growing mode ($\lambda_\text{max}$ and $\Omega$) we obtain the estimate $\varepsilon\sim \xi |N^2|$, and find simply
$$ \ell_\text{visc} \sim \text{Pr}^{1/2}\lambda_\text{crit}.$$
 We caution that the quasilinear approximation
fails when large-scale structures appear, such as elevator and zonal flows, in which case $\ell_\text{visc}$
will be shorter (see discussion in Section 5.4). It is, however, a useful first estimate.

In Table I, we list these characteristic
lengthscales at selected disc radii for reference. 
Within about 5 AU, the COS lengthscales are significantly less than $H$
thus justifying the vertically unstratified approximation. Further out,
however, this becomes increasingly a problem.
On the other hand, for $r<10 AU$ the main COS injection scale $\lambda_\text{max}$ always 
lies above the
photon mean free path. While the inertial range of
the turbulent cascade does fall within the optically thin regime,
radiative physics is
unimportant for the cascade. Together this justifies our use of the diffusion approximation. We stress, however, 
that these estimates are tied to a specific, rather massive, disc model: a ‘lighter’ disc
 (such as the template
used in Lesur \& Latter 2015) may find $ell_\text{mfp}$ and $\lambda_\text{max}$ comparable.
Perhaps more importantly: the surface density in our fiducial nebula falls off far more steeply than 
the observed discs discussed
in Section 2.2, especially those most susceptible to the COS. 
The number generated here are thus only illustrative 
and certainly not definitive.

\begin{table}
\begin{tabular}{ | l | l | l | l |}
 \hline
  Radius (AU) & 1  & 5  & 10  \\
 \hline
\hline
$H$ (cm) & $3.3\times 10^{11}$  & $2.6\times 10^{12}$ & $6.4\times 10^{12}$ \\
 \hline
$\ell_\text{mfp}$ (cm) & $1.3\times 10^{8}$ & $4.6\times 10^{10}$ &
                                                                    $5.7\times 10^{11}$ \\
\hline
$\lambda_\text{max}$ (cm) & $10^9$  & $2.1\times 10^{11}$ & $2.1\times 10^{12}$ \\
\hline
$\lambda_\text{crit}$ (cm) & $10^8$  & $7.5\times 10^{9}$ & $4.8\times 10^{10}$ \\
\hline
$\ell_\text{visc}$ (cm) & $7.8\times 10^{4}$  & $7.4\times 10^{5}$ & $1.9\times 10^{6}$ \\
 \hline
Re & $6.4\times 10^{7}$ & $4.0\times 10^{9}$ & $2.4\times 10^{10}$
  \\
\hline
\end{tabular}
\caption{Properties of the minimum mass PP disc of Chiang \& Youdin
  (2010)
  at three different
  radii. Symbols are defined in Sections 3.1 and 3.2. The Reynolds number is calculated assuming Pe=$4\pi^2$, thus taking $L=\lambda_\text{max}$.}
\end{table}

\subsection{Numerical methods}

\subsubsection{Code and set-up}

We perform numerical simulations with the code, SNOOPY
(Lesur \& Longaretti 2005, 2007), which solves the shearing box equations
using a pseudo-spectral method based on a shearing wave decomposition.
As we only calculate axisymmetric flow, the wavevectors do not
depend on time, and no remapping is required. Nonlinear terms are computed
in real space, but a $2/3$ aliasing rule is imposed in spectral space.
The time integration of non-diffusive terms is undertaken by the explicit
third-order Runge-Kutta method, while the diffusive terms are
integrated by an implicit procedure.

We employ a
rectangular domain of size $L_x\times L_z$. Typically $L_x=2L_z=2L$.
Our basic simulations employ a grid of $512\times256$ units.
Note that, being
spectral, SNOOPY develops hatched saw-tooth structure on the
grid-scale if a simulation is under-resolved (related to the Gibbs
phenomenon); 
though this does not always crash the run it is easy to detect, 
and when we do see it we stop the simulation and rerun it at a higher
resolution.

The domain is periodic
in both $x$ and $z$. Units are chosen so
that $L=1$, $\Omega=1$, and $\rho=1$. 
Simulations are normally initialised with white noise of a given
amplitude, or otherwise with an exact COS mode sometimes polluted with smaller amplitude white noise.

\subsubsection{Parameter values}

In all runs, the disc is
Keplerian and so $q=3/2$, which leaves three dimensionless parameters:
 $R$, Pe, and Re. Roughly speaking,
the greater the $R$ and Re, the faster growing the instability and the
more vigorous the ensuing activity. We mainly vary $R$ and Re,
though to `speed up' the simulations we generally take a large value of
$R$, setting it often to 0.1. In reality, the thermal gradient could
vary greatly depending on the radial structure that generated it.
This is our principle unknown. 

The Peclet number, Pe, sets the thermal length (and $\lambda_\text{max}$) with respect to the box size $L$. Thus, indirectly, Pe also
controls the box size relative to the disk scale height.
In almost all runs we let Pe$=4\pi^2\approx 40$, which means the vertical size of the box is the same as that of the fastest growing COS mode, i.e.\ $L=2\pi\sqrt{\xi/\Omega}=\lambda_\text{max}$. By setting the energy input scale near the box size, we allow ourselves the dynamical range to set realistic (molecular) Re. The downside is that coherent large-scale structures near the input scale or larger will be impacted upon by the numerical domain. Future work might compromise on the viscous scales, but set the box size much larger so as to mitigate such effects. 

 We let the Reynolds number in our simulations range between $10^3$
to $10^7$.  Given a global disc model (such as in Section 3.2) we can relate $L$ to
$H$, the disc semi-thickness. Then, by fixing Pe$=4\pi^2\approx 40$, we can
determine our Re as a function of disc radius (noting that it differs from
the usual definition by a
factor $(H/L)^2$). Representative values for Re are placed in Table I.
We see that the Reynolds number at 1 AU is just within the range
achievable by our simulations. Beyond 1 AU, however, the Reynolds number increases beyond what is numerically
possible. These limitation should be kept in mind when interpreting
our results.

Finally, for given $R$ and Pe, there is a critical
Reynolds number Re$_\text{c}$ below which the instability switches off entirely in our simulations. This
critical value can be obtained by setting $\lambda_\text{crit}$, the marginal
wavelength, to $L$, which yields a cubic for Re. 
For
example, when $R=0.1$ and Pe$=4\pi^2$ then we have Re$_c\approx 1580$.
A rough but useful approximation, assuming small Pr, is Re$_c\approx 2^5\pi^4/(\text{Pe}\,R)$.

\subsubsection{Diagnostics}

Our main diagnostics are
\begin{align}
E_K = \langle \rho|\u|^2 \rangle,\qquad 
F_H = \langle u_x u_y' \rangle, \qquad
F_\theta = \langle u_x\theta\rangle,
\end{align}
representing the box-averaged kinetic energy, angular momentum flux,
and heat flux, respectively.  Here the angle brackets now signify an
average over the spatial domain, and
$u_y'$ is the deviation from the background differential rotation. 
As SNOOPY is a spectral code, the spatial integrations can be
undertaken conveniently in spectral space, via Parseval's theorem.
Note that `alpha' parameters can be constructed from $F_H$ and
$F_\theta$; using our system of units we obtain the classical $\alpha=
(L^2/H^2)F_H$, for instance. We also make use of the directional kinetic energies $E_{Kx}= \langle \rho u_x^2 \rangle$, $E_{Ky}= \langle \rho(u_y')^2 \rangle$, and $E_{Kz}= \langle \rho u_z^2 \rangle$.

It is possible to pick out the various spectral components of a
field (indexed by their radial and vertical wavenumber) and to plot
the associated power. Thus, for example, $\widetilde{u}_x^{mn}$
represents the Fourier coefficient of $u_x$ with the $m$'th radial
wavevector and $n$'th vertical wavenumber. The power in a mode
is the modulus square of its associated coefficient. 

\section{Onset of the convective overstability}
\subsection{Growth rates}

We begin our investigation by testing the numerical code against the linear
theory of the COS. We do this by comparing the
numerical linear growth
rates from SNOOPY to those predicted by L16.
The simulations are initialised
with a clean COS eigenfunction. By keeping $L$ fixed and the
wavevector of the mode $=2\pi/L$, when we vary
the Peclet number we can effectively sample the COS dispersion
relation; this is because the growth rate $s$ depends on
$k_z(\xi/\kappa)^{1/2}=(k_zL)\text{Pe}^{-1/2}$. 

\begin{figure}
\includegraphics[width=\linewidth]{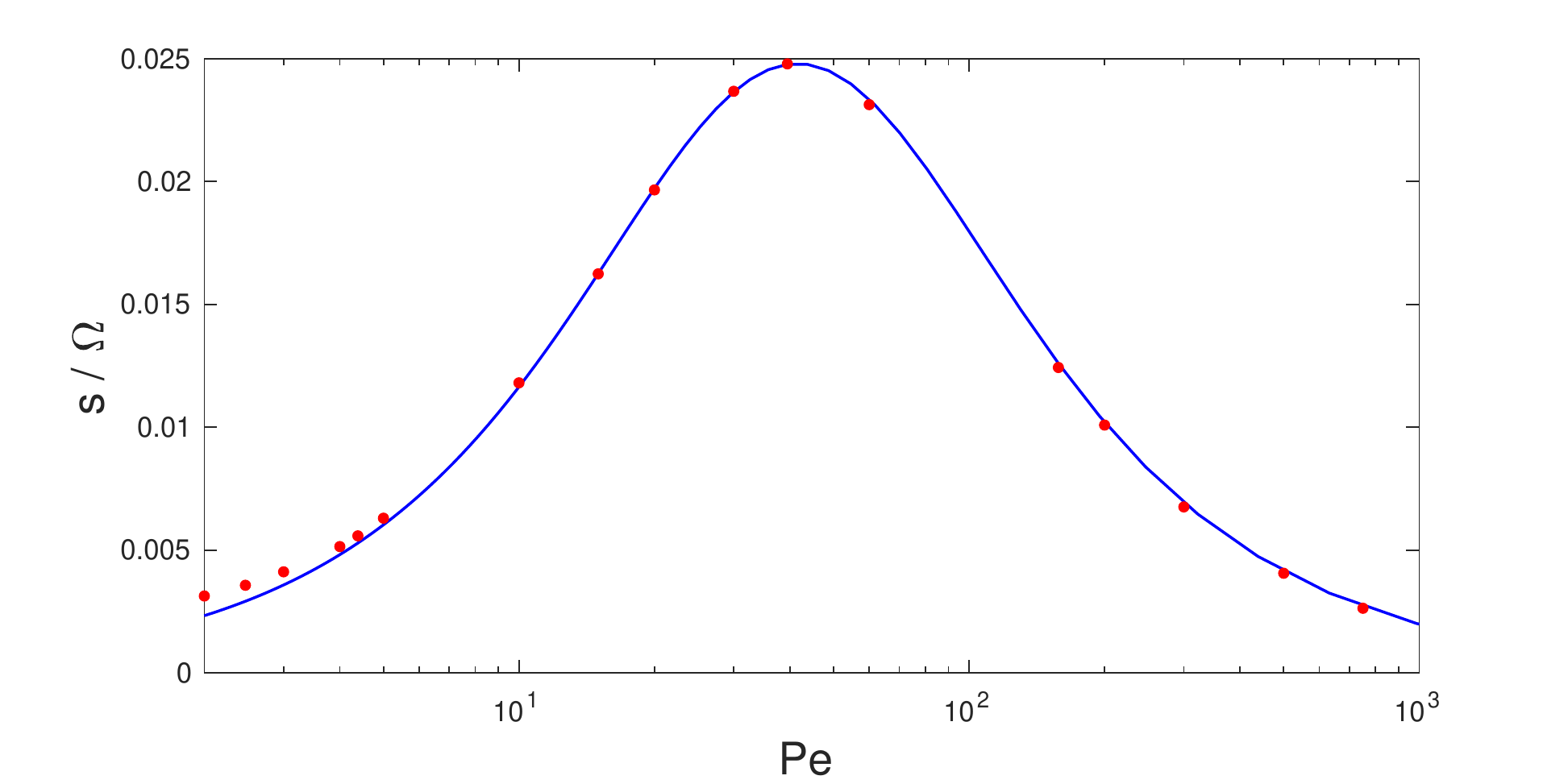}
\caption{Linear growth rate of the COS mode $s$ as a function of Pe, for $R=0.1$ and Re$=2\times10^5$. The solid blue curve represents the analytic solution; the red dots are numerically calculated values.}
\label{fig:growth}
\end{figure}

In Fig.~\ref{fig:growth} the analytical curve and numerical data
points are overlaid for the parameters $R=0.1$ and Re$=2\times 10^5$. 
The analytic curve is derived from the viscous dispersion relation,
Eq.~(A1) in L16. For most wavenumbers the agreement is excellent
(within 1\%), but for low Pe (large $k_z$) the growth rates diverge
somewhat. This is because these slower growing modes are usually overtaken by other modes before they can grow appreciably, and thus estimating their growth rates is more error prone.

\subsection{Maximum amplitudes}

Now we test the parasitic theory of L16 by considering
the amplitude of a COS mode at the point that its exponential growth
halts. At the same time we track the amplitude of the parasitic modes
that attack it. This is achieved
by comparing the sizes of the leading Fourier component of $u_x$
and of $u_z$ versus time in the initial stages of several runs:
the former corresponds to the COS, the latter to a parasite. It
 is possible to distinguish the two modes this way because the COS (the primary)
possesses no vertical velocity in its eigenfunction, while the
parasitic modes do (being inertial waves with $k_x\neq 0$). 

First, we find that if the
simulation is initialised with a clean COS mode, it will grow
indefinitely, or at least until the code crashes. 
This behaviour is in agreement with
Eq.~(33) in L16, which states that if the starting amplitude of the
parasite is very small (as when seeded by numerical error), then the
maximum COS amplitude will be extremely large on account of the
divergence of the Lambert W function near the origin. 

\begin{figure}
\includegraphics[width=\linewidth]{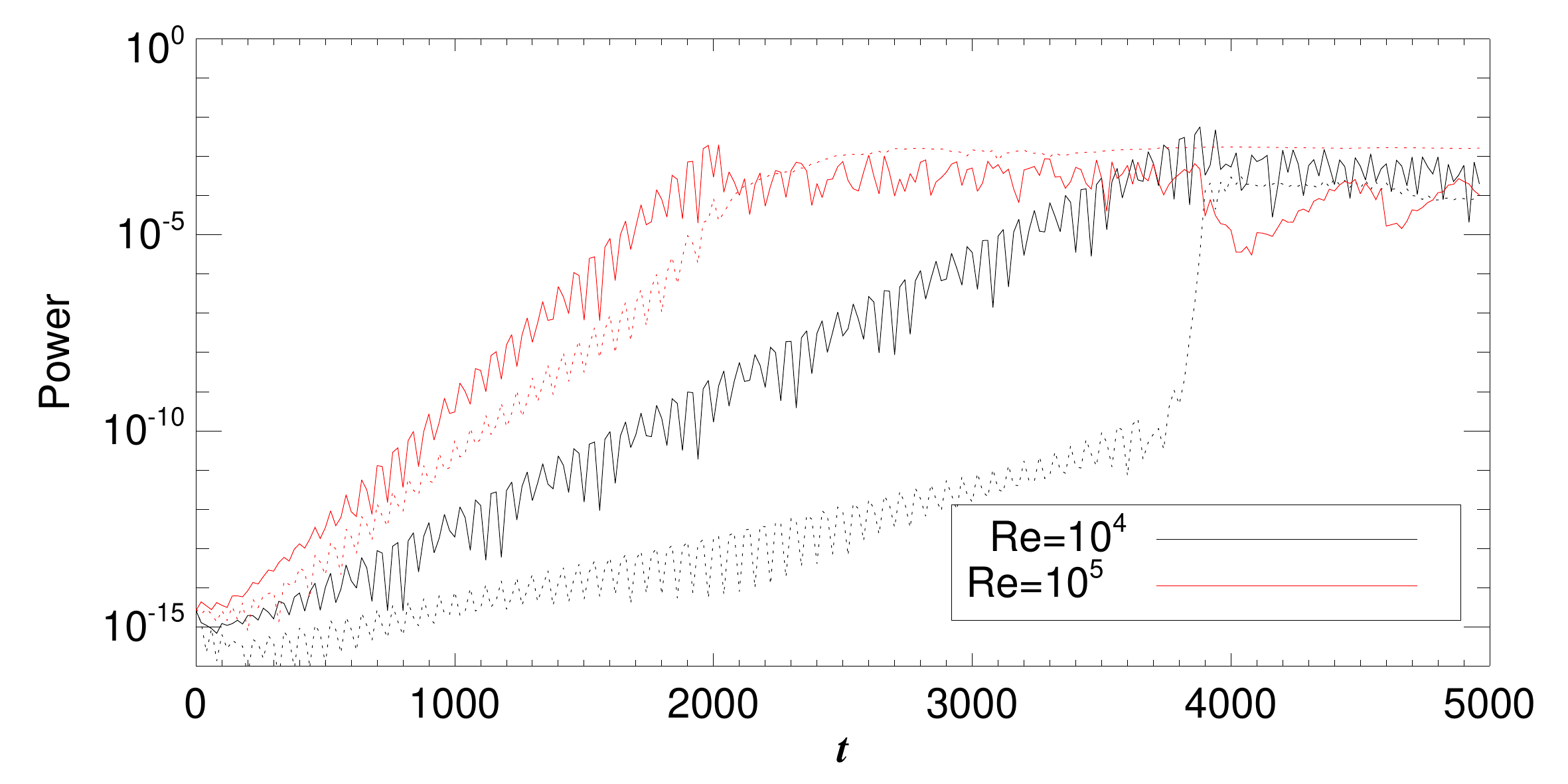}
\caption{Largest spectral components of $u_x$ (solid) and $u_z$ (dashed) as a function of time for two values of Re ($=10^4$ and $10^5$). Both runs have $R=10^{-1.5}$ and Pe$=4\pi^2$.}
\label{fig:specmax}
\end{figure}

We next seed the simulations with white noise, from which both the
parasite and COS can emerge. Now, after some time,
both modes grow and saturate at a similar order of magnitude.
Fig.~\ref{fig:specmax} demonstrates this behaviour in
two simulations performed with different values of the Reynolds
number. Power in the COS modes is represented by solid lines and power in
the parasites by dotted lines. Different colours indicate different
Re. In the linear growth phase the dominant spectral components are
$(k^m_x,k^m_z)=(0,1)$ for $u_x$ and $(k^m_x,k^m_z)=(1,1)$ for $u_z$.
In both cases the COS mode initially grows
exponentially fast, in accord with the linear growth rate, 
whilst any growth associated with the parasitic mode is marginal.
However, at a critical time, dependent on Re, the COS achieves a
sufficient amplitude for the parasite's growth rate to outcompete
the COS growth: the parasite increases rapidly until reaching the
COS amplitude and saturation occurs. 
At larger Re, growth in both modes is
stronger, and hence the saturation time occurs sooner. However, the final saturated states all possess a
similar power. L16 predicts an initial saturated state with kinetic
energy $\sim R^2 \sim 10^{-3}$, which provides a good estimate on the 
peak amplitude at the point of breakdown. Immediately afterwards the flow settles
down to a slightly less active level.

\section{Saturated states}

In this section we demonstrate
the possible long-term outcomes of the
system's evolution. The dynamics are parameter dependent
and we have identified at least three saturation routes, which are accessed consecutively as $R$ and/or Re increase,
i.e. as the system becomes more COS unstable. 
We introduce these three saturation routes separately
along with a
discussion of a typical case for each, and then present the results of a parameter
sweep showing
where in parameter space each state can be found.

\subsection{Weakly nonlinear regime}

At parameter values just above critical for the onset of the
instability, the system enters a low-order `weakly nonlinear' regime
which is controlled by a small number of modes.
For fixed $R$ and Pe, this regime occurs for Re near the critical
Re$_\text{crit}$ below which the linear COS fails to appear. 
If we equate $L$ with $H$ then Re$_\text{c}$ is generally too small
to apply to real PP discs. But we provide details of this regime for
completeness and also because it helps
illuminate the dynamics at larger,
and more realistic, Re.
 
 \begin{figure}
\includegraphics[width=\linewidth]{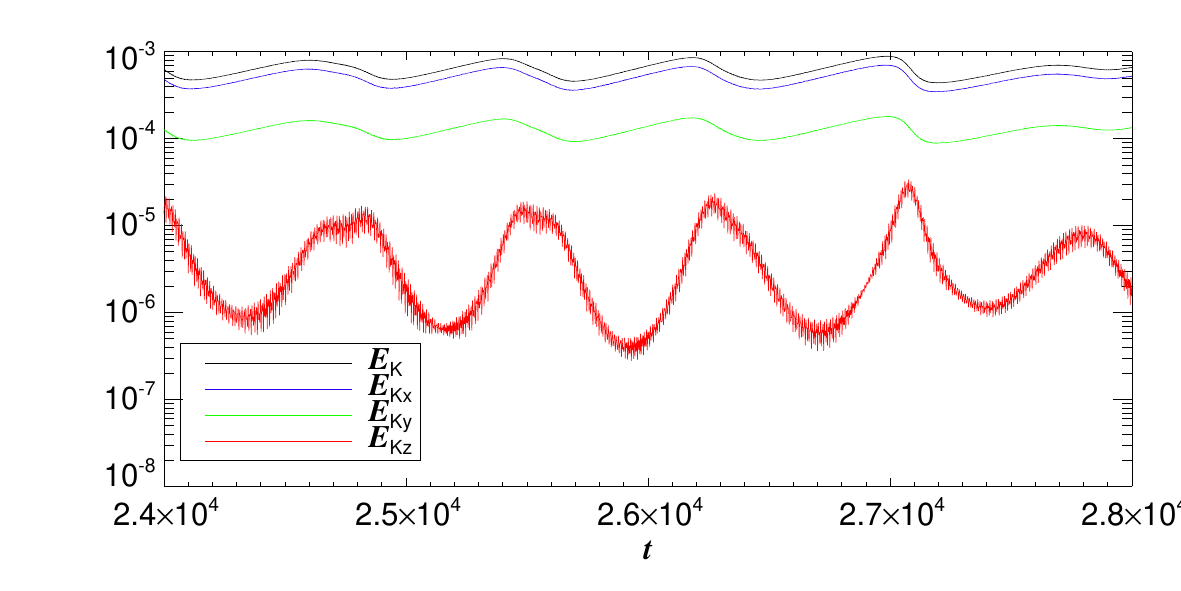}
\caption{Kinetic energy, and its separate components, as a function of time bandpass filtered to remove highest frequency oscillations.  Simulation parameters are $R=10^{-1.5}$, Pe=$4\pi^2$, Re=$10^{3.75}$.
}
\label{fig:rmsWNLfilt}
\end{figure}

\begin{figure}
\includegraphics[width=\linewidth]{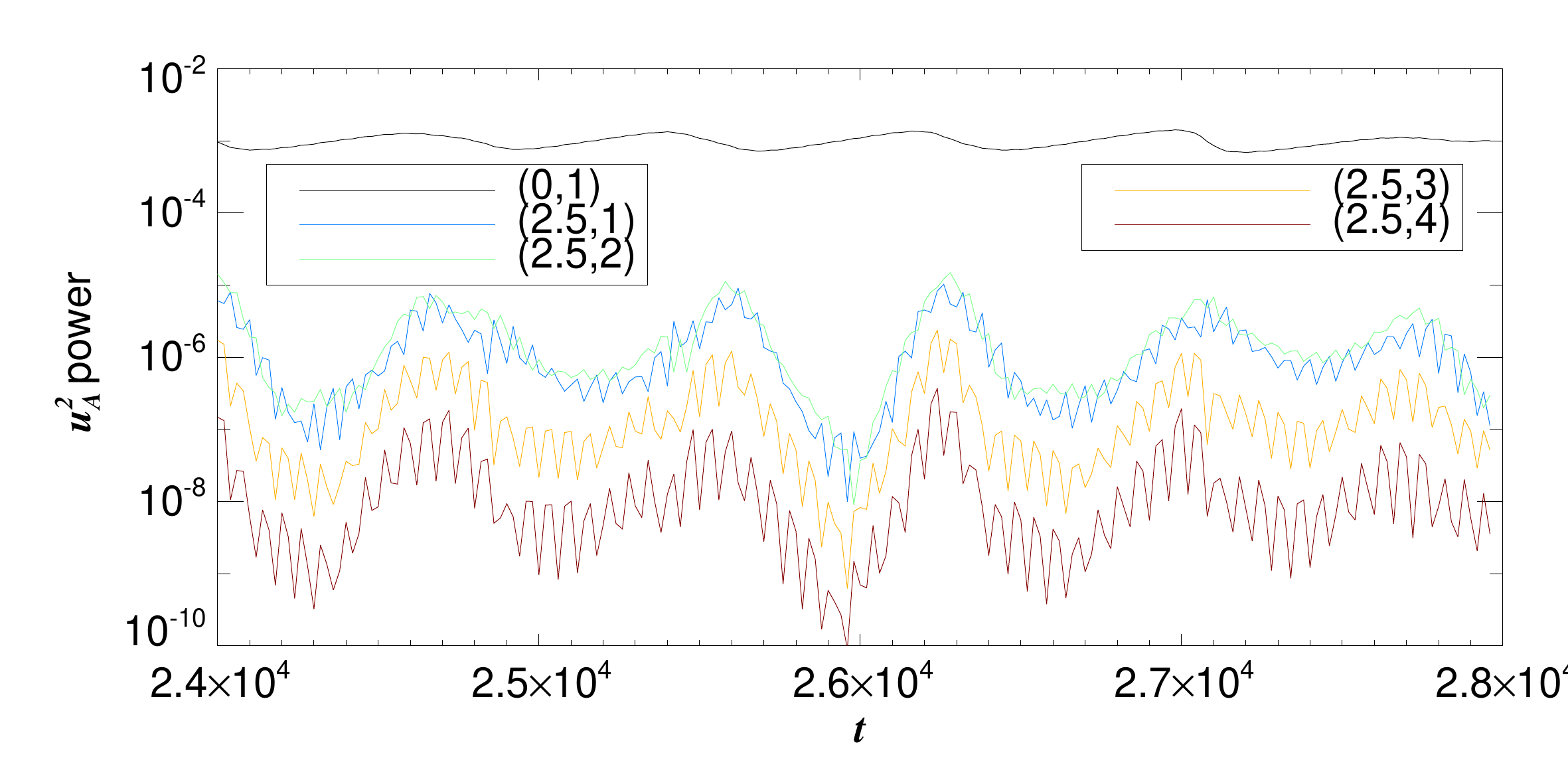}
\caption{Largest spectral components of $u_A$ as a function of time. Simulation parameters are $R=10^{-1.5}$, Pe=$4\pi^2$, Re=$10^{3.75}$.}
\label{fig:specuA}
\end{figure}

\begin{figure}

\subfloat[$t=26,000$]{\label{fig:WNLuz1}\includegraphics[width=0.49\linewidth]{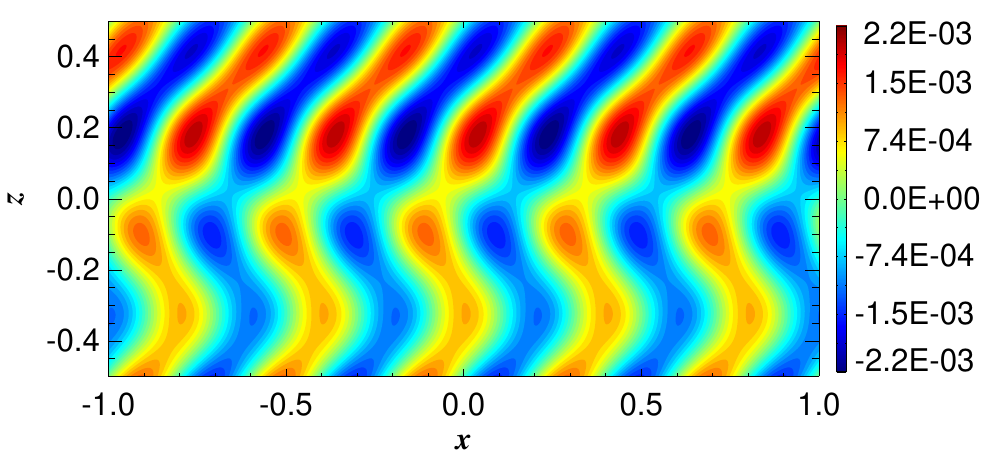}}
\subfloat[$t=26,000$]{\label{fig:WNLua1}\includegraphics[width=0.49\linewidth]{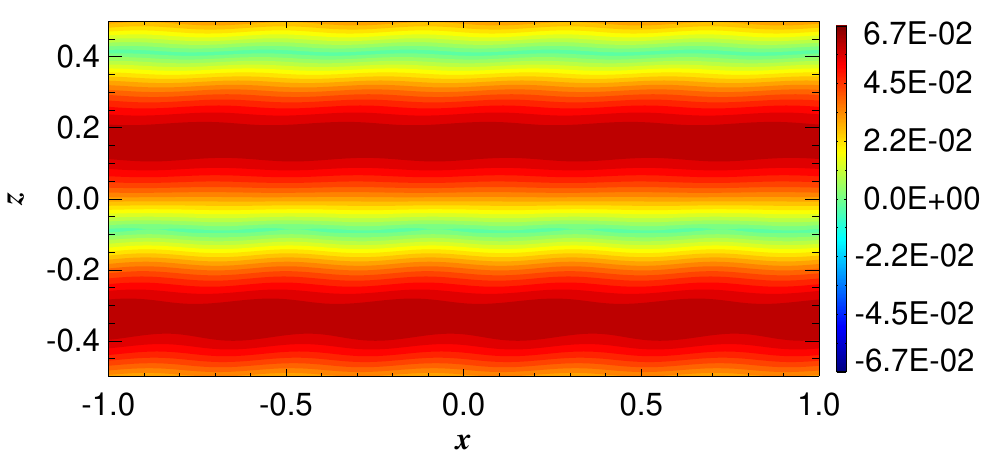}}

\subfloat[$t=26,200$]{\label{fig:WNLuz2}\includegraphics[width=0.49\linewidth]{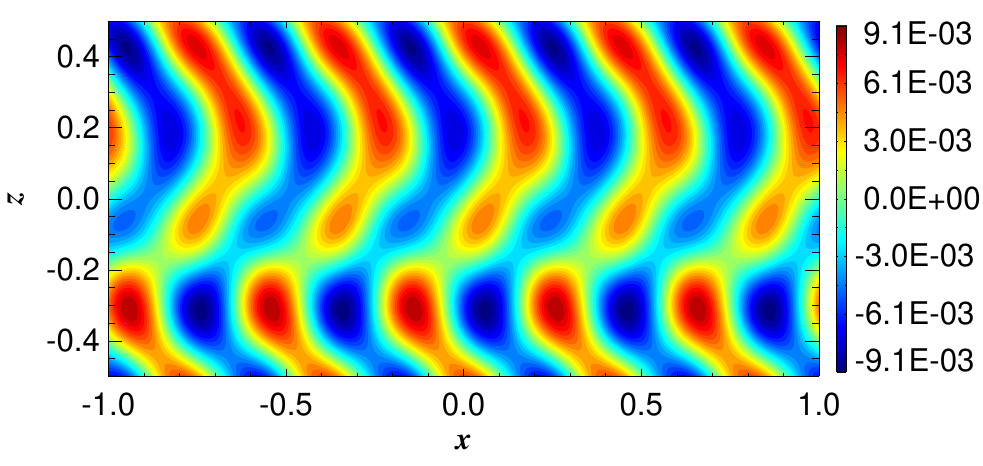}}
\subfloat[$t=26,200$]{\label{fig:WNLua2}\includegraphics[width=0.49\linewidth]{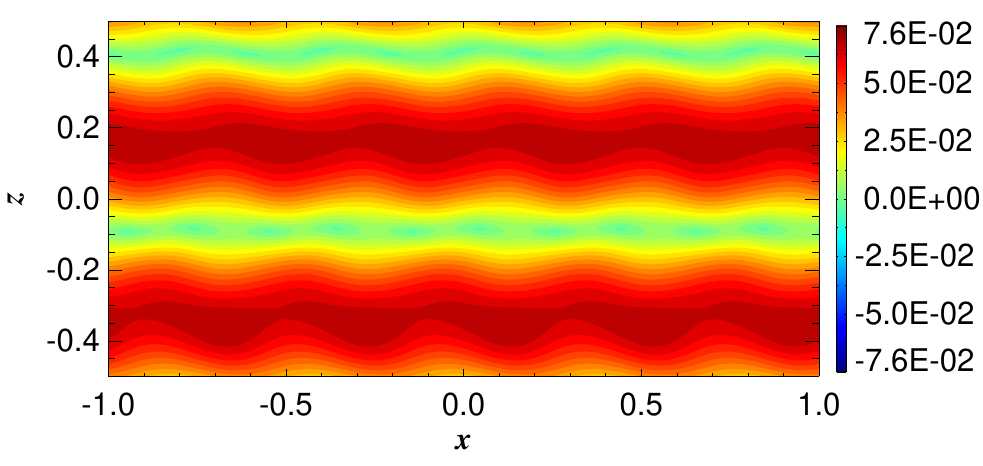}}

\subfloat[$t=26,400$]{\label{fig:WNLuz3}\includegraphics[width=0.49\linewidth]{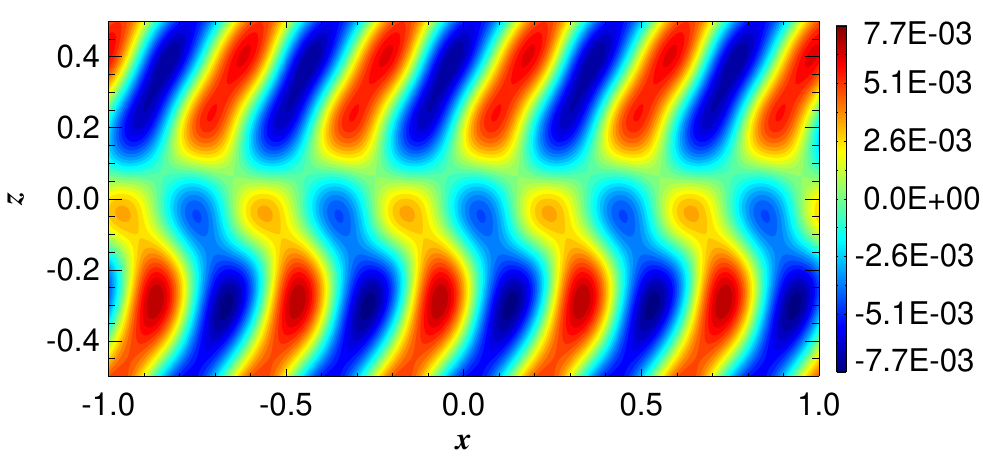}}
\subfloat[$t=26,400$]{\label{fig:WNLua3}\includegraphics[width=0.49\linewidth]{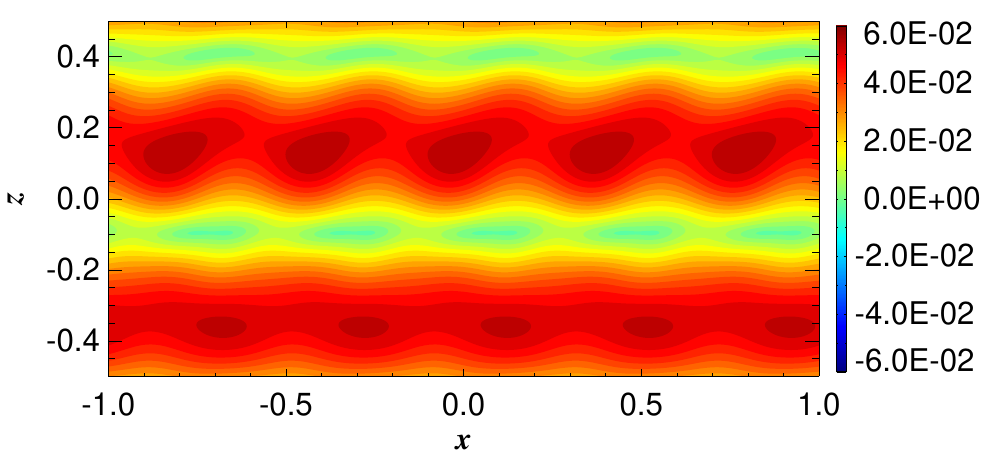}}

\subfloat[$t=26,600$]{\label{fig:WNLuz4}\includegraphics[width=0.49\linewidth]{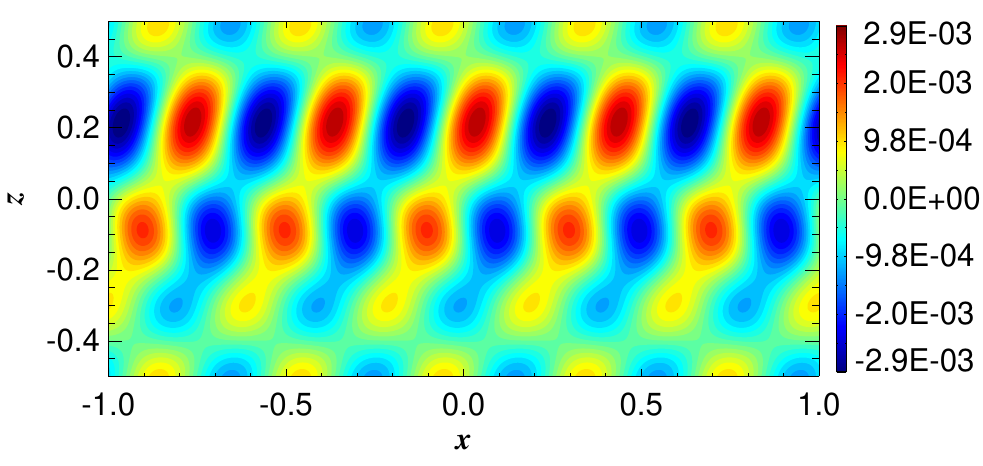}}
\subfloat[$t=26,600$]{\label{fig:WNLua4}\includegraphics[width=0.49\linewidth]{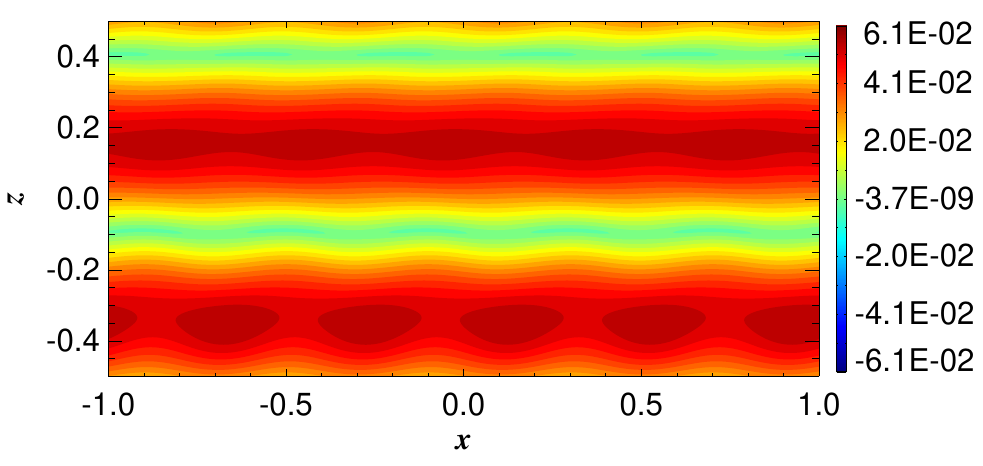}}

\caption{Plots of $u_z$ (left column) and $u_A$ (right columns) in $x$-$z$ space at three times for a run saturating as the weakly nonlinear regime. Simulation parameters are $R=10^{-1.5}$, Pe=$4\pi^2$, Re=$10^{3.75}$.}
\label{fig:WNLuzpanel}
\end{figure}

Our representative case possesses parameters
$R=10^{-1.5}\approx 0.032$, Pe$=4\pi^2\approx 40$, Re$=10^{3.75}\approx 5620$, which means the most unstable COS mode has wavelength equal to the vertical domain size $L$. We find that after saturation the system exhibits two timescales: the short period of the epicyclic oscillation $\sim \Omega^{-1}$ (associated with the primary COS mode), and a much longer timescale $\sim 1000\Omega^{-1}$ (of order the linear growth time of the COS). In order to bring out the longer variation we impose a band-pass filter to remove the shorter epicyclic frequency, and plot the filtered energies
in Fig.~\ref{fig:rmsWNLfilt} on an interval of size $4\times 10^3\Omega^{-1}$, well after the initial saturation and any transients associated with it. 

The first thing to note is that the energy is dominated by the horizontal components, and that the radial energy is roughly four times larger than the azimuthal, as expected from the linear eigenfunction of the dominant COS mode (and indeed any epicycle). The second thing is that the vertical kinetic energy is out of phase with the horizontal. We associate the vertical energy with the higher order inertial waves that attack the primary mode via the parametric instability. As a consequence, we interpret the long-time oscillations as a gentle predator-prey cycle: once the COS mode has attained a sufficiently large amplitude its energy is redistributed to the parasitic modes, which grow and peak shortly afterwards; next, with their source of energy diminished, the parasites' amplitudes decrease because of viscosity, letting the dominant COS rise again. Similar, albeit more violent, predator-prey dynamics characterises the MRI when near criticality (Lesaffre et al.~2009).

To directly verify that the cycles are controlled by the three-wave parametric coupling, in Fig.~\ref{fig:specuA} we plot the time-evolution of the filtered amplitudes of the strongest modes.  These are labelled by their horizontal and vertical wavenumbers $(k_x,\,k_z)$ in units of $2\pi/L$. 
Rather than plot the power in a given velocity component, we employ the horizontal epicyclic speed $u_A=\sqrt{u_x^2+4 (u_y')^2}$ because it screens out the fast oscillations of the primary mode, and partially smooths out the time-series of other modes.

Fig.~\ref{fig:specuA} shows that by far the most energy lies in the leading COS mode $(0,1)$. Next are the modes $(2.5,1)$ and $(2.5,2)$ which possess comparable energies that track each other in time: these are the two parasitic inertial waves in three-wave resonance with the primary. Theory predicts their wavenumbers are $k_x=2.49$ and $k_z=1,\,2$, which cements the identification (L16), and also indicates there is only minor detuning. In addition, there are higher order modes indicative of higher-order couplings.  

In Fig.~\ref{fig:WNLuzpanel} we show snapshots of the flow field at four different times, sampling a portion of a cycle. The vertical velocity and the horizontal epicyclic speed $u_A$ are plotted. The flow is relatively ordered, with the horizontal motion dominated by the primary COS mode structure, though at certain times one can discern shorter scale radial features associated with the resonant modes. The vertical velocity better represents these two modes, and we can see the clear signature of their slanted spatial structures.

It is possible to construct a reduced model that can adequately
describe this regime by deriving coupled evolutionary equations for the
complex amplitudes of the three most dominant modes: the COS (labelled `$A$') and the two parasitic inertial waves (labelled `$B$' and `$C$'). 
This we
undertake in Appendix A, with the final reduced system given by Eqs
\eqref{new1}-\eqref{new3}. Illustrative limit cycle solutions can be
reproduced that match qualitatively the behaviour shown in this
section (middle panels of Fig.~ A2), though the cycles exhibited by the simulations are less extreme, possibly because of the participation of additional modes. One interesting prediction of the reduced model is that when sufficiently detuned the oscillations converge on to a stable fixed point and the long-time cycles disappear (left panels of Fig.~A1). To check this, we ran simulations in a small box with $L_x=L$ thus limiting the radial wavenumber of the two parasitic modes to $k_x=2$, and thus relatively distant to the required 2.49. The simulated system indeed settled on a state exhibiting no long time dynamics.

\subsection{Wave turbulence}

\begin{figure}
\includegraphics[width=\linewidth]{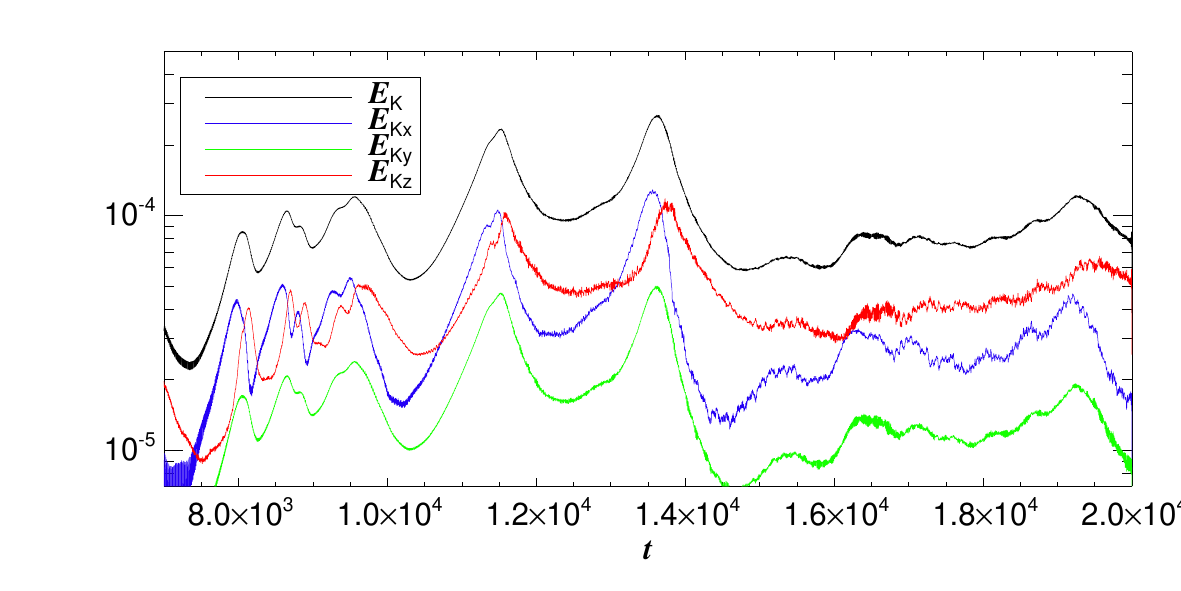}
\caption{Kinetic energies as a function of time in a regime displaying nonlinear waves and turbulence. The time series are filtered so that frequencies larger than $0.1$ are removed. Simulation parameters are $R=0.01$, Pe=$4\pi^2$, Re=$10^{5.25}$.}
\label{fig:rmsNLW}
\end{figure}

\begin{figure}
\subfloat[$u_x$ at $t=9,600$]{\label{fig:WTux1}\includegraphics[width=0.49\linewidth]{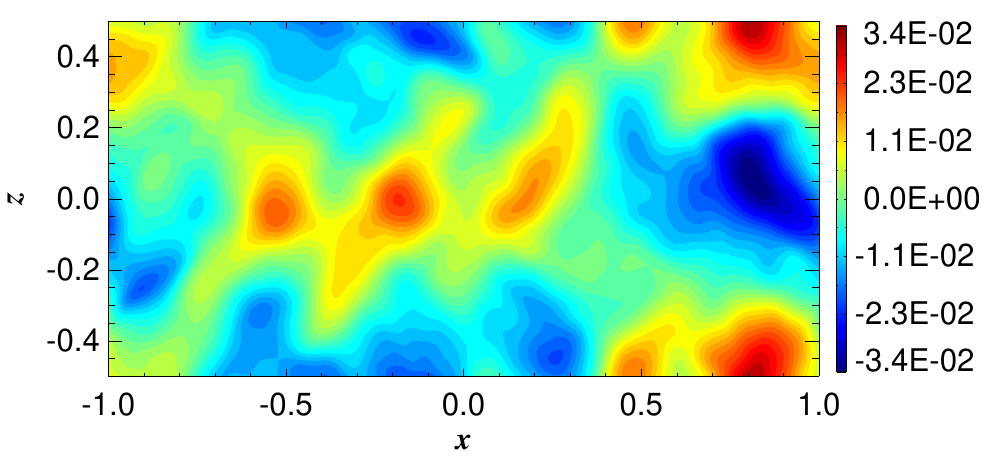}}
\subfloat[$u_x$ at $t=9,880$]{\label{fig:WTux2}\includegraphics[width=0.49\linewidth]{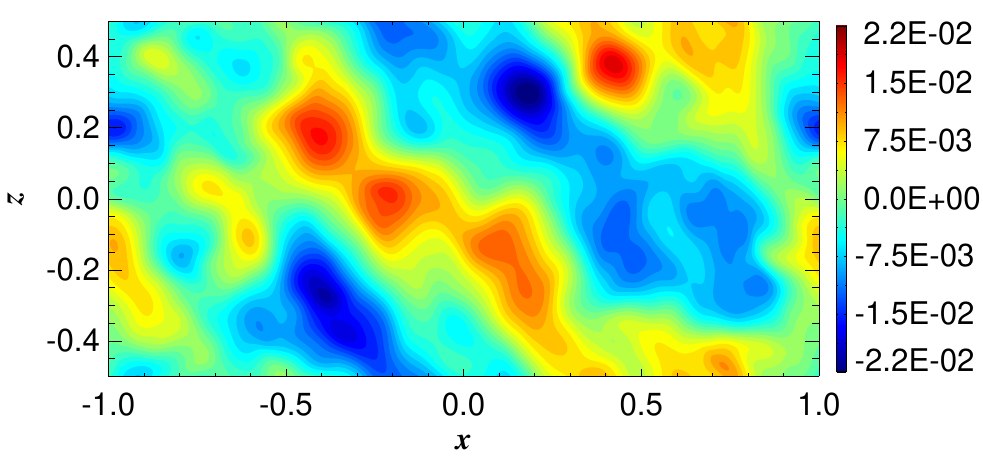}}

\subfloat[$u_y'$ at $t=9,600$]{\label{fig:WTuy1}\includegraphics[width=0.49\linewidth]{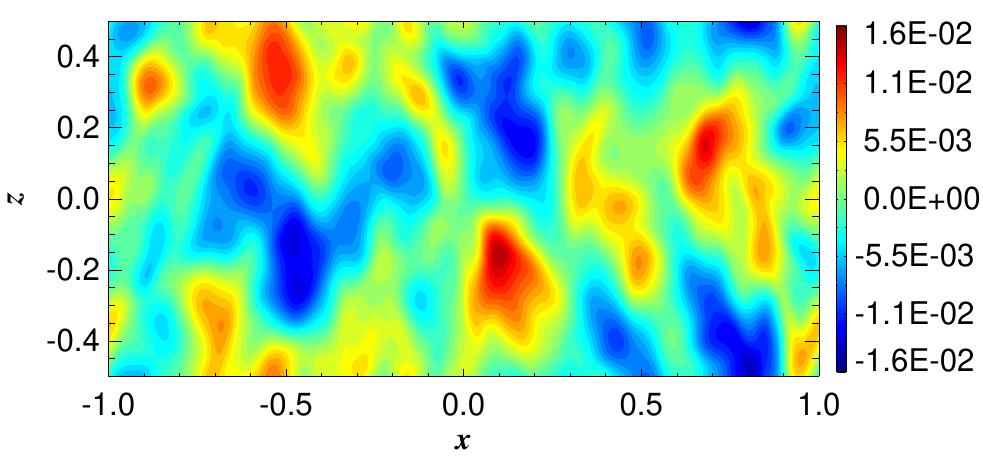}}
\subfloat[$u_y'$ at $t=9,880$]{\label{fig:WTuy2}\includegraphics[width=0.49\linewidth]{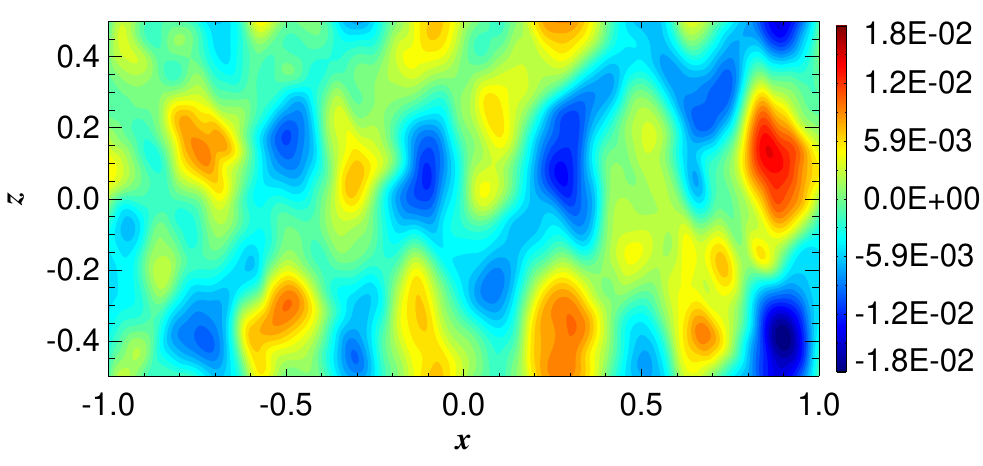}}

\subfloat[$u_z$ at $t=9,600$]{\label{fig:WTuz1}\includegraphics[width=0.49\linewidth]{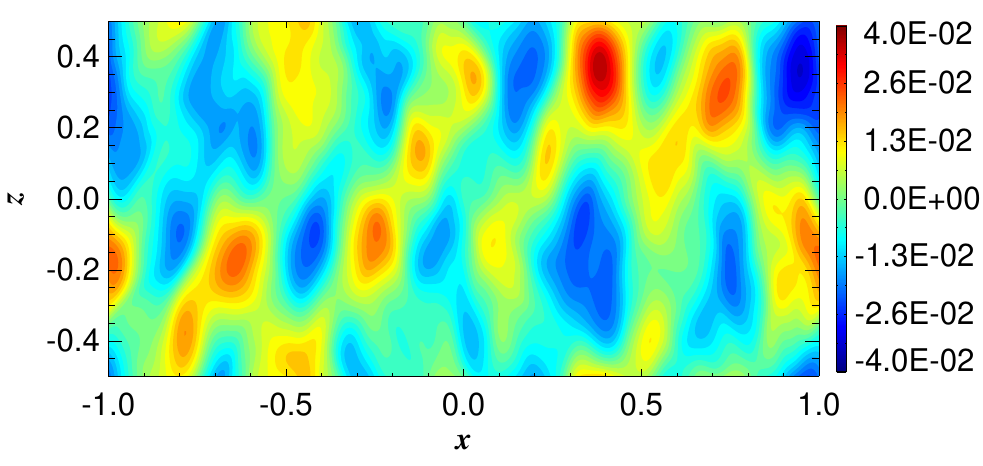}}
\subfloat[$u_z$ at $t=9,880$]{\label{fig:WTuz2}\includegraphics[width=0.49\linewidth]{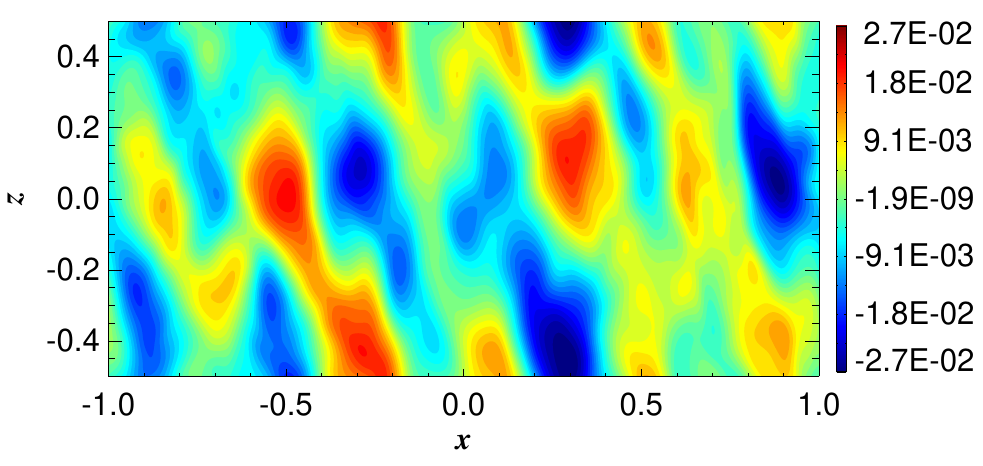}}

\subfloat[$\theta$ at $t=9,600$]{\label{fig:WTth1}\includegraphics[width=0.49\linewidth]{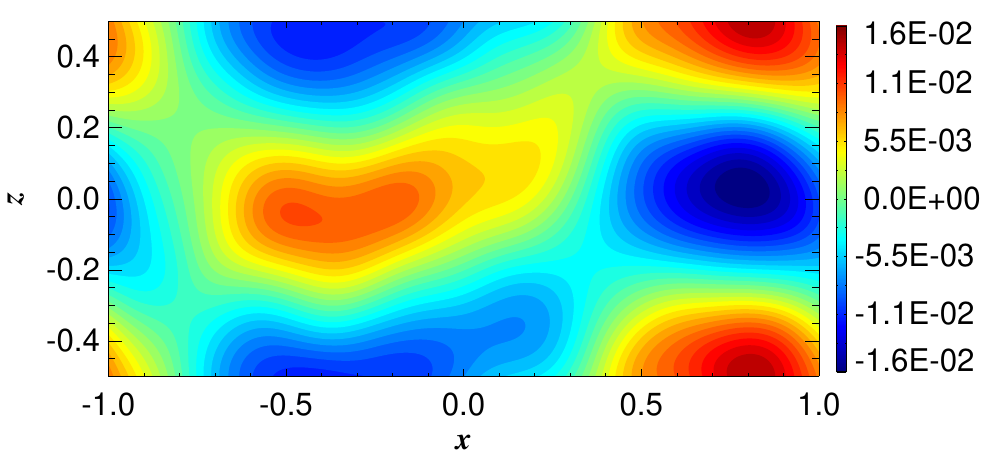}}
\subfloat[$\theta$ at $t=9,880$]{\label{fig:WTth2}\includegraphics[width=0.49\linewidth]{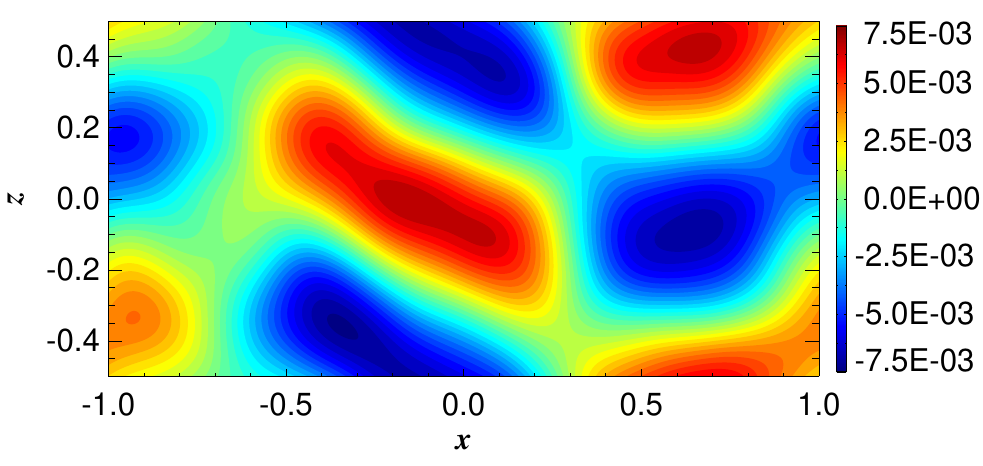}}

\caption{Plots of $u_x$, $u_y'$, $u_z$, and $\theta$ in $x$-$z$ space at two times for a run showing wave turbulence. Simulation parameters are $R=0.01$, Pe=$4\pi^2$, Re=$10^{5.25}$.}
\label{fig:WTpanel}
\end{figure}

\begin{figure}
\includegraphics[width=0.9\linewidth]{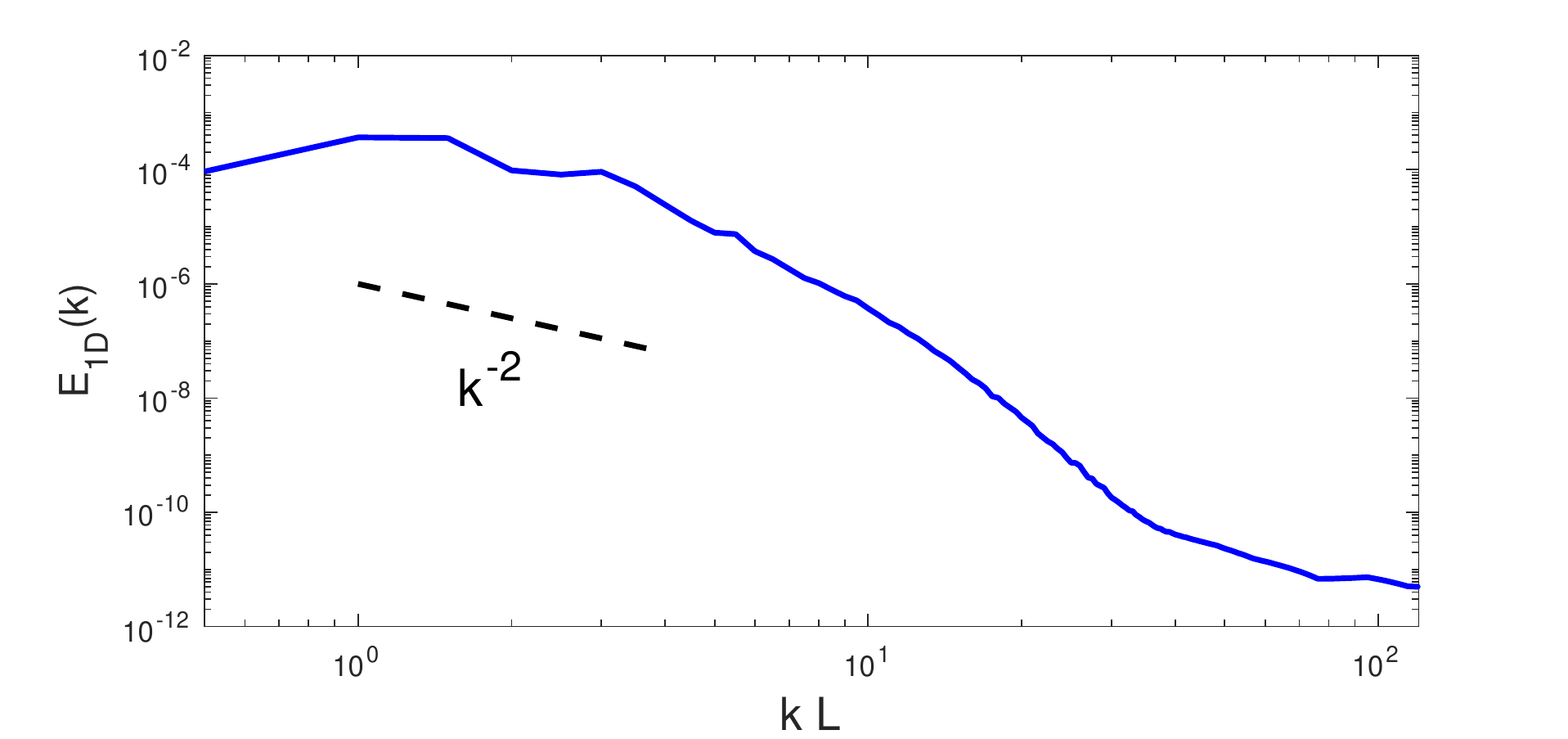}
\caption{Power spectrum of kinetic energy as a function of wavenumber, $k=\sqrt{k_x^2+k_z^2}$ in the wave turbulent regime. Simulation parameters are $R=0.01$, Pe=$4\pi^2$, Re=$10^{5.25}$.}
\label{fig:rmsps1DNLW}
\end{figure}

On either increasing the Reynolds or pseudo-Richardson numbers (Re or $R$), the relatively well-ordered weakly nonlinear state
is replaced by a more chaotic and richer flow field in which more modes participate. This state we term wave turbulence, as it consists of many interacting inertial waves (see, e.g., Nazarenko 2011), the longest driven by the COS while the others redistributing that energy to small scales.

We discuss a
representative run with parameters: $R=10^{-2}$, Pe$=4\pi^2$, and
Re$=10^{5.25}\approx 1.78\times 10^5$. For these parameters the COS driving is  localised to large-scales, and occurs on a range between $\lambda_\text{max}=L$ and $\lambda_\text{crit}\approx (\text{Pr}/R)^{1/4}L\approx 0.35 L$. An estimate for the viscous scale is $\sim \ell_\text{visc}\sim \text{Pr}^{1/2}\lambda_\text{crit}\sim 0.01 L $, and so a hint of an inertial range is possible.

In Figure \ref{fig:rmsNLW} we plot the time evolution of various components of the filtered 
kinetic energy after initial saturation. The system exhibits variability on long time-scales $\sim 1000\Omega^{-1}$ (similar to the linear COS growth time for these parameters)  but, in contrast to the previous subsection, the variation is irregular and exhibits additional frequencies. Nonetheless the total kinetic energy remains $\sim R^2\sim 10^{-4}$, in agreement with the L16 `parasitic theory' of saturation.

In Fig.~\ref{fig:WTpanel} we plot the velocity components and $\theta$ at two
different times. The velocity maps display considerably more disorder than in the weakly nonlinear case, but the signatures of various slanted inertial wave-fronts emerge aperiodically. In contrast, the $\theta$ field is rather structured and its 
variation large-scale. This is because the Peclet number is
low, and the $\theta$ dynamics are dominated by strong
thermal diffusion: any small-scale structures in $\theta$ generated by
the velocities are wiped away rapidly. The $\theta$ spectrum is hence monoscale and is slaved to the dominant COS mode(s), which in these plots is the $k_x=1/2,\,k_z=1$ mode.

To make contact with the theory of wave turbulence we plot the 1D kinetic energy spectrum in Fig.~\ref{fig:rmsps1DNLW}. The system is not especially anisotropic, so this is sufficient for our purposes. We also superimpose the $k^{-2}$ scaling predicted by a Kuznetsov-Zharakov turbulence theory (Galtier 2003, Nazarenko and Schekochihin 2011). As the plot shows, our simulation spectrum does not really follow a straightforward power law, probably on account of too narrow an inertial range (if one exists at all): the viscous scale is close to the input scale and thus steepens the spectrum from what it would be otherwise. As we show in Section 5.4, going to higher Re does not solve the problem, because at higher Re the system begins to develop coherent structure rather than pure wave turbulence.

Finally, though not apparent in the representative case given, wave turbulent states of larger Re and $R$ can generate elevator flows. These correspond to quasi-steady vertically-homogeneous jets in $u_z$, and are exact solutions to our governing equations. Generally these are superimposed on the field of inertial waves and are probably excited by their nonlinear mode couplings, though the details of that process is yet to be determined. Elevator flows are discussed further in the next subsection.


\subsection{Zonal flows}

\begin{figure}
\includegraphics[width=\linewidth]{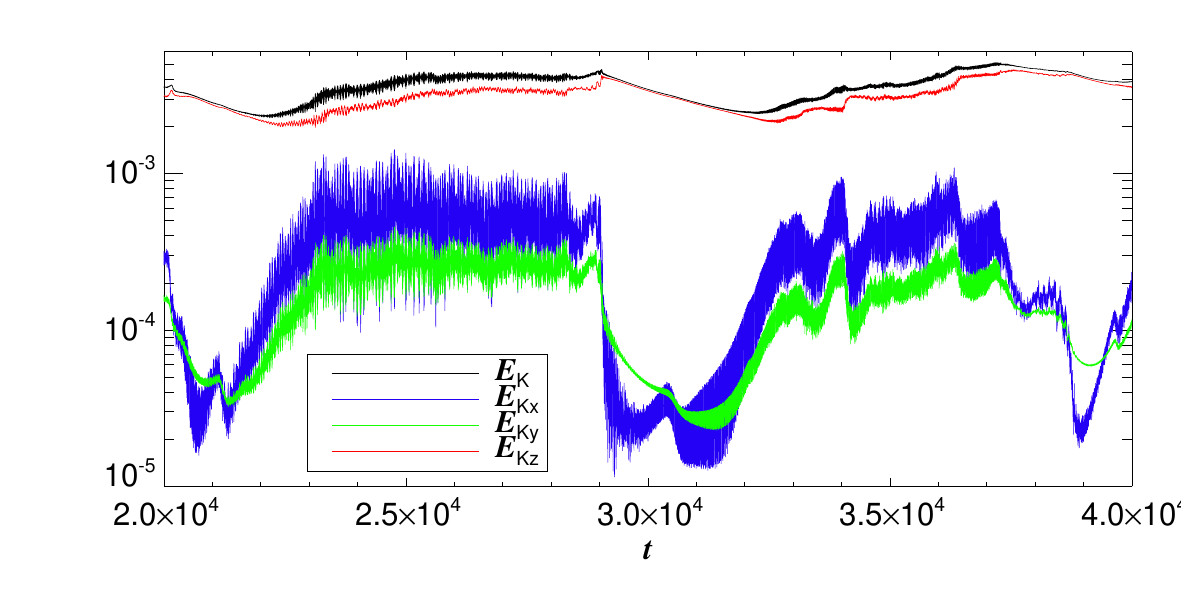}
\caption{Kinetic energies as a function of time in a regime that fluctuates between wave turbulence and zonal flows. Simulation parameters are $R=0.01$, Pe=$4\pi^2$, Re=$10^{5.5}$.}
\label{fig:WTZFfilt}
\end{figure}

\begin{figure}
\subfloat[$u_x$ at $t=30,000$]{\label{fig:WTZFux1}\includegraphics[width=0.49\linewidth]{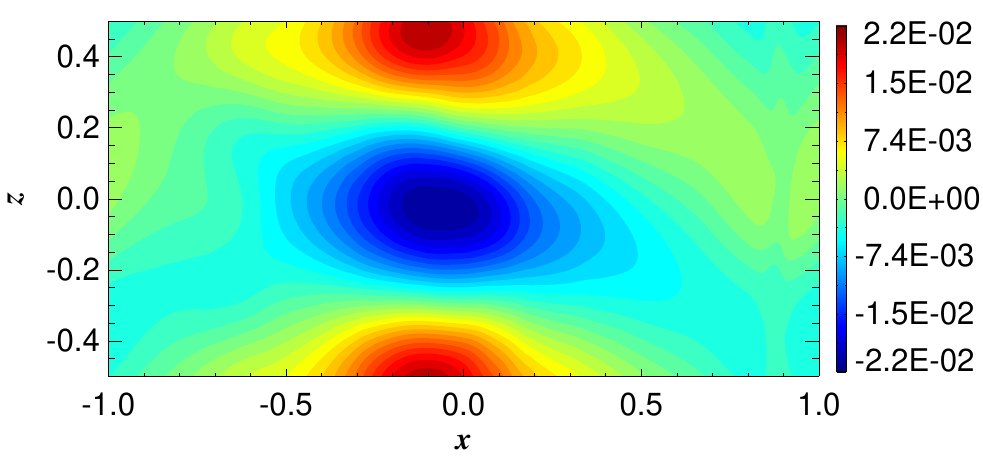}}
\subfloat[$u_x$ at $t=35,000$]{\label{fig:WTZFux2}\includegraphics[width=0.49\linewidth]{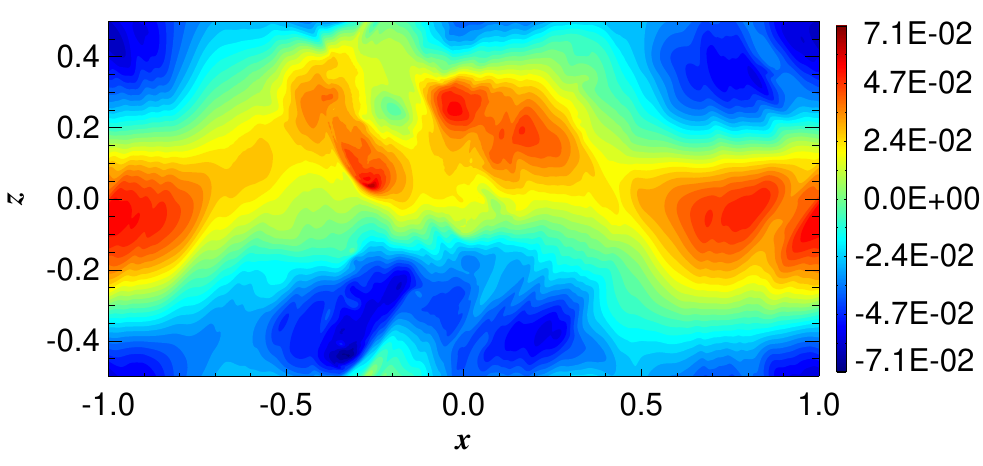}}

\subfloat[$u_y'$ at $t=30,000$]{\label{fig:WTZFuy1}\includegraphics[width=0.49\linewidth]{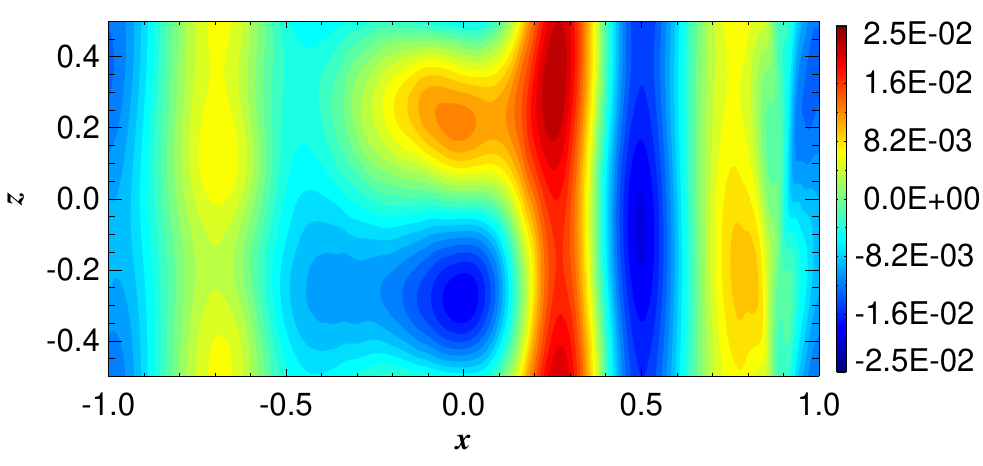}}
\subfloat[$u_y'$ at $t=35,000$]{\label{fig:WTZFuy2}\includegraphics[width=0.49\linewidth]{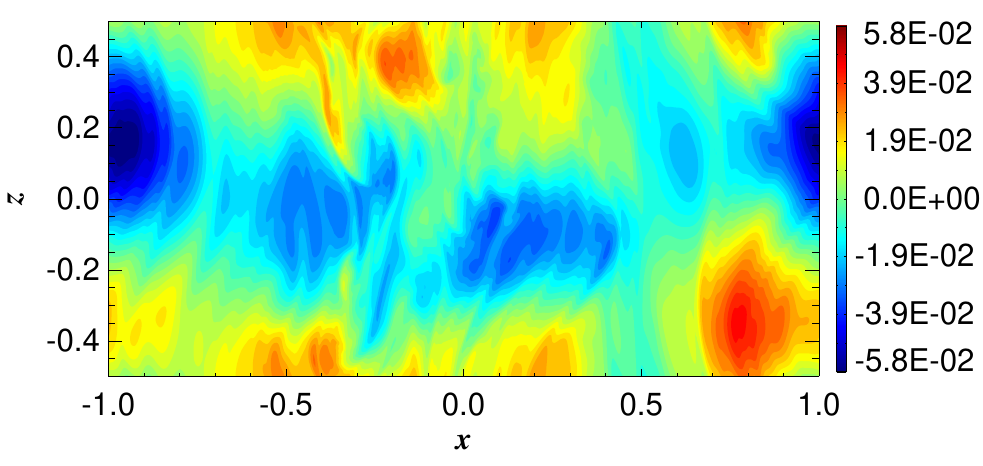}}

\subfloat[$u_z$ at $t=30,000$]{\label{fig:WTZFuz1}\includegraphics[width=0.49\linewidth]{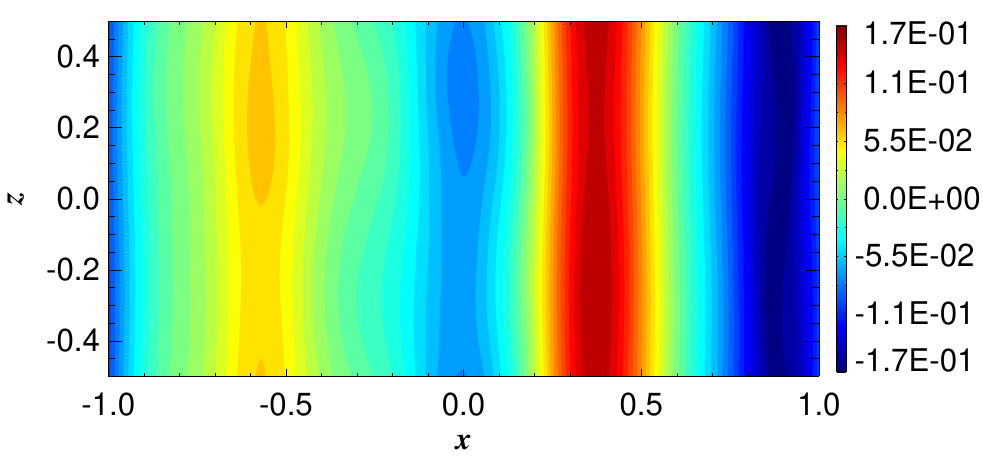}}
\subfloat[$u_z$ at $t=35,000$]{\label{fig:WTZFuz2}\includegraphics[width=0.49\linewidth]{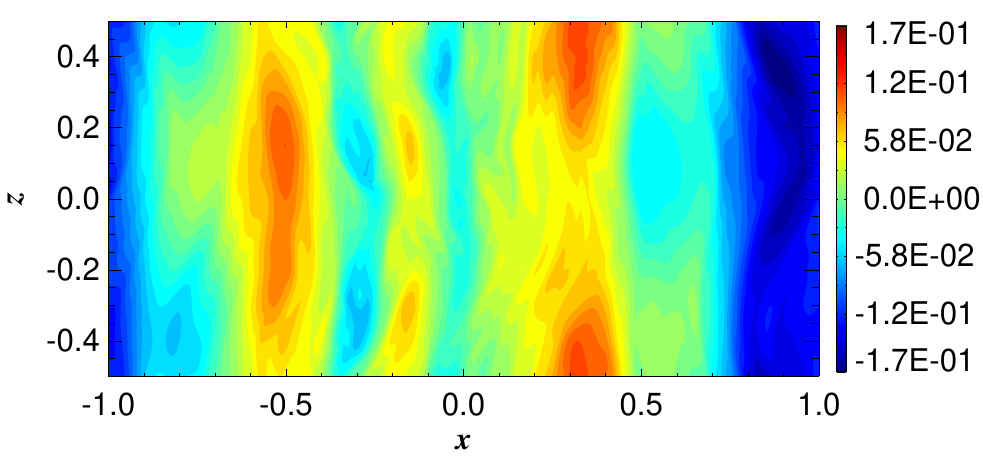}}

\caption{Plots of $u_x$, $u_y'$, and $u_z$ in $x$-$z$ space at two times for a run exhibiting intermittent zonal flows (left panels) that emerge from periods of pure wave turbulence (right panels). Simulation parameters are $R=0.01$, Pe=$4\pi^2$, Re=$10^{5.5}$. }
\label{fig:WTZFpanel}
\end{figure}

\begin{figure}
\includegraphics[width=\linewidth]{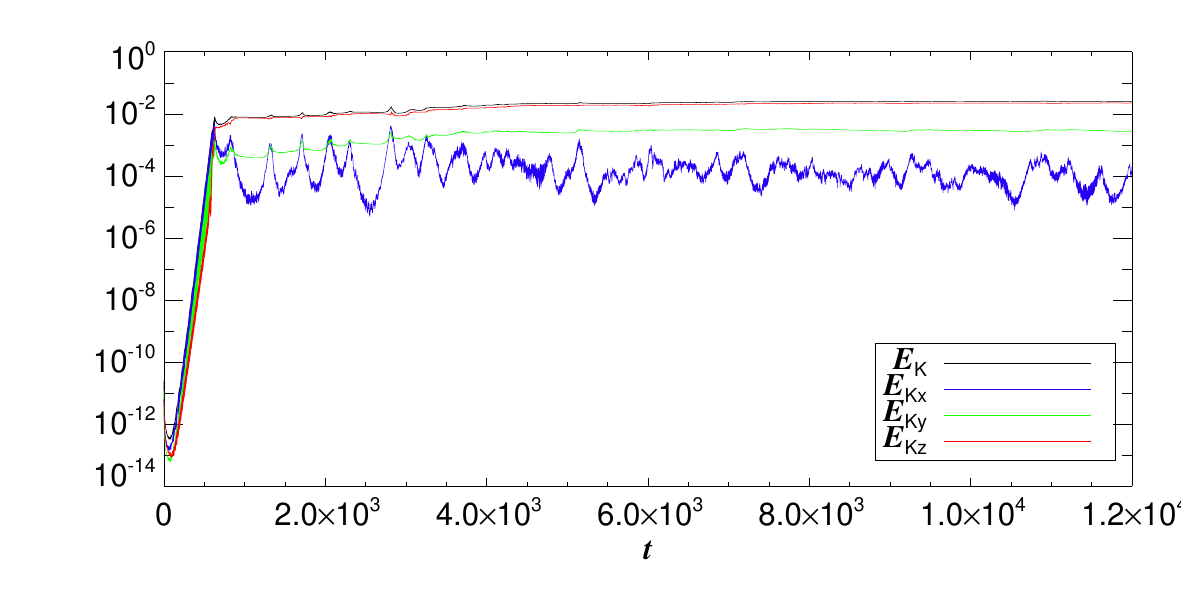}
\caption{Kinetic energies as a function of time in a regime containing strong, persistent zonal flows. Simulation parameters are $R=0.1$, Pe=$4\pi^2$, Re=$10^{5.5}$.}
\label{fig:ZFa}
\end{figure}

\begin{figure}
\subfloat[$u_x$ at $t=9,280$]{\label{fig:ZF1x}\includegraphics[width=0.49\linewidth]{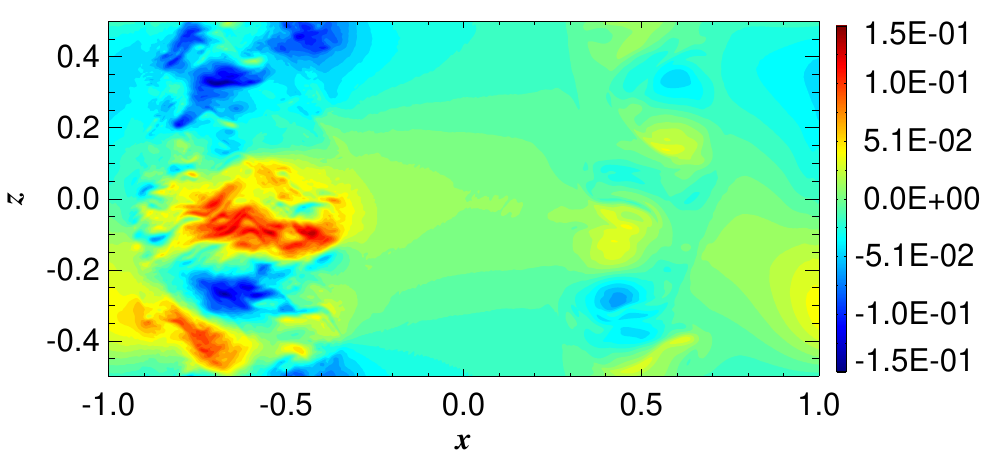}}
\subfloat[$u_x$ at $t=9,800$]{\label{fig:ZF2x}\includegraphics[width=0.49\linewidth]{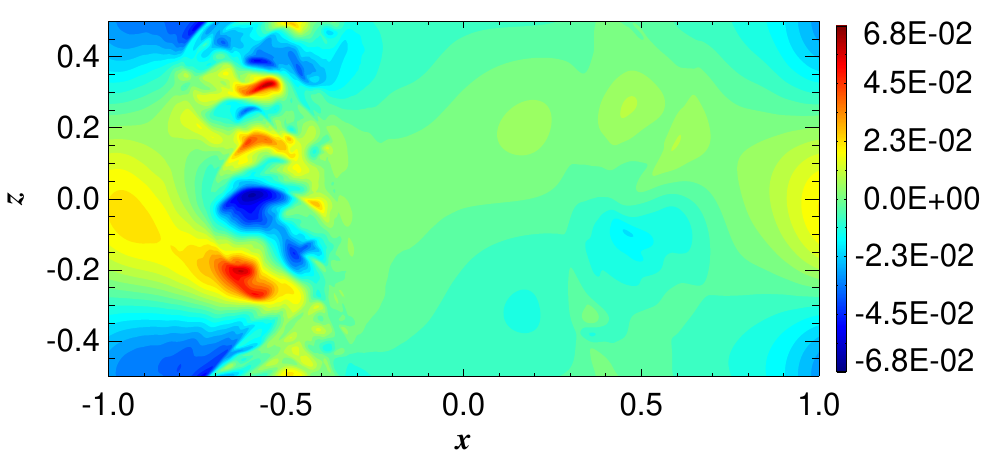}}

\subfloat[$u_y'$ at $t=9,280$]{\label{fig:ZF1}\includegraphics[width=0.49\linewidth]{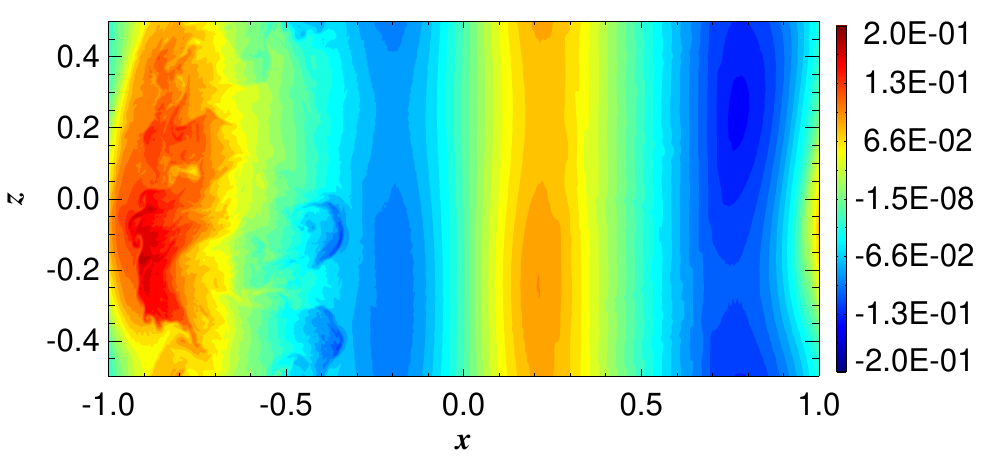}}
\subfloat[$u_y'$ at $t=9,800$]{\label{fig:ZF2}\includegraphics[width=0.49\linewidth]{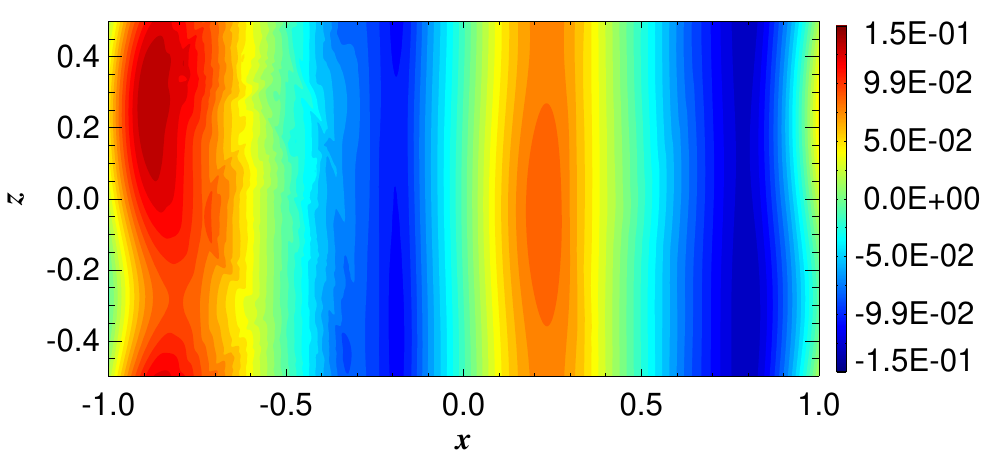}}

\subfloat[$u_z$ at $t=9,280$]{\label{fig:ZF1z}\includegraphics[width=0.49\linewidth]{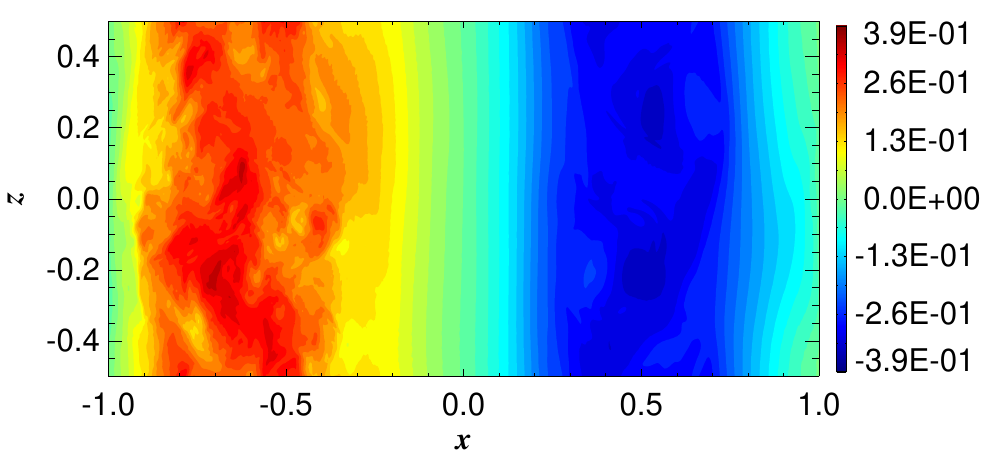}}
\subfloat[$u_z$ at $t=9,800$]{\label{fig:ZF2z}\includegraphics[width=0.49\linewidth]{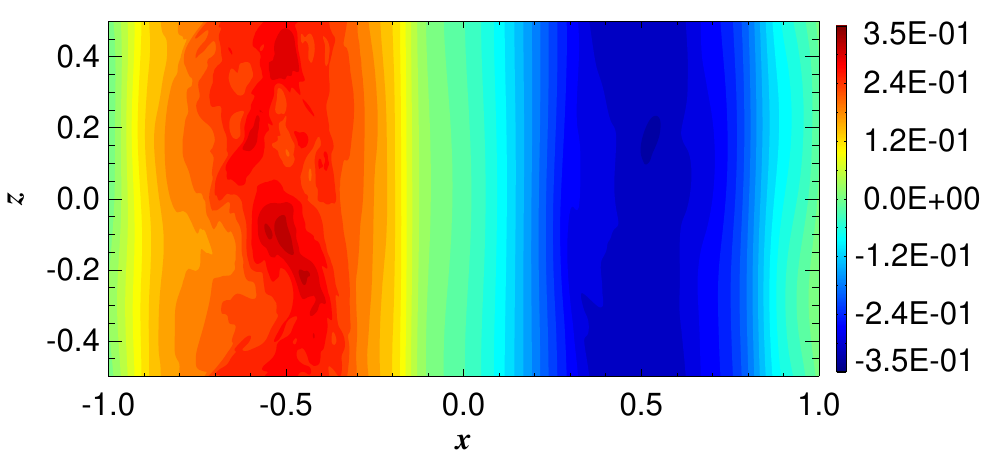}}

\subfloat[$p'$ at $t=9,280$]{\label{fig:ZF1p}\includegraphics[width=0.49\linewidth]{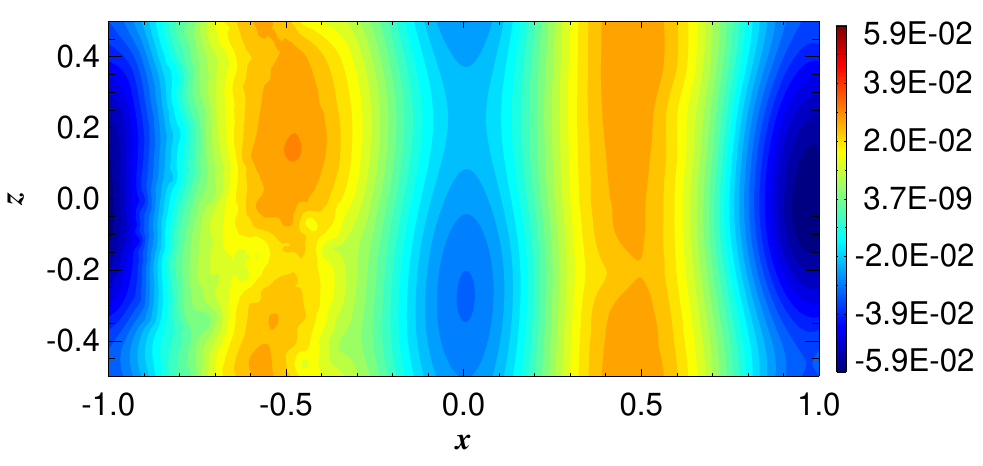}}
\subfloat[$p'$ at $t=9,800$]{\label{fig:ZF2p}\includegraphics[width=0.49\linewidth]{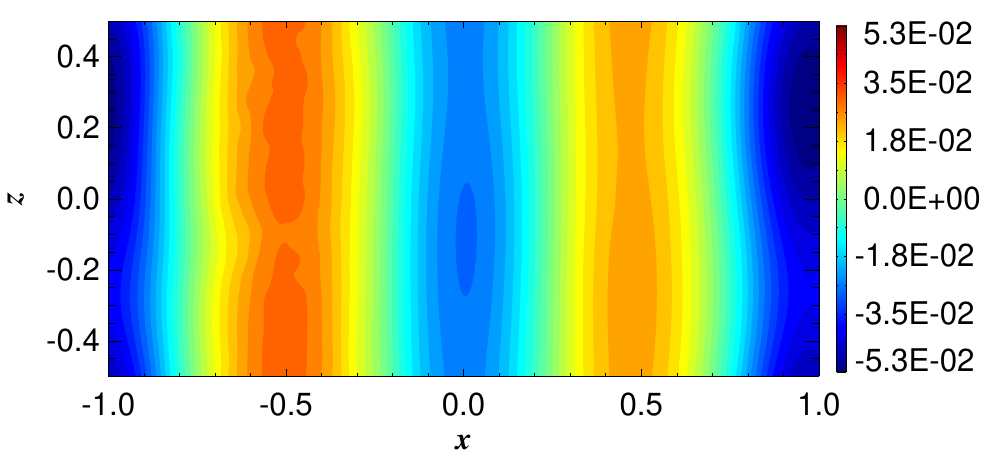}}

\caption{The perturbed velocity components and the perturbed pressure $p'$ in $x$-$z$ space at two times for a run showing persistent zonal flows. Simulation parameters are $R=0.1$, Pe=$4\pi^2$, Re=$10^{5.5}$.}
\label{fig:ZFgeostrophy}
\end{figure}
\begin{figure}
\includegraphics[width=\linewidth]{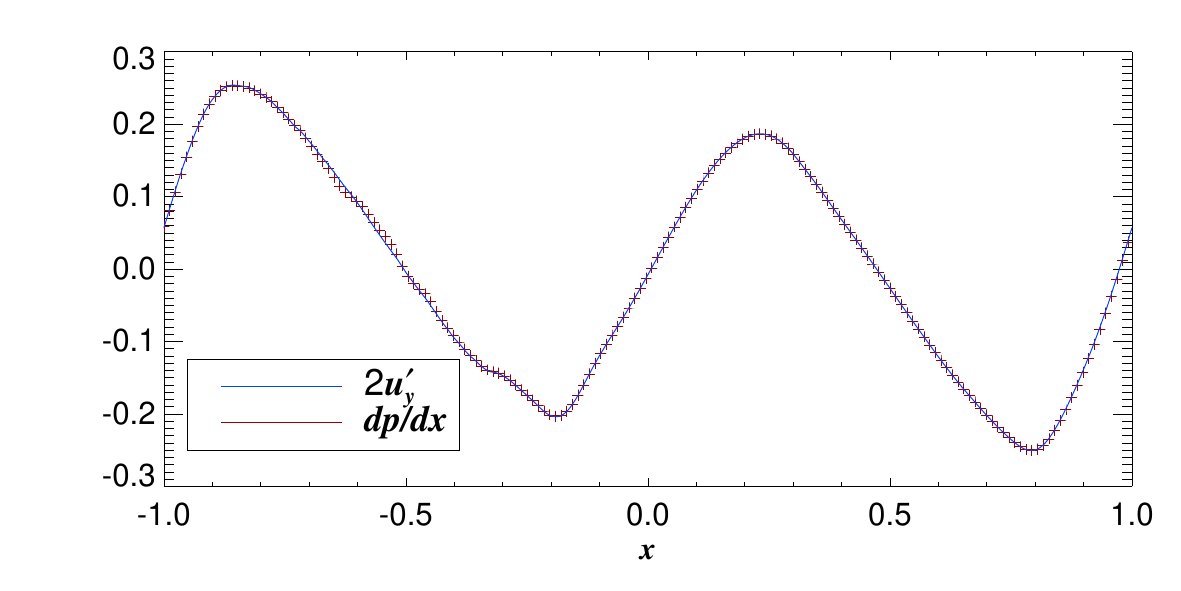}
\caption{$u_y'$ and $dp/dx$ averaged over $z$ at $t=9,800$ (corresponding to the snapshots seen in panels \ref{fig:ZF2} and \ref{fig:ZF2p}
). Simulation parameters are $R=0.1$, Pe=$4\pi^2$, Re=$10^{5.5}$.}
\label{fig:ZFzav}
\end{figure}

As we increase $R$ and/or Re, zonal flows emerge from the wave turbulence, first intermittently, and then persistently at larger values of the parameters. A zonal flow is understood here to consist of a sequence of radially varying (but vertically homogeneous) bands in $u_y'$ and $P$, where the leading order force balance is between the radial pressure gradient and the Coriolis force (`geostrophic balance'). In the accretion disc context, these flows consist of a radial sequence of super-Keplerian and sub-Keplerian motion, and are exact steady solutions of our governing equations. Their excitation mechanism has been briefly discussed in Section 2.3 and we will take this up again in Section 6. 

\subsubsection{Intermittent flows}

Initially the zonal flows emerge periodically from the inertial wave-turbulence in bursts that significantly impede the workings of that turbulence. The period of these bursts can be as long as $10^4\Omega^{-1}$. In Fig.~\ref{fig:WTZFfilt} we plot the filtered energies as a function of time showing two long periods of wave turbulence interrupted by three shorter bursts of zonal flows, in which the kinetic energies actually drop, including the azimuthal kinetic energy. But note that during a turbulent episode $E_{Ky}< E_{Kx}$, while during a zonal-flow episode $E_{Ky}\gtrsim E_{Kx}$. It is also worth pointing out that elevator flows persist throughout, and in fact dominate the kinetic energy budget.

The velocity components are plotted in Fig.~\ref{fig:WTZFpanel} at two times, the first during a zonal-flow burst, and the second during wave turbulence.
In the former case, the system exhibits considerable order and a clear signature of zonal flows in $u_y'$ (the vertically homogeneous bands).
The elevator flows appear in $u_z$, also as vertically homogeneous bands.

The emergence and collapse of zonal flows from inertial wave turbulence has been witnessed in local simulations of eccentric discs by Wienckers and Ogilvie (2018), who model the phenomena in detail with a predator-prey style of dynamical system.\footnote{Models of planetary interiors or atmospheres also develop similar cycles, termed `bursts of convection' (e.g.\ Teed et al., 2012).} Something similar appears to be going on in our simulations, though we only sketch out the main features. The basic cycle consists of (a) the growth of a zonal flow out of the sea of inertial waves driven by the COS (discussed in Section 6), which (b) acts to scatter/impede the leading COS modes, and thus reduce the input of energy from the thermal gradient, and hence the strength of the COS turbulence itself; as a result, (c) the zonal flows are no longer excited/sustained and decay due to residual turbulent motions or viscosity, and (e) the leading COS modes are free to grow once more and instigate inertial wave turbulence, allowing the cycle to repeat. The key difference to eccentric disc simulations is that our zonal flows degrade the primary oscillation (the dominant COS mode) rather than detune the parametric resonance attacking the primary (eccentric) oscillation (which is fixed in the simulations of Wienckers and Ogilvie). 

\subsubsection{Persistent flows}

\begin{figure}
\includegraphics[width=\linewidth]{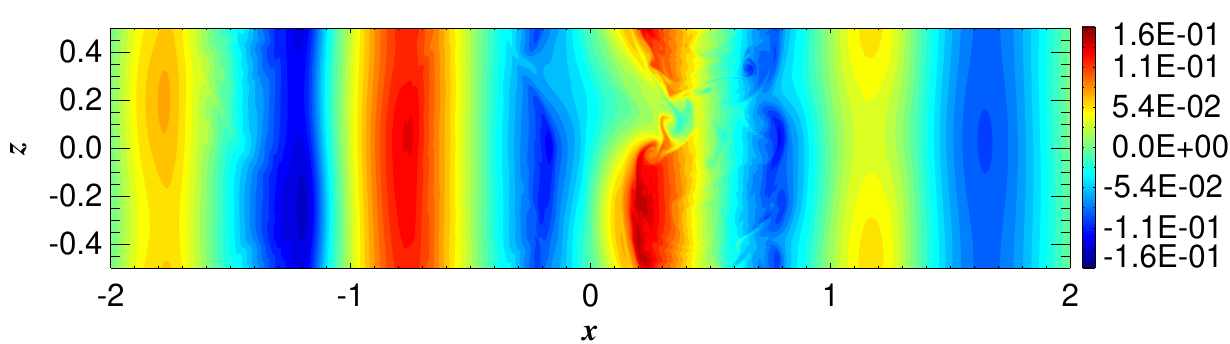}
\caption{Plot of $u_y'$ in $x$-$z$ space for a run with $L_x=4L_z$ that displays persistent zonal flows in its saturated state. Simulation parameters are $R=0.1$, Pe=$4\pi^2$, Re=$2\times10^{5}$.}
\label{fig:ZFarpanel}
\end{figure}

\begin{figure}
\includegraphics[width=\linewidth]{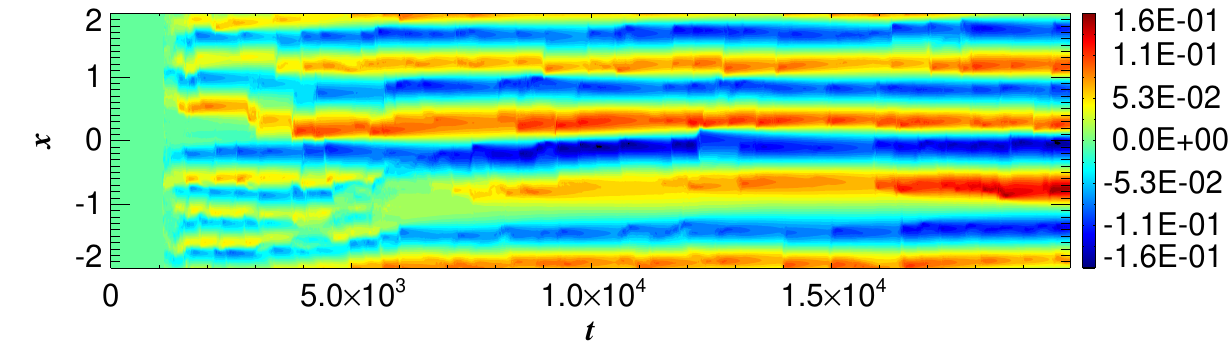}
\caption{Space-time plot of $u_y'$, averaged over $y$ and $z$, in $t$-$x$ showing the evolution of persistent zonal flows. Simulation parameters are $R=0.1$, Pe=$4\pi^2$, Re=$2\times10^{5}$ with $L_x=4L_z$.}
\label{fig:txplots}
\end{figure}

As $R$ and/or Re increase further the zonal flows become stronger and the periods of wave turbulence shorter. Ultimately we pass through a bifurcation and the system achieves a state characterised by a quasi-steady balance between the excitation and degradation of the leading COS mode(s) and of the zonal flow. The time evolution of the energies is plotted for a fiducial example in Fig.~\ref{fig:ZFa}, where now clearly the azimuthal kinetic energy dominates the radial kinetic energy. The latter, however, is not zero and provides evidence of COS activity working away in the background to sustain this state. 

In addition, we plot the velocity components and pressure in Fig.~\ref{fig:ZFgeostrophy} at two different times to indicate the relatively ordered configuration achieved, at least in the dominant $y$ and $z$ velocities. Interestingly, the weaker $x$-component of the velocity tends to localise in certain radial regions: the zonal flow appears to push this inertial wave activity into narrow bands, though it cannot be fully suppressed or else the zonal flows themselves would decay to zero. Similar dynamics is witnessed in semiconvection and the zonal flows/fields supported by the Hall-MRI (Mirouh et al.~2012, Kunz and Lesur 2013).

The quasi-steady nature of the flows allows us to check whether they obey geostrophic balance, as claimed. We take the data from Fig.~\ref{fig:ZFgeostrophy} at $t=9800$, vertically average, and then plot $2u_y'$ and $\partial p/\partial x$ in Fig.~\ref{fig:ZFzav}. As is clear, the two fields lie over each other almost perfectly, verifying that geostrophic balance holds to a very high level of approximation and that these are indeed zonal flows.

The ordered flows exhibit a characteristic radial wavelength that increases over time as the flows gradually merge or disappear. This `inverse cascade' halts once a characteristic lengthscale is achieved; in our $2L\times L$ boxes this lengthscale equals the vertical lengthscale of the fastest growing COS mode $2\pi \sqrt{\xi/\Omega}$. But for our choice of Pe, it also happens to be equal to the vertical box size and half the horizontal box size. As a consequence, one might conclude that the inverse cascade is only halted by the box, and would otherwise continue indefinitely (as in staircase formation in semi-convection; e.g. Rosenblum et al.~2011, Mirouh et al.~2012, Zaussinger and Spruit 2013). 
To check this, we ran a simulation with double the horizontal box size, and show these results in Figs \ref{fig:ZFarpanel} and \ref{fig:txplots}. The latter space-time plot demonstrates the merging of zonal flows between $t=0$ and $t=5\times 10^3$, but also shows that the process stalls from that point on. For the rest of the simulation, the system is in a `glassy' state, with the same wavelength as in the smaller box. This simulation provides some evidence that there is a well-defined characteristic scale of variation that the zonal flows converge towards.

\subsubsection{Secondary instability}

We now discuss the possibility that the elevator and zonal flows are subject to secondary shear instability that could break them up and/or form vortices. We consider the elevator flows first. 

As the vertical velocity exhibits considerable $x$-periodic shear we might expect a form of Kelvin-Helmholtz instability (KHI) via the inflection point theorem. If the elevator flows have a radial wavelength of $\lambda_\text{el}$, then quite generally the instability grows on vertical wavelengths longer than this. Given that $\lambda_\text{el}=L_z$, one might assume that the KHI modes may not fit into our numerical domain. However, this fails to account for the stabilising influence of rotation, which imparts an `elasticity' to the shear flow that resists its deformation. Instability only occurs for sufficiently strong velocities and/or gradients: $V_z \gtrsim \lambda_\text{el}\Omega$, where $V_z$ is the maximum elevator flow amplitude (Latter and Papaloizou 2018). Our simulations never support vertical flows this strong, and thus we do not expect KHI even in larger boxes. We conclude that elevator flows are a robust and unavoidable by-product of the COS in local models.

Zonal flows, on the other hand, are subject to non-axisymmetric instability and encounter no equivalent resistance from the rotation. Generally, if the flow exhibits an extremum in the potential vorticity (or a related quantity), shear instability sets in (Papaloizou and Lin 1985, Papaloizou and Lin 1989, Papaloizou and Savonije 1991, Lovelace et al. 2000). Often it is called `Rossby wave instability', though the link to actual Rossby waves is somewhat tenuous. In any case, our $x$-periodic lattice is surely unstable to non-axisymmetric modes that will generate vortices once they reach sufficient amplitudes. Of course, being axisymmetric our simulations are unable to capture this shear instability. But in a fully three-dimensional set-up we anticipate the zonal flow regime to naturally produce non-axisymmetric structure (as witnessed in previous work; Lyra 2014, Raettig et al.~2021).
In fact, we expect that it is precisely via shear instability that
the COS can break its inherent axisymmetry and give rise to disordered
three-dimensional flow, in particular vortices (which may then be sustained against instability via the subcritical baroclinic instability perhaps; Lesur and Papaloizou 2010). Future simulations will confirm this.


\subsection{Parameter survey}
\label{sec:paramsurv}

\subsubsection{Regimes}
\begin{figure*}
\includegraphics[width=14cm]{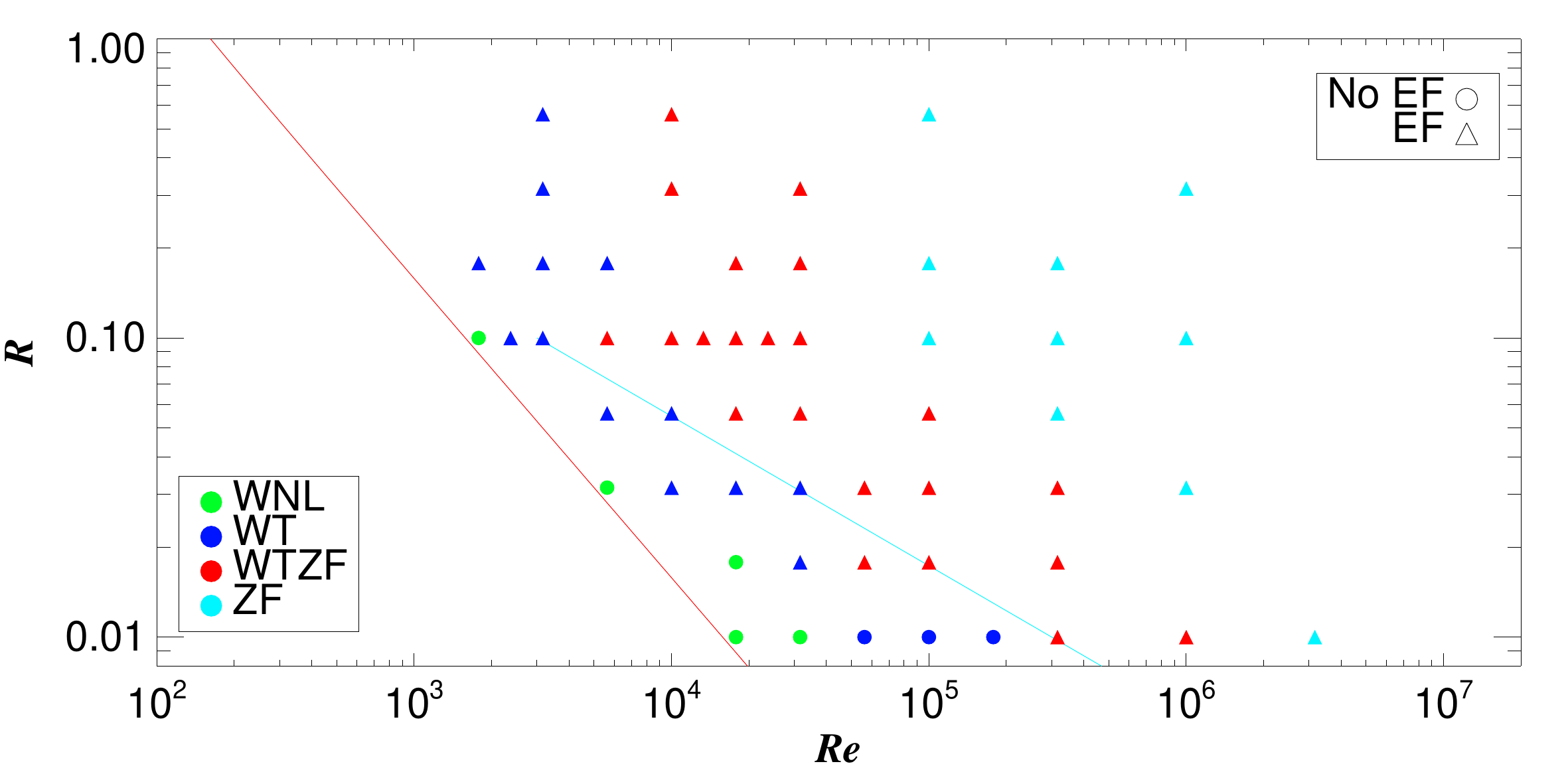}
\caption{Plot depicting the different saturation routes open to the system in Re-$R$ space. The red line indicates marginal COS stability (when $\lambda_\text{crit}=L$). The cyan line loosely denotes the boundary between pure wave turbulence and zonal flows and is given by $R=0.05\times \left(\text{Re}/10^4\right)^{-1/2}$. }
\label{fig:sur}
\end{figure*}

In this subsection we vary both $R$ and Re and plot out the boundaries
between the various saturation outcomes in this two-dimensional
parameter space. The demarcation is often
somewhat loose, with one state `blurring' into the other; nonetheless,
it is possible to construct a relatively reliable plot, which we 
show in Fig.~\ref{fig:sur}. Here the green markers represent the weakly nonlinear state (WNL), the blue markers represent the pure wave turbulent state (WT), the red markers indicate a regime of alternating wave turbulence and zonal flows (WTZF), and finally cyan denotes  persistent zonal flows (ZF). A triangular marker indicates the presence of elevator flows. The red line is the stability boundary for the COS; regions to its left are stable.

The trend in the figure is rather clear, larger $R$ and Re favour zonal flows. A rough boundary between pure wave turbulence and these flows (for sufficiently large Re) is given by the cyan line, $R\propto \text{Re}^{-1/2}$ the form of which we motivate in later sections. In fact, most of the parameter space supports such flows whether they are intermittent or quasi-steady. Realistic PP discs exhibit Re $>10^7$ (according to our definition), and thus would appear to support the steady zonal flow regime unless the thermal driving $R$ is very low. But if $R$ is much less than 0.01 the timescales of the COS become unfeasibly long, and the COS irrelevant. It follows that if the COS features in PP disc dynamics it will always generate zonal flows.

\subsubsection{Fluxes and energies}

Associated with the various states is the transport of angular momentum and heat.
It is important to note that, while inviscid axisymmetric inertial waves are unable to move angular momentum, unstable COS modes can in fact produce a small flux
on account of the small modification to their period by buoyancy accelerations (see Fig.~2 in L16). From linear theory we find that the
flux of angular momentum is
\begin{align}
F_H= 2\text{Re}[u_x' (u_y')^*]\approx -|u_x'|^2 (\text{Re}(s)/\Omega),     
\end{align}
where $u_x'$ and $u_y'$ are the linear velocity perturbations, and $s$ is the real part of 
the COS growth rate for that particular mode. As is clear the flux is non-zero and \emph{negative}. 

\begin{figure}
\includegraphics[width=8cm]{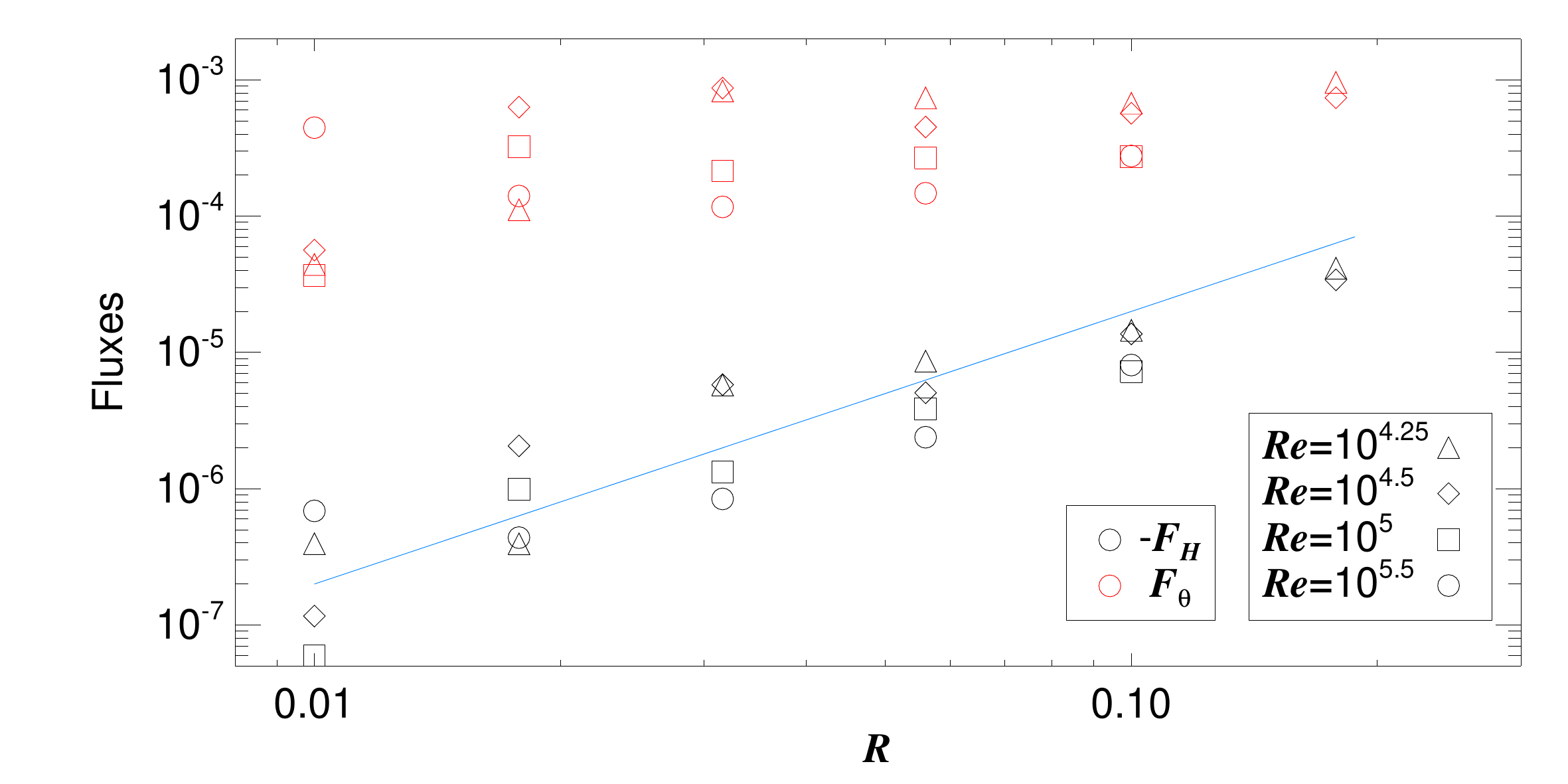}
\caption{Turbulent fluxes, $F_H=\langle u_xu_y'\rangle$ and $F_\theta=\langle u_x\theta\rangle$, as functions of $R$ for Pe$=4\pi^2$. The blue fitting line is given by $2\times 10^{-7}(R/0.01)^2$ }
\label{fig:tf}
\end{figure}

We plot the mean thermal and angular momentum fluxes, $F_\theta$
and $F_H$  in
Fig.~\ref{fig:tf}. The angular momentum flux remains small and
negative, and apart from its role in causing the emergence of zonal flows
(cf.\ Section 6), its impact on the large-scale evolution of
the disc should be minimal. The thermal flux is outward and is much larger. It
acts to equalise the unstable gradient that gave rise to the COS, and takes values, in our units, as high as $\sim 10^{-3}$. However, it is never on par with the radiative diffusive flux, which may be estimated to be $\sim \text{Pe}^{-1}\sim 2.5\times 10^{-2}$ (for an order one background variation in entropy varying on the box scale). 
Both fluxes increase in magnitude with $R$, as the system moves through the different states. Putting to one side variations associated with these transitions, the angular momentum flux appears to follow the rough scaling $F_H = -2\times 10^{-7}(R/0.01)^2 $, in our units. 

\begin{figure}
\includegraphics[width=8cm]{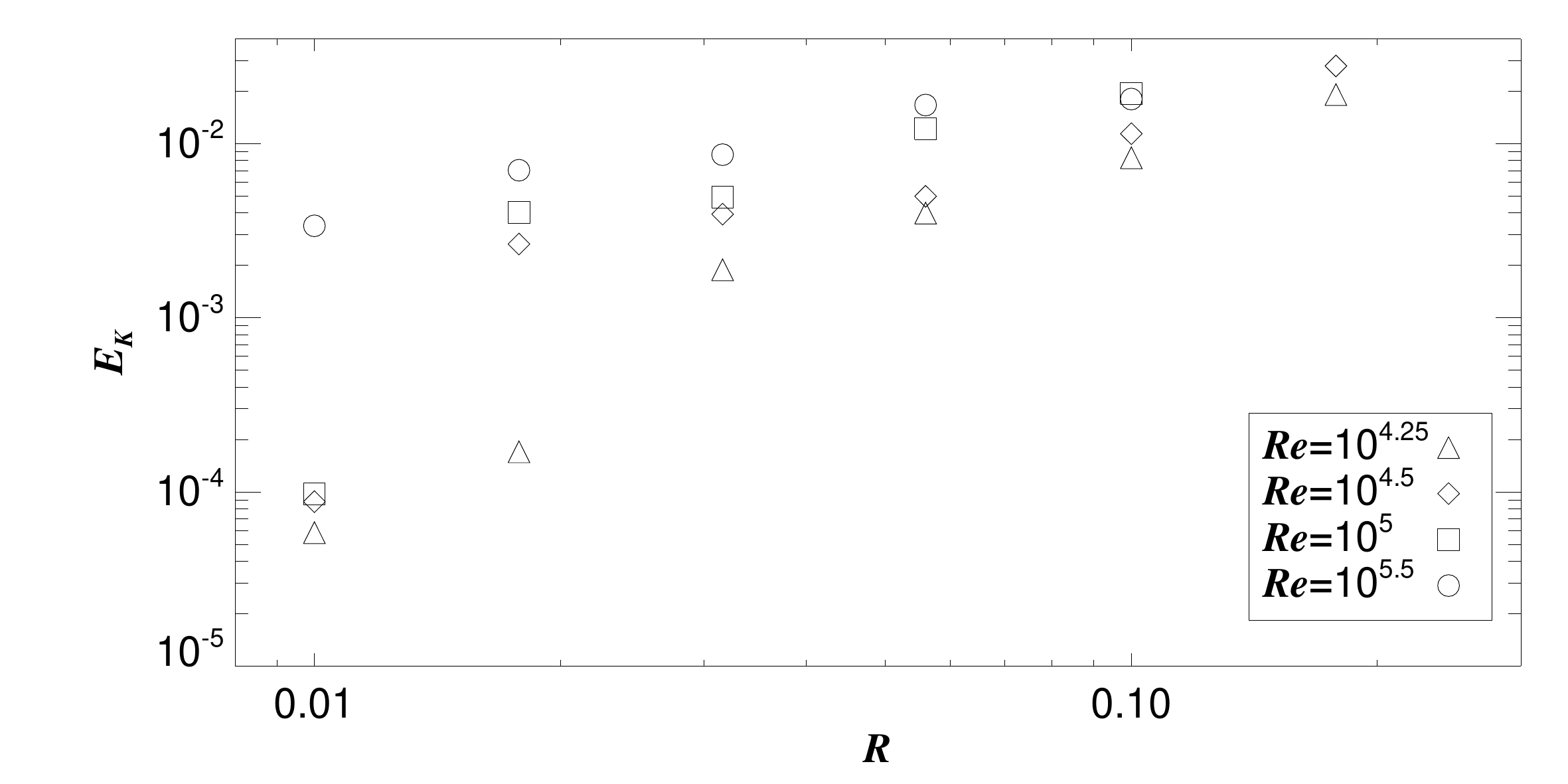}
\caption{Kinetic energy averaged in space and time (during saturated state) as a function of $R$ for Pe$=4\pi^2$, for selected simulations.} 
\label{fig:ke}
\end{figure}

For completeness we plot the kinetic energies in Fig.~\ref{fig:ke}. As the thermal driving increases so does the energy, though this is more pronounced for lower Re, mainly because these runs exhibit more state transitions. The WNL and WT states possess significantly weaker energies than those states that exhibit elevator and zonal flows. Velocities in the latter states can be an order of magnitude or larger, and thus easily violate the estimate in Eq.~(35) in L16 derived from the `parasitic' theory of saturation. Finally, we point out that the jumps in energy as one progresses to these ordered `layered' states is also a notable feature in semiconvection.  

\section{Physical model for layer formation}

In this section we expound a physically intuitive theory for why zonal flows might develop in COS turbulence. It is well established that 3D rotating turbulence manifests large-scale coherent structures via an inverse cascade (e.g. Cambon and Joaquin 1989, Minnini et al. 2009), but our simulations are not 3D, and the details of the nonlinear transfer amongst the wave modes are difficult to unpick (but see Smith and Waleffe, 1998 Kerswell 1999, Le Reun et al.~2020). 
Here, we sketch out some principles and arguments that help with our physical intuition. These ideas are fleshed out with a mean field model similar in spirit to Radko 2003, Rosenblum et al. (2011), and Mirouh et al. (2012), the details of which we package away in Appendix B. The predictions of this model we confront with our numerical simulations at the end of the section. 

\subsection{Basic principles: competing gradients and fluxes}

Ours is a story of two opposed gradients. There is an outwardly decreasing entropy gradient ($N^2<0$), which is destabilising, and an outwardly increasing angular momentum gradient ($\kappa^2>0$), which is stabilising. 
The COS uses thermal diffusion to circumvent the angular momentum gradient in order to grow, and yet it is still sensitive to $\kappa^2$ because its (maximum) growth rate is proportional to the ratio $-N^2/\kappa^2=R$. While, it is clear that a larger $|N|$ exacerbates instability (because the the entropy gradient is the source of free energy), interestingly a steeper angular momentum gradient (larger $\kappa^2$) inhibits instability, and conversely, a flatter gradient (smaller $\kappa^2$) enhances it\footnote{This point could be rephrased in terms of `wave elasticity'. See McIntyre and Dritschel (2008)}. In fact, in the limit of a constant angular momentum disc ($\kappa=0$), the COS ceases to be an overstability, as epicycles and inertial waves vanish; we then obtain standard convection growing at the significantly \emph{larger} rate of $|N|$ (albeit modified by the thermal diffusion). In summary, the angular momentum gradient, while not precluding instability, does get in the way; it obliges unstable modes to undergo oscillatory motion, which are unnecessary to the transport of heat. If these superfluous motions are minimised, instability works better.

These concepts can be extended beyond linear theory if we consider how the ensuing turbulence transports both $\theta$ and angular momentum locally. By virtue of this transport, small but significant variations in the local distribution of both entropy and angular momentum can develop. These variations can then feed back on the driving of the instability in the nonlinear regime.

 As we have seen in Section 5.5, and in Figure \ref{fig:tf}, the turbulent heat flux generated by the COS is outward, and thus augments (slightly) the laminar diffusive flux. This is as expected: the COS is trying to eliminate the unstable state from which it arose; by mixing entropy, $|N|$ can be reduced. If we define an effective radially and temporally varying $R$, then locally this $R$ will decrease and the linear driving will  weaken.

 But we also observe a small \emph{inward} transport of angular momentum. Because angular momentum increases outward, the turbulence works to flatten this gradient: thus reducing $\kappa^2$. Consequently, the locally varying $R$ will increase in magnitude. This transport, though weak, \emph{exacerbates} instability (the COS growth rate is $\propto |N^2|/\kappa$). Though not a strong effect in relative terms, we shall find it is critically important for the local distribution of angular momentum - and may lead to anti-diffusion.

 \subsection{An anti-diffusive angular momentum flux}
 
Let us work through the consequences of these basic ideas in a wave turbulent state. Suppose that this turbulent state supports an inward turbulent flux of angular momentum $F=-\langle u_x u_y'\rangle$,  with $F>0$, and that viscous diffusion may be ignored for the moment. Let us assume, reasonably, that this flux $F$ depends locally on the strength of the COS turbulence at that location: the stronger the turbulence the more transport takes place. Next, suppose that this turbulent strength is an increasing function of the local $R$ parameter: thus we may write $F=F(R)$, and $dF/dR>0$. (Figure \ref{fig:tf} certainly helps justify these statements.) Finally, let us define the local effective $R$ parameter as a ratio of the total entropy gradient and the total angular momentum gradient:
\begin{equation}
R_\text{eff}= \frac{N^2 \d_x\theta_x}{2\Omega\d_x h},
\end{equation}
which returns to the constant $R=-N^2/\kappa^2$ in the unperturbed state. The local COS intensity will be tuned to the magnitude of this locally varying $R$.

 \begin{figure}
\center
\includegraphics[width=0.4\textwidth]{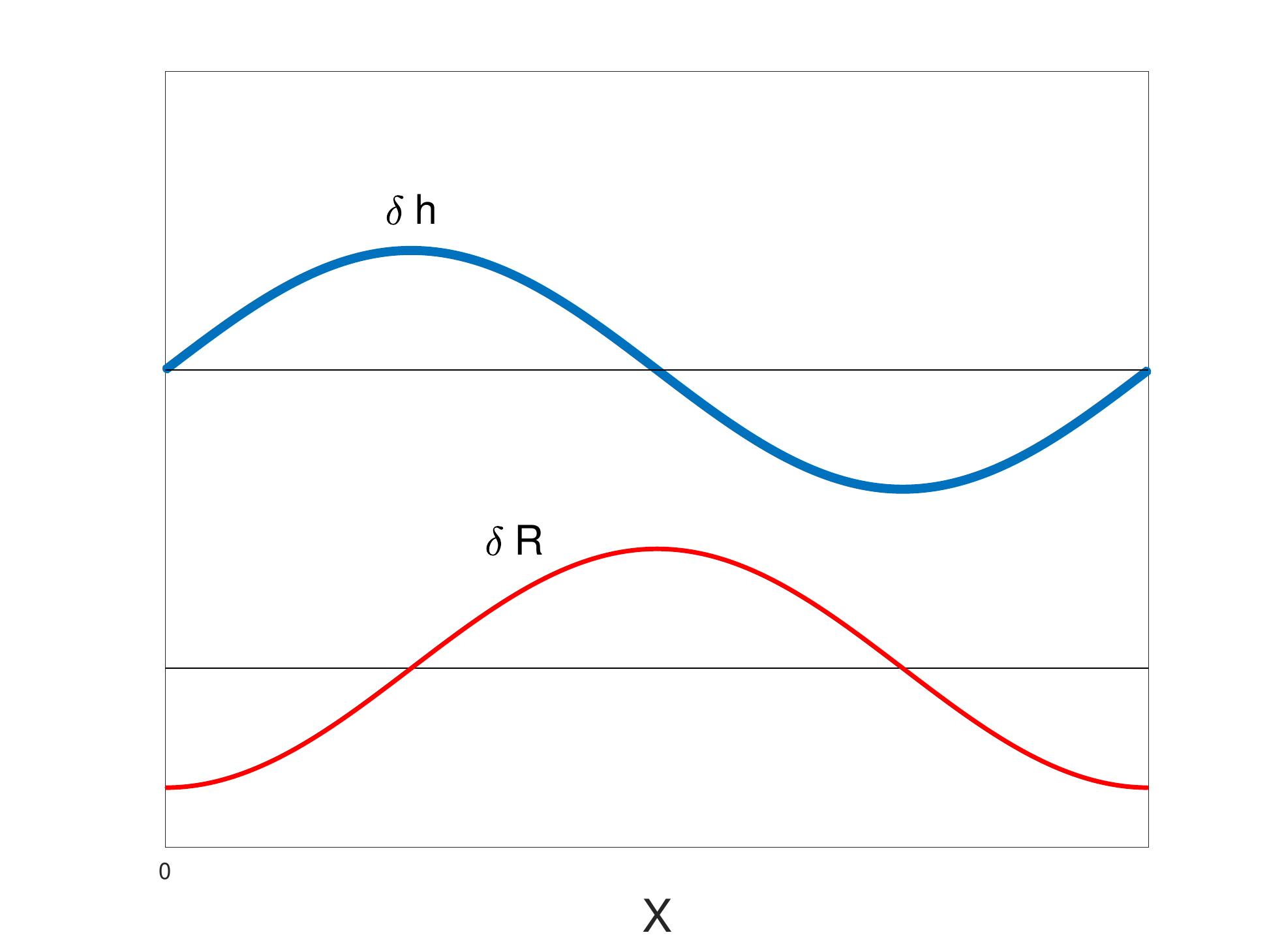}
\includegraphics[width=0.4\textwidth]{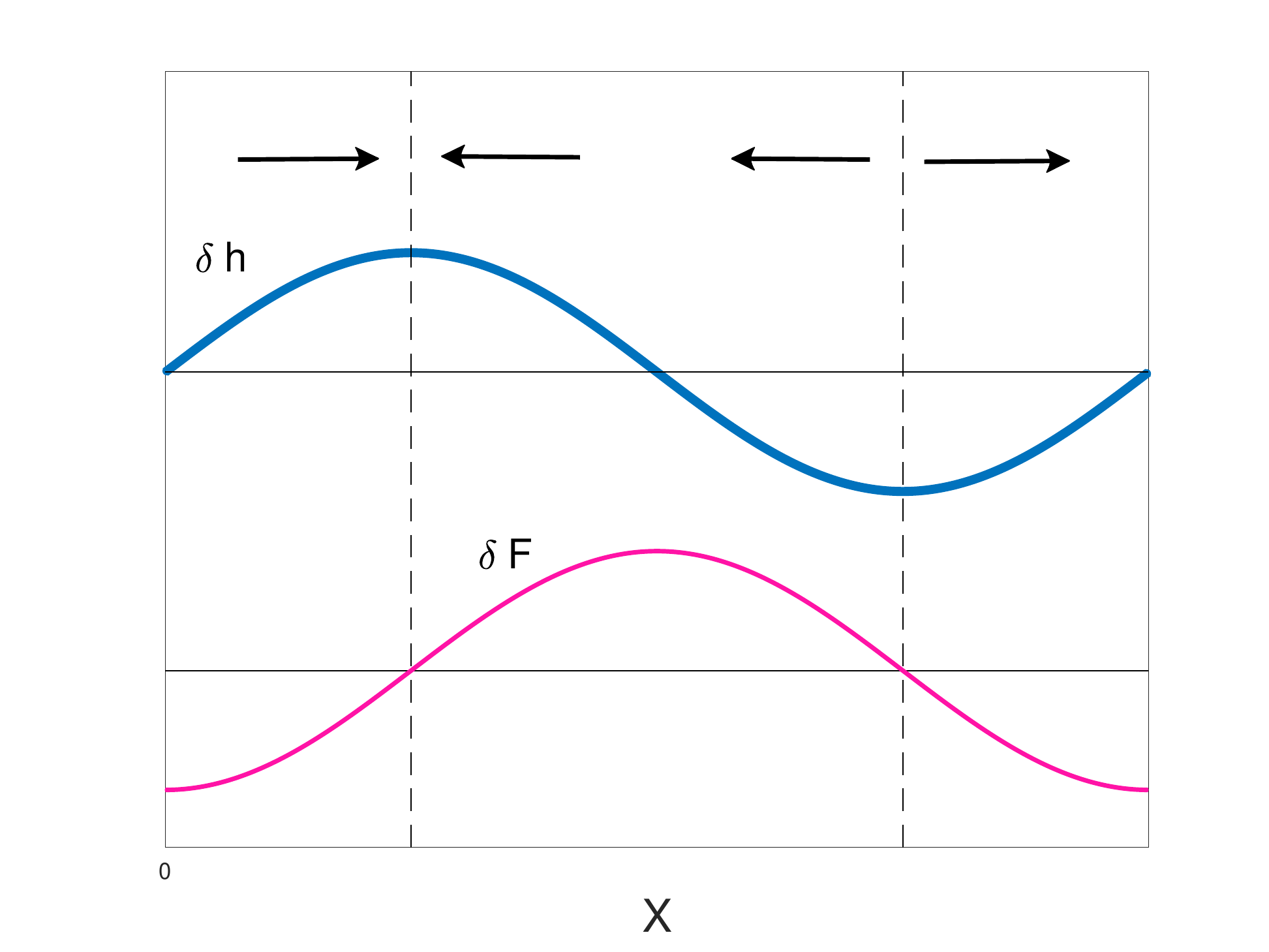}
\caption{Sketch of how zonal flows may develop in COS turbulence. See text for an explanation. }
\label{sketch}
\end{figure}
 
 Suppose there is a large-scale sinusoidal $x$-dependent perturbation to the background angular momentum $h_0(x)$. We call it $\delta h$ and plot it in the top panel of Fig.~\ref{sketch}. Because of the sinusoidal shape of $\delta h$, the total angular momentum gradient is slightly flatter in the middle of the box, and slightly steeper at the edges of the box. As a consequence of this flattening and sharpening, the disturbance drives a perturbation in the local $R_\text{eff}$ parameter, which we denote by $\delta R$. It will be $\pi/2$ out of phase with $\delta h$. Next, if the turbulent intensity, and hence any turbulent flux, depends on $R_\text{eff}$, the perturbation in the inward flux of angular momentum, denoted $\delta F$ will be correlated with $\delta R$, as shown in the bottom panel of Fig.~\ref{sketch}. Now radial regions in which $\delta F$ is positive experience an additional momentum flux, and where it is negative a smaller momentum flux: the arrows in the bottom panel indicate the direction of the perturbed flux (note that the total flux will be inward for all $x$). As is clear, this means that peaks of $\delta h$ will increase, and troughs decrease, leading to a runaway process, and the development of radial layers of angular momentum: i.e.\ zonal flows. 
 
 Working against this tendency is viscous diffusion, which, given the weakness of the angular momentum flux, is not necessarily negligible. Hence any criterion of zonal formation must tension the perturbed turbulent angular momentum flux against the perturbed viscous flux. If we neglect buoyancy perturbations, the former can be written as $-(dF/dR)\delta R= 2(\Omega/\kappa^2)(RdF/dR)(\d_x \delta h)$, while the latter is $-\nu\d_x\delta h$. Clearly, the turbulent flux overpowers the viscous flux when
 \begin{align}\label{ZFcriterion}
 \frac{d F}{d\ln R} > \frac{\kappa^2\nu}{2\Omega},
\end{align}
which is our criterion for zonal flow formation.
 In principle, accompanying variations in the turbulent flux of entropy works against this anti-diffusion and should also enter the criterion, but the effect is small when the
 laminar radiative flux dominates the turbulent flux, as is the case in our simulations. In Appendix B we construct a mean field model that accounts for this additional physics, and puts some of the ideas in this section on a more mathematical footing. (In the notation of Appendix B, $F=-F_H$.)

\subsection{Comparison with simulations}



We next apply criterion \eqref{ZFcriterion}, a posteriori, to our simulation data, though we must heed some caveats. Perhaps the greatest issue is that our arguments have relied on a separation of scales that the simulations do not generally exhibit. Is there enough space in our simulation domain for a mean flux to be defined as we have done, when the eddies are only a bit smaller than the box size? Moreover, can $R$ exhibit meaningful variations on scales so close to the characteristic turbulent lengthscales, and thus provide the associated $x$-dependent COS driving? Though it is difficult to answer these questions in the affirmative, our mean field model can capture in some sense the underlying physics taking place, and as we shall see is roughly consistent with the results.

In our code units the criterion can be reframed as $dF/d\ln R> (1/2)\text{Re}^{-1}$. We may then use our numerically determined scaling for $F$,
obtained in Section 5.7: $F\approx 2\times 10^{-7}(R/0.01)^2$, which holds within the regime of weak turbulence (but not outside). The revised zonal flow criterion becomes simply
$ R \gtrsim 0.1 \left(\text{Re}/10^4\right)^{-1/2}.$
This should be compared to the cyan line in Figure \ref{fig:sur}, which is given by $R= 0.05\,(\text{Re}/10^4)^{-1/2}$. The two curves differ by an order one factor, which (given the many approximation involved) is not too bad.  
(At lower Re it appears that the system is just too viscous for the theory to be applicable.) 
 In summary, the comparison does seem to justify the mean field theory, and most importantly provides support for our physical explanation of zonal flow formation.

\section{Conclusion}

In this paper we have investigated the nonlinear development of the convective overstability (COS) in a local model of a protoplanetary disc.
Our main aim has been to categorise and determine the underlying physics of the several dynamical regimes the instability supports. Of particular interest is the onset of the coherent structures known as zonal flows, which comprise a radial sequence of vertically homogeneous azimuthal jets. They are important because they provide a route by which the COS can break its inherent axisymmetry (via a non-axisymmetric shear instability) and thence develop fully three-dimensional flow, vortices most importantly. Though our simulations are axisymmetric, and thus cannot describe vortex production, what they can do is establish the critical parameters required to obtain zonal flows; and being only two-dimensional we can push our parameters to values nearly representative of real PP discs.

The nonlinear dynamics of the COS, even in axisymmetry, is remarkably rich. Our main parameters are the Reynolds number Re and a number describing the unstable entropy gradient $R$ (which we sometimes call the `pseudo-Richardson number'). For values of Re and $R$ near criticality, the system supports interesting nonlinear waves, which involve a three-way resonance linking the primary COS mode and two daughter inertial waves (cf. Section 5.1). The principle features of this state can be captured analytically by a weakly nonlinear analysis (Appendix A). As we push Re and $R$ to larger values away from criticality, the system enters a more disordered state that shares some features with inertial wave turbulence (Galtier 2003), though its inertial range is too short to make more than passing contact with weak turbulence theory (cf. Section 5.2). We expect the nonlinear wave and wave turbulent states to remain axisymmetric even when simulated in full three-dimensions. 

On increasing either or both Re and $R$ further, zonal flows begin to emerge intermittently and enter a predator-prey cycle with the wave turbulence. The latter, when sufficiently strong, drives the growth of the coherent structures but is then impeded by them, leading to oscillatory dynamics (cf. Section 5.3). For larger Re and/or $R$ the zonal flows become persistent and the COS turbulence and the coherent structures agree on a quasi-steady state. Concurrently, the system exhibits elevator flows, which consist of a radial pattern of upward and downward velocities; they appear to be forced by inertial wave turbulence through a process yet to be understood (see also Dewberry et al.~2020). Elevator flows are robust features in local models of discs because the Kelvin-Helmholtz instabilities that might otherwise break them down are suppressed by rotation. The development of zonal flows, on the other hand, is a generic feature of any rotating flow, and similar features appear in semi-convection, which shares many mathematical and physical details (e.g. Cambon and Joaquin 1989, Waleffe 1993, Mirouh et al.~2012). We construct a mean-field theory in Section 6 and Appendix B that illuminates some of the underlying physics behind their onset in COS unstable discs. But how zonal flows saturate, be it through a cyclical predator-prey dynamics with inertial wave turbulence or a steady balance with the same, is not entirely clear to us and forms the basis of future work.  

Astrophysically, the most important boundary in the parameter space is that separating the wave-turbulent state from the intermittent zonal-flow state, as the latter is a possible site of vortex production. According to our simulations, this boundary curve can be fitted by $R= 0.05\times \left(\text{Re}/10^4\right)^{-1/2}$ (cf. Section 5.4). In realistic PP discs, Re takes values $\sim 10^7$ at 1 AU to $\sim 10^{10}$ at 10 AU (using our definition of Re; Section 3.2), which means that zonal flows only fail to appear when $R$ is smaller than $R_\text{crit}\sim 10^{-3}$ (1 AU) or $10^{-4}$ (10 AU). As discussed in Section 2.2, it is difficult to determine what values $R$ realistically adopt; but what can be said is that the characteristic timescale of the COS is $\sim R^{-1}\Omega^{-1}$ and thus when $R<R_\text{crit}$ the COS is certainly operating too slowly at 10 AU to play a role in the disc dynamics, and is on the sluggish side at 1 AU. We conclude that if the COS is functioning on a reasonable timescale in PP discs it will probably be in the regime of zonal flows and hence of potential vortex formation. 

The COS, being small-scale, will generate vortices that will also be small-scale, initially with lengths $\sim \sqrt{\xi/\Omega}$, and hence not observable. Even if small, vortices can collect solids and actively take part in planet formation, and while they will certainly be subject to secondary instability (Lesur and Papaloizou 2009, Railton and Papaloizou 2014), they may also be protected from complete dissolution by the subcritical baroclinic instability mechanism once they have formed (Lesur and Papaloizou 2010). 

As with any project of numerical simulation, compromises have had to be made. We adopted the diffusion approximation for radiative cooling, which may not be suitable at larger radii and for less massive discs; though some of the fine details may need revision, we believe that our picture of zonal flow production should carry across. Our simulations are also local and ideally we would want a separation of scales between the box size and the energy input size (and indeed any larger-scale structure that might develop), and between the energy input size and the viscous length. Given our resources, we cannot achieve both and so have chosen to explore the latter separation of scales, thus allowing us the chance to simulate discs approaching realistic molecular viscosities at the inner radii of PP discs, a novelty that we could not resist. This choice does mean that the large-scale coherent structures that develop in our simulations (zonal and elevator flows) are possibly impacted upon by the numerical box. Additional simulations might explore the opposite regime, where the box is much larger than the COS input scale, so as to check that box-size effects are not critical to what we show here (cf. convergence issues in Lyra 2014). Different vertical boundary conditions could also be trialled, such as impermeable walls (e.g. Barker et al.~2019). 

Future work includes three-dimensional simulations to observe how zonal flows wrap up into vortices, and whether this is possible in the intermittent zonal flow regime. Such a numerical program can also determine how these vortices self-sustain once formed (using the background entropy gradient), how large they get, how long they live, etc. Forthcoming studies might explore the influence of important physical processes omitted so far. While a stable vertical entropy gradient has no effect on the fastest growing modes, it will alter their ensuing nonlinear wave resonances, and consequently the wave turbulence; its impact on the elevator flows will be even more pronounced. Similarly, vertical shear will not impede the fastest COS modes, but could modify their nonlinear saturation, as might the VSI if present. Finally, the non-ideal MHD element of the problem could be assessed: not only how magnetic tension impedes instability, but how the vorticity dynamics inherent in the COS evolution combines with the Hall effect (cf. Kunz and Lesur 2013). Such a project might also explore the nonlinear development of the resistive double diffusive instability (Latter et al.~2010), which is fueled from the same energy source as the COS, and may compete with the COS under certain circumstances. 

\section*{Acknowledgements}
The authors thank the reviewer for a useful set of comments, and Gordon Ogilvie, Wlad Lyra, Min-Kai Lin, and Tobias Heinemann for helpful conversations and/or feedback on the manuscript. This work was partially funded by STFC grant ST/L000636/1. 

\section*{Data Availability}

The data underlying this article will be shared on reasonable request to the corresponding author.

\bibliographystyle{mnras}

\begin{appendix}
\section{Reduced dynamical model: three-wave coupling}

In this appendix we derive a simple dynamical
system for the three-wave interactions governing our simulations
in the weakly nonlinear regime. The analysis
here
extends that of L16 by allowing for the feedback of the instability on the primary COS mode, in addition to energy dissipation by viscosity. 

\subsection{Asymptotic expansions}

Suppose our disc is Keplerian and consider the nonlinear equations for the perturbations $\u'$,
$h'=P'/\rho$, and $\theta'$ in units so that $\Omega=1$ and $\sqrt{\xi/\Omega}=1$:
\begin{align}
&\left(\d_t + \u'\cdot\nabla\right)\u'= -\nabla h' +(2u_y'+R\theta')\ex\\ &\hskip3cm -\tfrac{1}{2}u_x'\ey+\text{Pr}\nabla^2\u'=0, \\
&\left(\d_t + \u'\cdot\nabla\right)\theta'= u_x'+\nabla^2\theta', \quad
  \qquad \nabla\cdot\u'=0, 
\end{align}
We introduce a small parameter $0<\epsilon\ll 1$ and consider a regime in which $\text{Pr}\sim R \sim \epsilon$. Given the stability criterion of the COS, this scaling indicates that we are near criticality. As a consequence, the saturation of the instability takes place at relatively low amplitudes, of order $\epsilon$. 
In addition, we assume that nonlinear solutions evolve on a timescale much longer than the fast orbital time; we thus introduce a slow time variable $T=\epsilon t$. 
Finally, we expand the
perturbations in $\epsilon$ so that
\begin{align*} 
&\u'=\epsilon\u_1(\mathbf{x},t,T)+\epsilon^2\u_2(\mathbf{x},t,T)+\dots, \\
&h'=\epsilon h_1(\mathbf{x},t,T) +
\epsilon^2 h_2(\mathbf{x},t,T)+\dots, \\
&\theta'=\epsilon \theta_1(\mathbf{x},t,T)+\dots, \qquad \text{Pr}=p\epsilon, \qquad R= r\epsilon,
\end{align*}
where $p$ and $r$ are order-one `tuning' parameters, and $\u_i$, $h_i$, and $\theta_i$ must remain of order one. This ansatz is thrown into the nonlinear equations and terms in the
various orders of $\epsilon$ are collected.

\subsection{Structure of the solution at order $\epsilon$}
 At leading order
$\mathcal{O}(\epsilon)$ we obtain the linear problem governing
incompressible inertial waves:
\begin{align}\label{nleq}
\d_t\u_1 = -\nabla h_1 + 2u_{1y}\ex -\tfrac{1}{2}u_{1x}\ey, \qquad \nabla\cdot\u_1=0.
\end{align}
This can be reduced to the convenient $\mathcal{L}u_{x1} =0$ where 
$\mathcal{L}=\d_t^2\nabla^2+\d_z^2$, is the `inertial wave operator'. This equation admits solutions of
the form $\propto \text{exp}(\text{i}k_x x + \text{i}k_z z -
\text{i}\omega t)$, where $k_x$ and $k_z$ are wavenumbers and the
frequency is $\omega= \pm k_z/k$, with $k^2=k_x^2+k_z^2$. In our simulations' periodic domain, the wavenumbers must be discretised. 

There are an infinite number of wave solutions to the problem at
this order and the most general solution comprises a linear
combination of them all. We consider only three: the $k_x=0$ mode
associated with the COS, and two `daughter' modes that can couple to it
via a resonance. The primary (COS) mode we denote with a subscript
`$A$': it has a wavevector $\mathbf{k}_A=(0,0,k_{Az})$, and frequency
$\omega_A=1$. We select the fastest growing COS mode, which means $k_{Az}=1$ in our dimensions.

The two daughter waves are denoted by `$B$' and `$C$'.
In order to obtain (near) resonance we must have 
\begin{align*}\pm \mathbf{k}_A\pm
\mathbf{k}_B\pm \mathbf{k}_C=\mathbf{0}, \quad \text{and}\quad
\pm \omega_A\pm\omega_B\pm\omega_C= \Delta. \end{align*}
So as to best compare with
L16 in the following the signs are chosen in the order `--\,--\, +'. 
Because of the finite size of the numerical
domain (and the consequent discretisation of the wavenumbers)
it may not be possible to achieve perfect resonance, in which case 
there will be some degree of detuning, represented in the above by the quantity $\Delta$. This will be assumed small, and to ease the asymptotic ordering we set $\Delta=\delta \epsilon$, where $\delta\sim 1$ is a new parameter. This frequency mismatch comes in only at higher order. In addition, we only consider vertical wavenumbers that are discrete multiples of the primary's; thus we set $k_{BZ}=n$, where $n$ is an integer, and so $k_{Cz}=n+1$. Lastly, we assign $\omega_B=-n/k_B$ and $\omega_C=(n+1)/k_C$,
from the dispersion relation for inertial waves.

In summary, at this order our solution is
\begin{align*}
\u_1 &= \u_A A(T)E_A(\mathbf{x},t)+
 \u_B B(T)E_B(\mathbf{x},t) \\
 & \hskip3.5cm +\u_C C(T)E_C(\mathbf{x},t)+ \text{c.c.}
\end{align*}
where `c.c.' indicates the complex conjugate of the preceding, $A$, $B$, and $C$ are complex amplitudes (to be determined), $E_A(\mathbf{x},t)=\text{exp}(\text{i}\mathbf{k}_A\cdot\mathbf{x}-\text{i}\omega_A
  t)$, etc., and the constant velocity vectors
  are given by $\u_A=\left[1,-\text{i}/(2\omega_A),-k_{Ax}/k_{Az}\right]$, etc.
  
  The resonance condition ensures $k_{Bx}=k_{Cx}\equiv k_x$, so we can write $\mathbf{k}_A=(0,0,1)$,
  $\mathbf{k}_B=(k_x,0,n)$, $\mathbf{k}_C=(k_x,0,1+n)$, and given that $\omega_A=1$, we have at leading order, in small $\epsilon$, that $\omega_C=1+\omega_B$, which can be expressed in the remaining parameters:
  \begin{align} \label{rezzy}
\frac{n}{\sqrt{k_x^2+n^2}}+\frac{n+1}{\sqrt{k_x^2+(n+1)^2}}=1.
  \end{align}
   Given a fixed $n$, this condition yields a distinct $k_x$ at which exact resonance can occur. For $n=1-5$, we obtain $k_x\approx 2.49, 4.26, 6.02, 7.76, 9.50$, (see L16). 
  
  An expression for $h_1$ is not needed, but we do require the leading order buoyancy variable $\theta_1$ in what follows. The buoyancy equation at leading order is a forced diffusion equation:
  $(\d_t-\nabla^2)\theta_1= u_{x1}.$
  We neglect the decaying complementary function and retain only the particular integral. Thus $\theta_1= u_{x1}/(k^2-\text{i}\omega)$, for each of the three wave components introduced above. At leading order, the thermal physics is slaved to the inertial waves, but feeds back critically on the problem at higher order (via the buoyancy acceleration) to produce the convective overstability's growth (see physical arguments in Section 3.3 in L16).
  
\subsection{Solvability conditions at order $\epsilon^2$}

At the next order the Navier-Stokes equation can be boiled down to the
relatively simple
\begin{align}\label{ep2}
\mathcal{L}u_{x2} = \d_z^2\d_t N_x + 2 \d_z^2 N_y - \d_{xzt} N_z,
\end{align}
where the right hand side terms involve only solutions of the preceding order and are written using
$$\mathbf{N}= -\d_T\u_1 - \u_1\cdot\nabla\u_1+r\ex \theta_1+p\nabla^2\u_1.$$
A solvability condition for Eq.\eqref{ep2} is that the right hand
side possesses no component proportional to the eigenfunctions of
$\mathcal{L}$, i.e. $E_A$, $E_B$, and $E_C$, in our problem. To ensure this we simply zero the coefficients of these three factors, recognising that $E_B^*E_C=E_A e^{-\text{i}\delta\,T}$, $E_A E_B=E_C e^{ \text{i}\delta\,T}$, and $E_A^*E_C=E_B e^{-\text{i}\delta\,T}$.  
Doing so obtains three evolution equations for the mode amplitudes $A$, $B$, and $C$:
\begin{align}\label{red1}
&\frac{dA}{dT} = (\sigma_A-p)A + \text{i} c_1 B^*C\,e^{-\text{i}\delta\,T} , \\
&\frac{dB}{dT} = (\sigma_B-p k_B^2)B - \text{i} c_2 A^*C e^{-\text{i}\delta\,T}, \\
&\frac{dC}{dT} = (\sigma_C-p k_C^2)C + \text{i} c_3 ABe^{\text{i}\delta\,T}. \label{red3}
\end{align}
Here the linear terms combine (a) the asymptotic COS growth rates of each mode in the limit of small $R$ (see Section 3.1 in L16), i.e. $\sigma_A= \frac{1}{4}r(1+\text{i})$, $\sigma_B=\tfrac{1}{2}r\omega_B^2/(k_B^2-\text{i}\omega_B)$, and 
$\sigma_C=\tfrac{1}{2}r\omega_C^2/(k_C^2-\text{i}\omega_C)$, and (b) the viscous damping terms proportional to $p$. Note that for modes $B$ and $C$ the viscous damping easily dominates the growth due to the COS (which can be omitted), while in mode $A$ we can control the rate of COS growth via the size of $\tfrac{1}{4}r-p$.

Expressions for the nonlinear coefficients are
\begin{align}
&c_1= \frac{k_x\left[\omega_B(2+\omega_B)+n(2\omega_B^2+2\omega_B-1) \right]}
{2n(1+n)\omega_B(1+\omega_B)}, \\
&c_2 = \frac{k_x\omega_B\left[(k_B^2+2n)\omega_B^2+(k_B^2+n)\omega_B+n^2\right]}
{2n(1+n)(1+\omega_B)},\\
&c_3 = -\frac{k_x\omega_C\left[(k_B^2-1)\omega_B^2+(k_B^2-n-2)\omega_B+ n(n+1) \right]}
{2n(1+n)\omega_B},
\end{align}
where we recall that $\omega_B=-n/k_B$, $k_B=\sqrt{n^2+k_x^2}$, $\omega_C=1+\omega_B$, and $n$ and $k_x$ are related via the leading order resonance condition Eq.\eqref{rezzy}. For the first few resonances we consider, $c_i>0$. Note that the detuning introduces extra complex exponential factors into the quadratic terms.

\subsection{Simplifications, rescalings, and analysis}

For the rest of the appendix we omit the subdominant growth rates of the $B$ and $C$ modes, i.e. $\sigma_B$ and $\sigma_C$. These have little to no impact on the dynamics. Also to simplify the equations somewhat, without altering their main features, we set $k_C=k_B$, and thus the damping rate of the two daughter modes are the same.

\subsubsection{Energetics and parametric instability}

Though the system is open, the nonlinear transfer terms must conserve the kinetic energy $K=\frac{1}{2}(|\u_A A|^2+|\u_B B|^2+|\u_C C|^2)$. On differentiating $K$ with respect to $T$ and using Eqs \eqref{red1}-\eqref{red3}, but only with the nonlinear terms active, we 
obtain the identity $c_1 |\u_A|^2-c_2|\u_B|^2-c_3|\u_C|^2=0$, which provides a useful check on the algebraic expressions for the $c_i$. 
It follows that the total energy of the system is controlled by the linear terms, namely energy input by the primary COS mode (A) and viscous dissipation of the two daughter modes (B and C):
$$dK/dT= (\tfrac{1}{4}r-p)|\u_A A|^2- pk_B^2(|\u_B B|^2 +|\u_C C|^2).  $$
 For a quasi-steady state the right side need not be zero (for most of the oscillations we find it varies between positive and negative values), but it must integrate to zero on sufficiently long times or over the period of a cycle.

In the case of a constant $A$ and an exact resonance ($\delta=0$), Eqs \eqref{red1}-\eqref{red3} provide the growth
rate of the parametric instability discussed earlier; it is simply $\sqrt{c_2 c_3}|A|$. This expression agrees with the growth rate derived in
L16 once it is recognised that we can identify $|A|=4S$: in L16 a real standing wave was used for the primary while in this paper we have assumed that the primary is a complex travelling wave. 

\subsubsection{Simplified, rescaled system}

The dynamical system can be simplified by the following transformation:
 \begin{align*}
     A\to \frac{p k_B^2\,A}{\sqrt{c_2 c_3}},\quad
     B\to \frac{\text{i} p k_B^2\,B}{\sqrt{c_1 c_3}}, \quad
     C\to \frac{p k_B^2\,C}{\sqrt{c_1 c_2}},
     \quad T\to \frac{ T}{pk_B^2}.
    \end{align*}
We then obtain
\begin{align} \label{new1}
&\frac{dA}{dT} = \lambda A +  B^*C\,e^{-\text{i}\overline{\delta}\,T} , \quad 
\frac{dB}{dT} = -B - A^*C e^{-\text{i}\overline{\delta}\,T}, \\
&\hskip2cm \frac{dC}{dT} = -C -  ABe^{\text{i}\overline{\delta}\,T},\label{new3}
    \end{align}
where the scaled primary's growth rate is $\lambda=(\sigma_A-p)/(p k_B^2)$ and the scaled detuning factor is $\overline{\delta}=\delta/(p k_B^2)$. The timescale of the new system is pinned to the (fast) viscous decay of the daughter modes, against which the primary's growth may be considered slow.

For $\lambda$ real, this system has enjoyed considerable attention, most notably in Vyshkind and Rabinovich (1976), Wessinger et al.~(1980), Bussac (1982a, 1982b), and Hughes and Proctor (1990, 1992), who plot out its various behaviours and bifurcations. Aside from plasma physics, where the system first appeared, analogous dynamics in astrophysics occurs in the `r-mode
instability' in neutron stars, where an unstable Rossby wave transfers
its energy to other smaller scale inertial waves (Arras et al.~2003),
and in overstable gravity modes in ZZ Ceti stars (Wu and Goldreich 2001), amongst other 
applications (Moskalik 1985).
In addition to unbounded solutions, the system exhibits a range of sometimes chaotic oscillations, some akin to predator-prey bursts (combining the slow and fast timescale), and others far more regular.  In the following we briefly describe the main features of these behaviours, as they impact on our particular system.

\subsubsection{Amplitude-phase dynamics and their fixed points}

Equations \eqref{new1}-\eqref{new3} appear to be sixth order but can be reduced to a third order system. First, it is easy to show that $\left||C|^2-|B|^2\right|\propto \text{exp}(-2T)$, and thus on the longer timescales of interest the moduli of the two daughter modes are the same. We hence set $|B|=|C|$ and derive the following evolutionary equations:
\begin{align}
&\hskip0.5cm\dot{a}=\lambda_r a + b^2 \cos\psi, \qquad\qquad \dot{b}=-b-ab\cos\psi, \\
&\hskip2cm \dot{\psi}=\overline{\delta}+\lambda_i + \frac{2a^2-b^2}{a}\sin\psi,
\end{align}
where an overdot signifies a $T$ derivative, $a=|A|$, $b=|B|$, and $\psi=\text{Arg}(A)+\text{Arg}(B)-\text{Arg}(C)+\overline{\delta}\,T$ (Vyshkind and Rabinovich 1976). A feature that distinguishes our equations from those derived in other physical applications is the imaginary part of the growth rate $\lambda_i$, which functions as an additional detuning. 

These equations support two fixed points, the trivial state $a=b=0$ of Keplerian shear, which we know is convectively overstable, and a second state determined from
$$\psi=\tan^{-1}\left(\frac{\overline{\delta}+\lambda_i}{2-\lambda_r}\right),\quad a=-\sec\psi, \quad b^2=\lambda_r\sec^2\psi, $$
in which $\psi$ must lie in the 2nd or 3rd quadrant. Though it appears as a fixed point in the amplitude-phase dynamics, in $(A,B,C)$ space, this invariant object corresponds to an orbit of constant $|A|$, $|B|$, and $|C|$, with a fixed kinetic energy. 

\begin{figure}
\center
\includegraphics[width=0.5\textwidth]{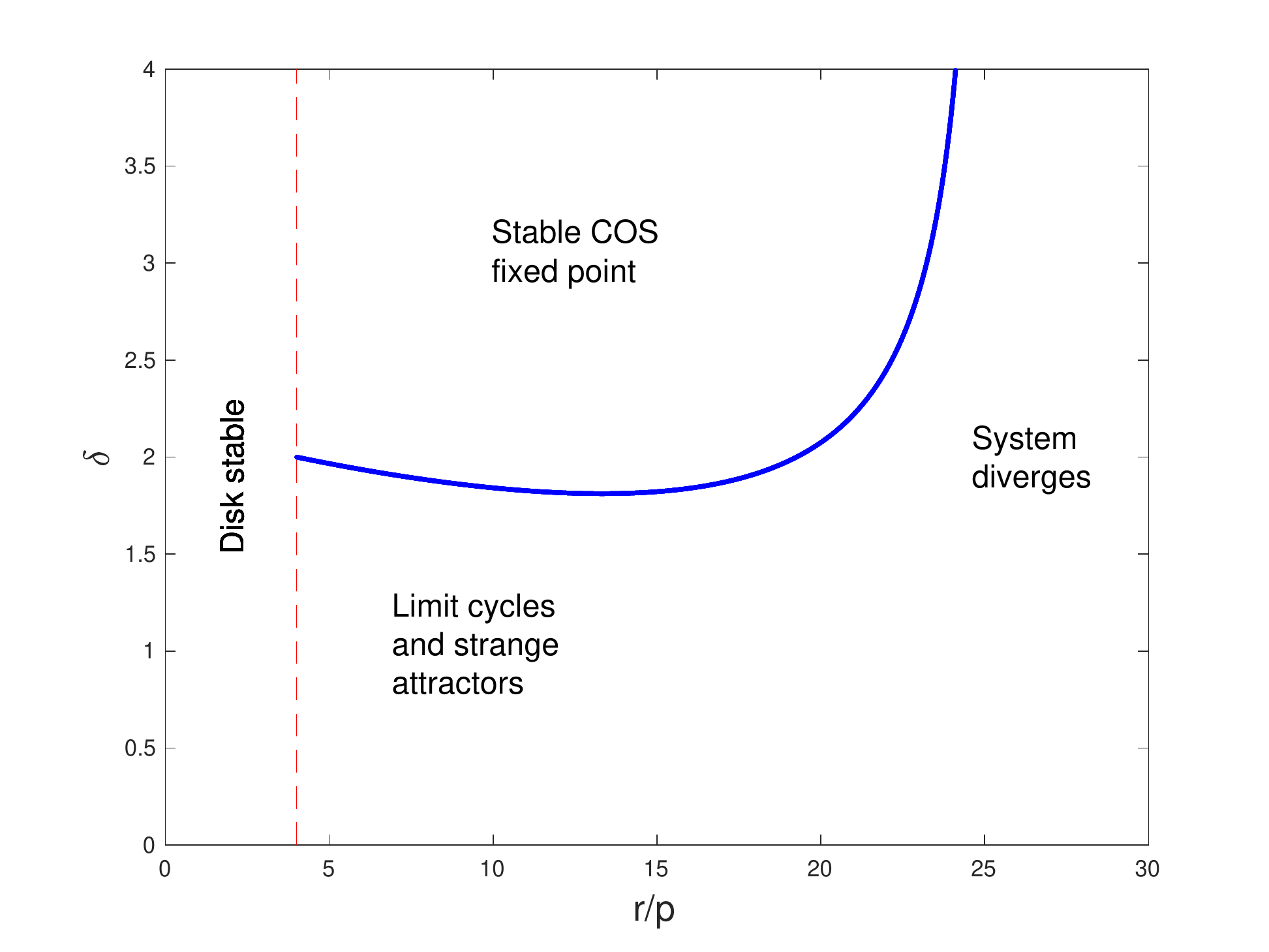}
\caption{Following Bussac (1982a), the basic structure of the dynamical system as described in the $(r/p,\,\overline{\delta})$ parameter plane for a $n=1$ resonance.}
\label{Bifurc}
\end{figure}

The linear stability of this non-trivial fixed point is straightforward to check. Skipping all the algebra (see Wersinger et al.~1980, Hughes and Proctor 1990), we find that we have stability for sufficiently large detuning:
\begin{equation} \label{nlcrit}
\overline{\delta}+\lambda_i> |2-\lambda_r|\sqrt{\frac{\lambda_r^2-2\lambda_r+2}{2-2\lambda_r-\lambda_r^2}},
\end{equation}
and as long as $\lambda_r<\sqrt{3}-1$.
 To leading order in small $\lambda_r$, the stability criterion simplifies to $\overline{\delta}+\lambda_i> 2$. 
  The curve of marginal stability is plotted in Fig.~\ref{Bifurc}.

As we decrease $\overline{\delta}$ and pass through the critical stability threshold \eqref{nlcrit}, the fixed point undergoes a Hopf bifurcation and at first is encased in a stable limit cycle. As $\overline{\delta}$ and/or $\lambda_r$ decreases the cycle undergoes a sequence of period-doubling bifurcations and then transitions to a set of mildly chaotic relaxation oscillations. On the other hand, when $\lambda_r>\sqrt{3}-1$ the fixed point is subject to a monotonically growing instability and the system tends to blow up. For a fuller account of these dynamics, the reader is directed to the numerical surveys described in
Vyshkind and Rabinovich (1976), Wersinger et al. (1980) and Bussac (1982a), and their analytic reduction to simple one-dimensional maps, such as in Bussac (1982b) and Hughes and Proctor (1990, 1992). Note that on account of the non-negligible imaginary part of the COS's growth rate (which acts as an additional detuning), the dynamics we witness never plunge into the full bouts of chaos some of these authors discover.  

\begin{figure*}
\center
\includegraphics[width=0.33\textwidth]{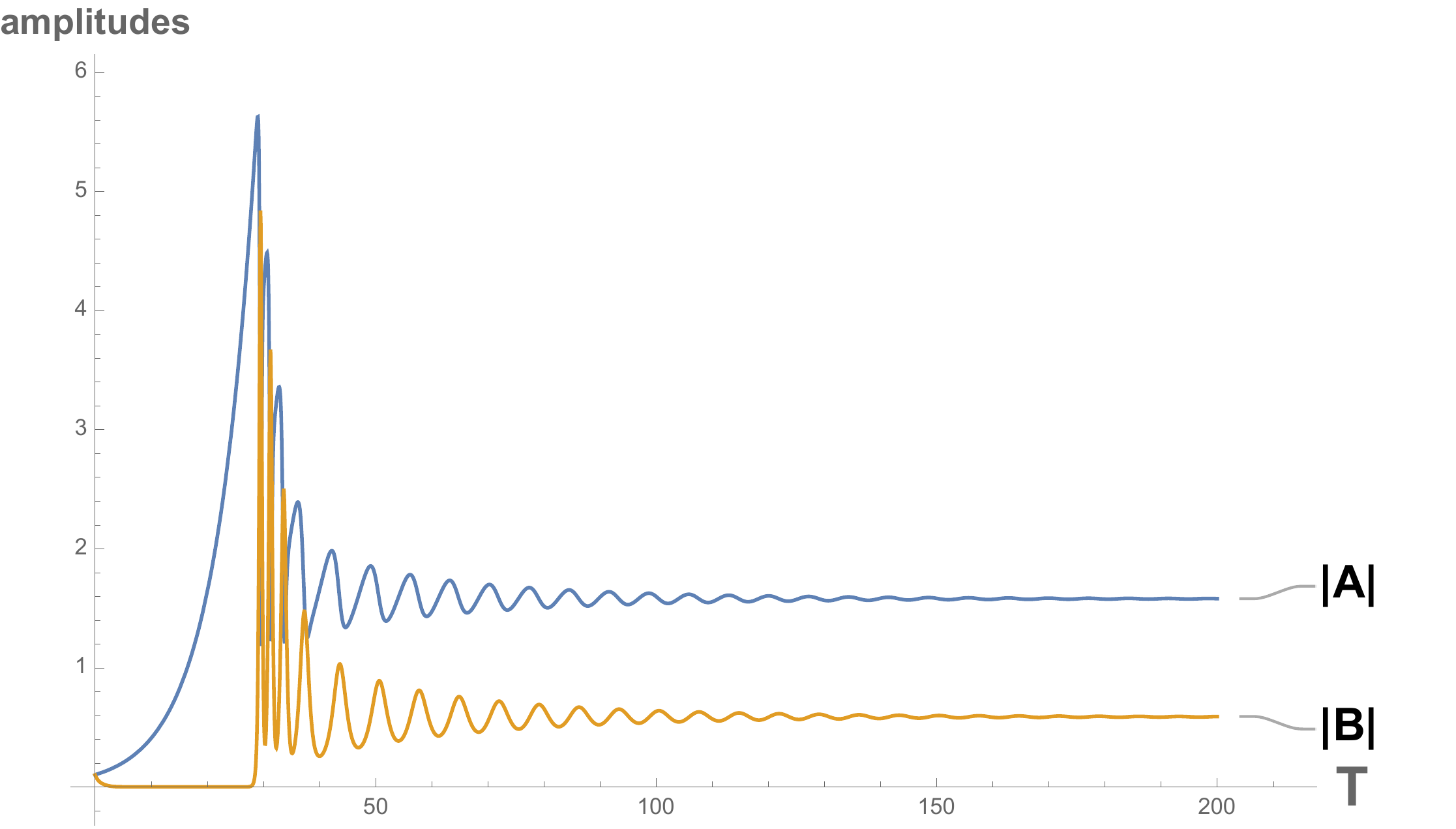}
\includegraphics[width=0.33\textwidth]{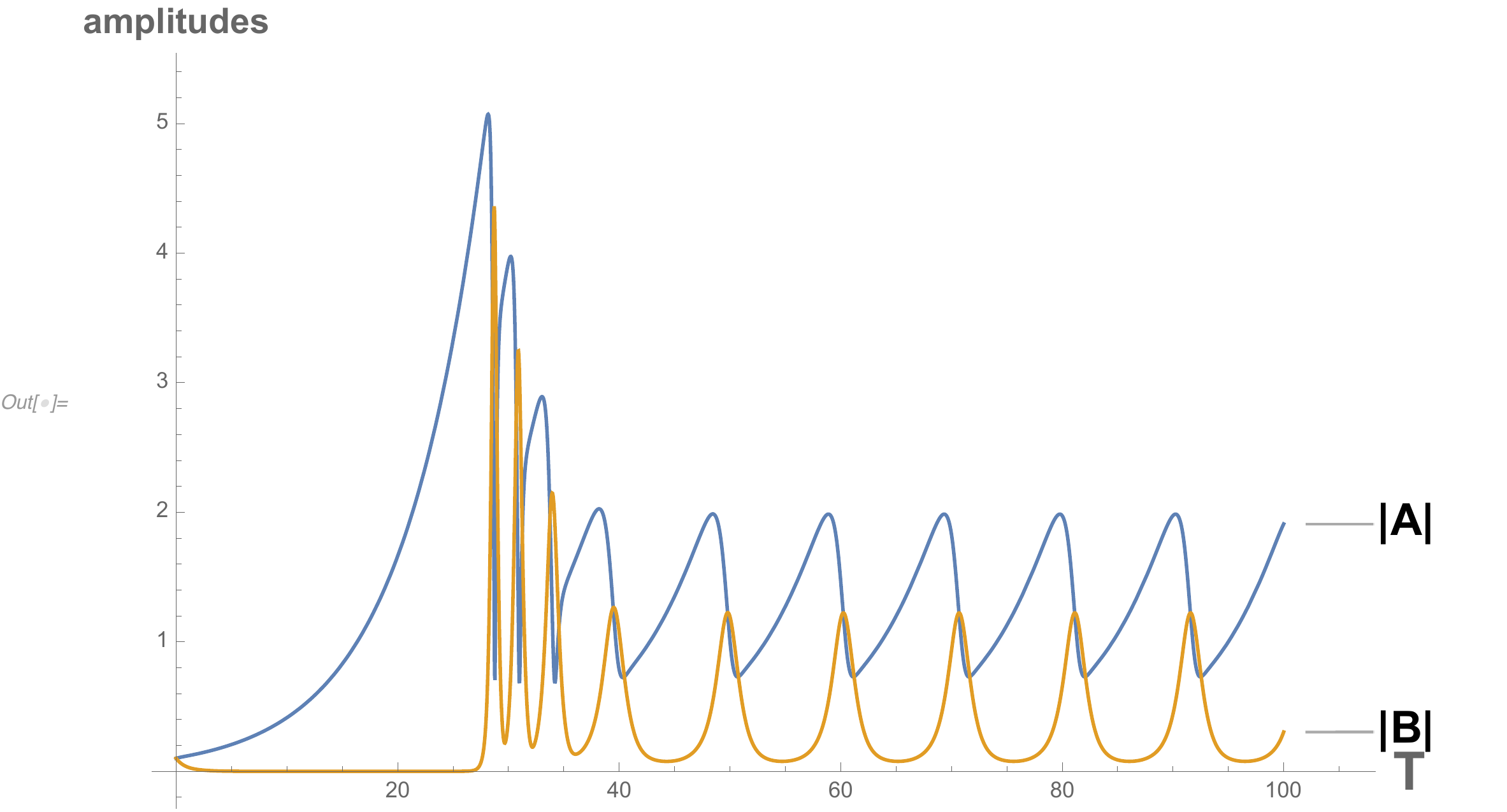}
\includegraphics[width=0.33\textwidth]{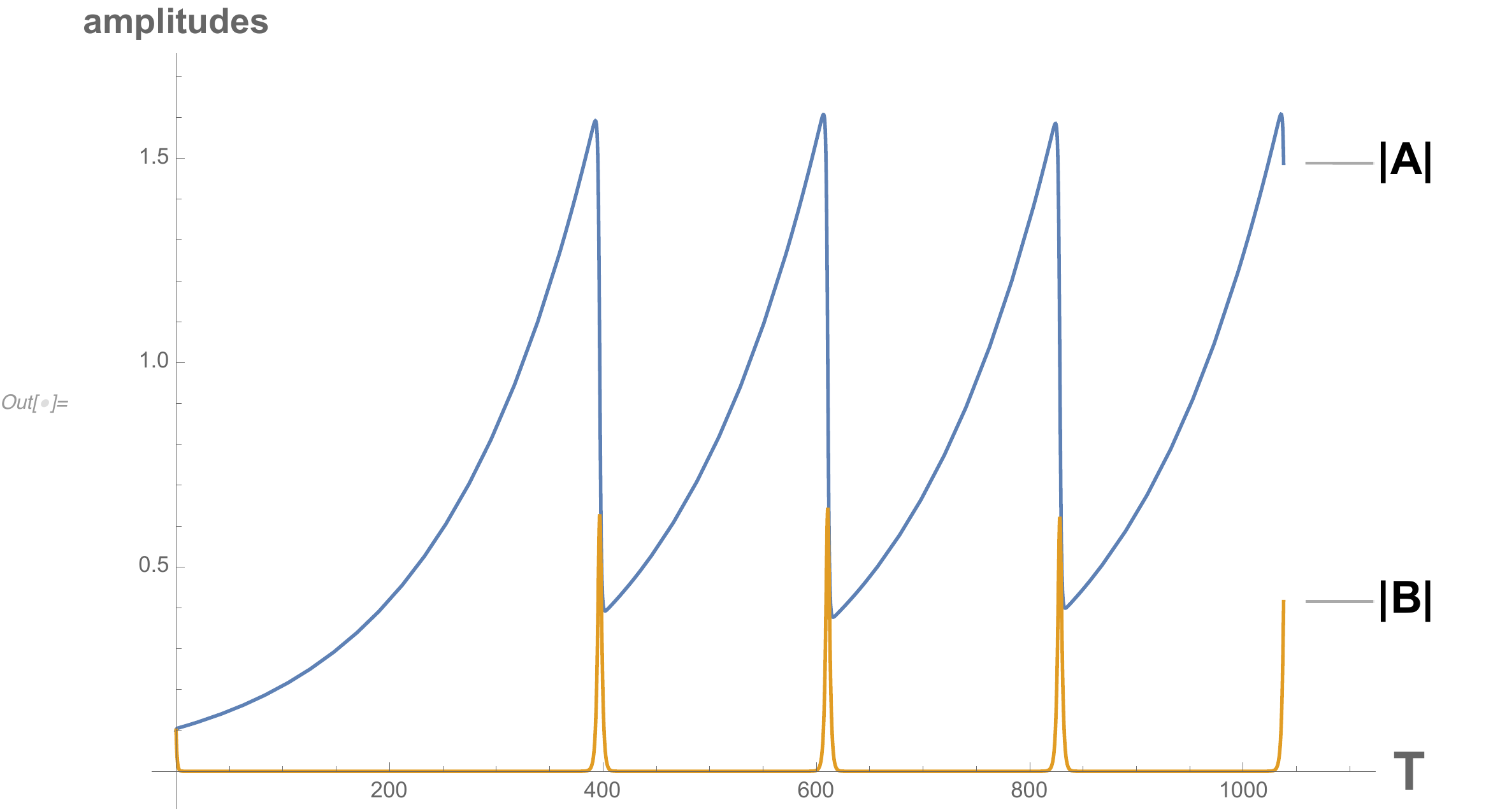}\\
\includegraphics[width=0.3\textwidth]{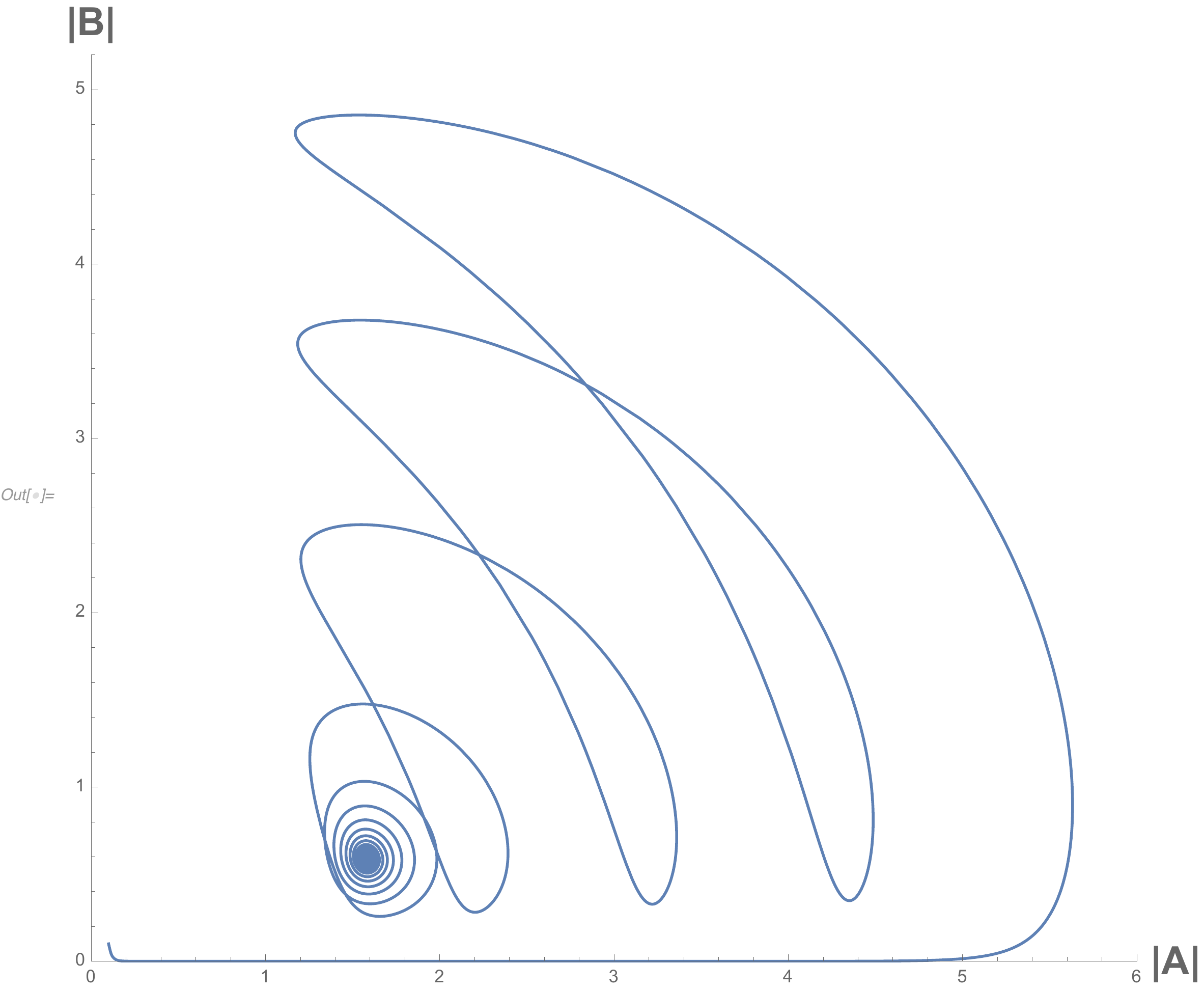}
\qquad \includegraphics[width=0.3\textwidth]{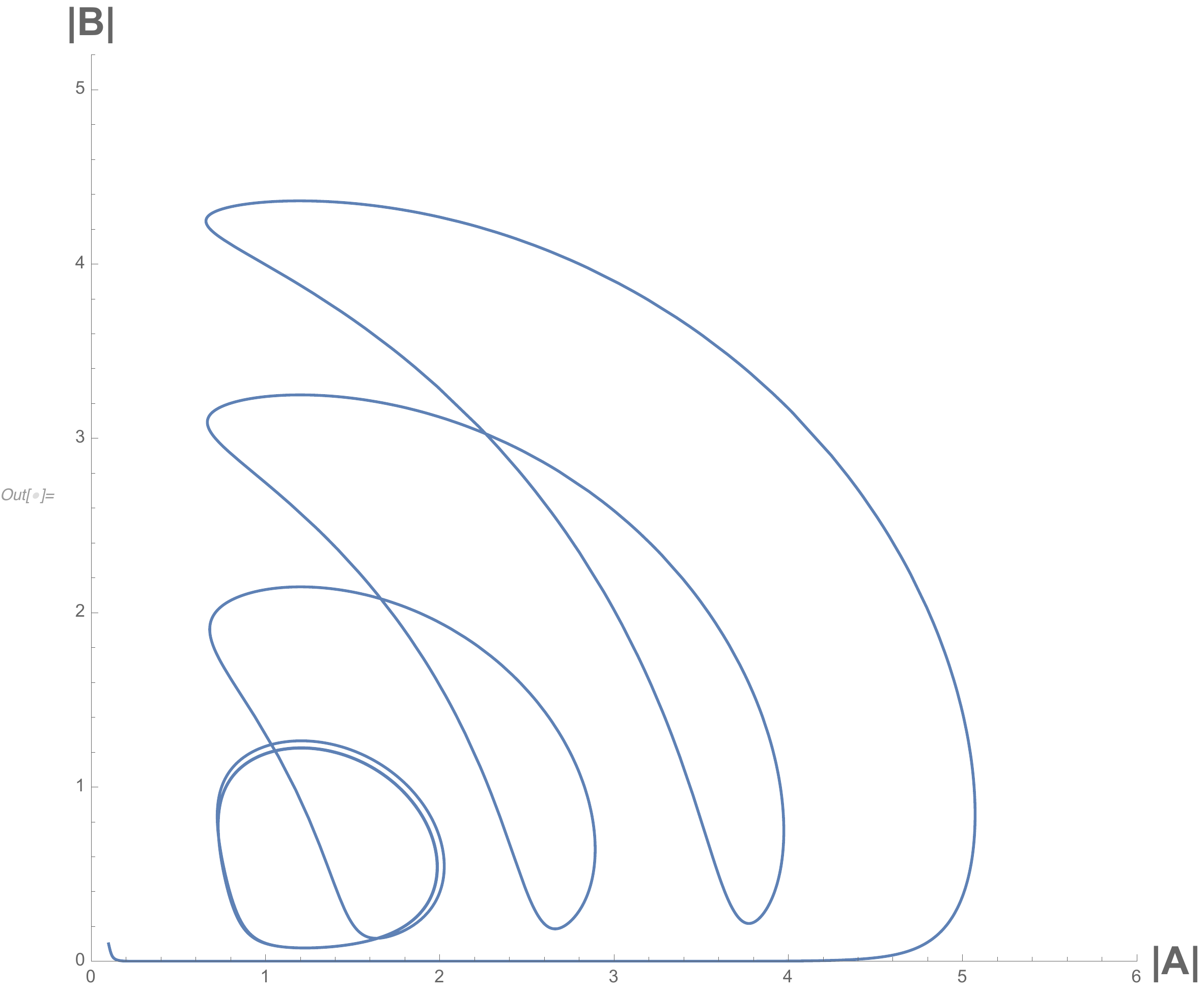}
\quad\quad\includegraphics[width=0.31\textwidth]{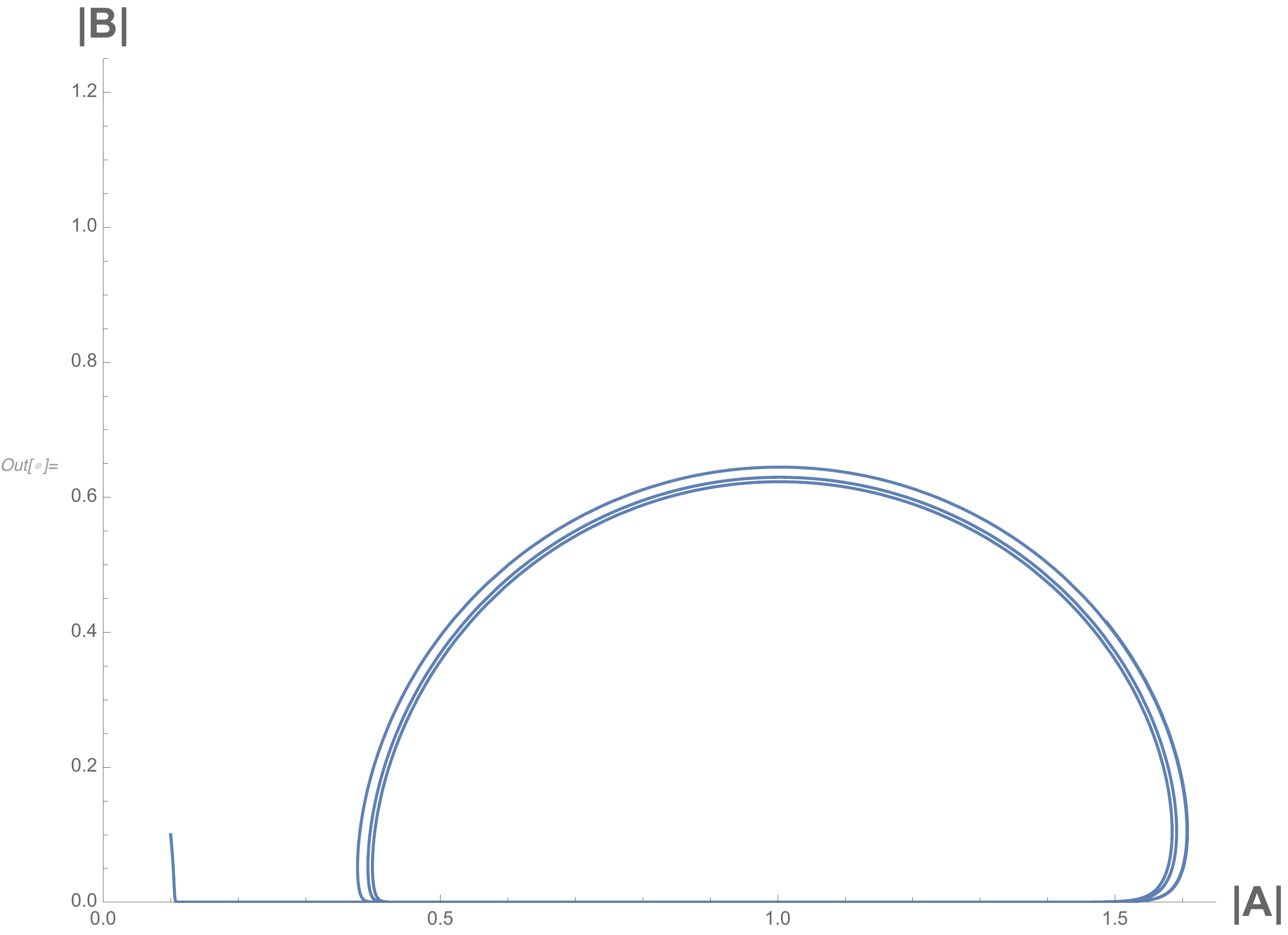}
\caption{Three numerical examples of different dynamical behaviours. In the left column we plot a solution associated with the parameter set $r/p=8$ and $\overline{\delta}=2$, with the time evolution of the wave amplitudes on the top, and their phase portrait on the bottom. The middle column corresponds to $r/p=8$ and $\overline{\delta}=1$, and the right column to $r/p=4.2$ and $\overline{\delta}=0$.  }
\label{NLsolns}
\end{figure*}
Before moving on, a final point worth making is that, even if unstable, the fixed point above organises the dynamics and to some extent determines the mean amplitude of the nonlinear saturation. As all the parameters that appear in the above analysis are order one or small, we obtain $a\sim 1$
at saturation, which yields (for $n=1$) $|\u'|\sim \epsilon p k_B^2 \sim \text{Pr} k_B^2\approx 7 R$, where we note that $k_B^2\approx 7$ for $n=1$. This estimate is similar, if slightly larger, than the saturation estimate given in L16 when adopting a `parasitic theory'. It thus extends and places such a theory on a sounder footing, at least in the regime of near criticality.

\subsection{Illustrative solutions}

To give a flavour of the types of behaviour exhibited near COS criticality, we numerically solve Eqs \eqref{new1}-\eqref{new3}
for fixed $n=1$ but different $r/p$ and $\overline{\delta}$, the two main control
parameters. For this choice of resonance, $k_x=2.49$ and thus $c_1=5.52$,
$c_2=0.287$, $c_3=1.36$, and $k_B^2=7.20$. 

In Figure \ref{NLsolns} we plot three representative solutions. In the left panels we set $r/p=8$ (thus $\lambda=0.139+0.278\text{i}$) and adopt a moderate level of detuning $\overline{\delta}=2$. The fixed point in this case is a stable focus, and after an initial transient the system spirals on to it, as illustrated in the bottom left phase portrait. In the middle plots we retain the same $r/p$ but reduce the detuning so that $\overline{\delta}=1$; as predicted, the fixed point is now unstable but centres a stable limit cycle that controls a regular nonlinear oscillation, with the primary mode and the daughter modes somewhat out of phase. The predator-prey alluded to in these dynamics is brought out best at lower $r/p$ and $\overline{\delta}$, and we show this in the right panels, for which $r/p=4.2$ (thus $\lambda=0.00694+0.146\text{i}$)
and $\overline{\delta}=0$. Here in the upper panel we witness bursty dynamics comprising the slow exponential growth of the primary, and its abrupt destruction by the parasitic daughter modes on a much faster timescale. Once they dissipate, the primary is free to grow again and the cycle repeats. Similar behaviour is exhibited in several biological systems involving multiple predators (e.g. Muratori and Rinaldi 1989), and in the net-vertical-flux MRI near criticality (Latter et al.~2009, Lesaffre et al.~2009). These `fast-slow' dynamics may be captured by a two-scale formalism, as in Hughes and Proctor (1990, 1992), though we decline to engage with that here. Of interest, however, is the somewhat chaotic nature of the relaxation oscillation, best illustrated in the lower right panel: there is no stable limit cycle, but rather a strange attractor which imparts some disorder to the various peaks of $|A|$, though this variability is rather minor in this case, and not evident in the top right panel.

\section{Mean field model of zonal flow formation}

Here we provide a somewhat more quantitative approach to the process described in Section 6, by developing a crude mean field model, in a similar spirit to Garaud and coworkers (see also Frisch 1989, and Latter and Balbus 2012). 

\subsection{Preliminaries}

We are mostly concerned with the angular momentum in our local model, which corresponds to the canonical $y$-momentum: $h=2\Omega x + u_y$. It obeys the conservation equation:
\begin{align}
\d_t h +\nabla\cdot\left(h\u -\nu \nabla h \right)=0.
\end{align}
In laminar equilibrium $u_y=-q\Omega x$ and thus $h=h_0=(2-q)\Omega x = [\kappa^2/(2\Omega)]x$. Clearly, the angular momentum increases outward (at least for $\kappa^2>0$ as assumed here), and the disc is Rayleigh stable. 

In addition, it is useful (though not necessary) to consider the total buoyancy in the shearing box: $\theta_x= -x + \theta$, which obeys our other conservation equation:
\begin{align}
\d_t\theta_x+\nabla\cdot\left(\theta_x\u-\xi\nabla\theta_x \right)=0.
\end{align}
Recognising that $\theta=-(\d_Z S)_0^{-1} S'$, where $S'$ is the dimensionless entropy perturbation, we identify $\theta_x$ with the total entropy in our box. In our laminar equilibrium $\theta_x=\theta_{x0}=-x$, and thus the entropy decreases outward, indicating the disc is Schwarzchild unstable. 

The $R$ parameter, the determinant of COS instability and strength, can hence be generalised to a space (and time dependent) quantity, sensitive to the particulars of the turbulence at any location and how effectively it has counteracted the equilibrium gradients presented above. We define 
\begin{align}
R_\text{eff} = \frac{N^2(\d\theta_x/dx)}{2\Omega(\d h/\d x)},
\end{align}
which returns to $R_\text{eff}\equiv R_0=-N^2/\kappa^2$, in our basic state, as required. 

\subsection{Turbulent fluctuations and mean field equations}

Suppose the COS grows and saturates in inertial wave turbulence, characterised by short length and time scales, $x$ and $t$, etc. Let us consider long radial wavelengths and slow temporal variations atop these fluctuations using the slow radial and temporal variables $X$ and $T$, so that $h=h(x,z,t,X,T)$ and $\theta_x=\theta_x(x,z,t,X,T)$. Next we introduce an averaging procedure over intermediate radial and time scales and all vertical scales, $\langle\cdot \rangle_{x,z,t}$, which removes the small-scale fluctuations and thus isolates mean components $H$ and $\Theta$, so that 
\begin{align*}
 &H(X,T)\equiv \langle h(x,z,t,X,T) \rangle_{x,z,t}, \\ 
 &\Theta(X,T)\equiv \langle \theta_x(x,z,t,X,T) \rangle_{x,z,t},
 \end{align*}
these can then be used to define an additional fluctuating part to $h$ and $\theta_x$, which we denote with primes, and which average to zero. It is next assumed that there is no appreciable mean radial velocity, which may be justified if we take the mean quantities to be generally small in amplitude. Note that the averaging retains the laminar equilibrium background, so that $H$ includes the component $\kappa^2/(2\Omega)X$ and $\Theta$ the component $-X$.
Finally, we adopt units so that $\Omega=1$ and $\sqrt{\xi/\Omega}=1$.

The mean conservation laws for angular momentum and entropy are hence:
\begin{align}
&\d_T H +\d_X\left(F_H - \text{Pr}\,\d_X H\right) =0,\label{Hev}\\
&\d_T \Theta +\d_X\left(F_\Theta - \d_X \Theta \right) =0,
\end{align}
where Pr is the (small) Prandtl number, the angular momentum flux is $F_H=\langle u_x' p_y'\rangle_{x,z,t}$, in other words the Reynolds stress, and the thermal flux is $F_H=\langle u_x' \theta_x'\rangle_{x,z,t}$. 

From our numerical experience, the turbulent thermal flux $F_\Theta$ is usually much smaller than the laminar radiative flux and if dropped permits the two equations to decouple. Then $\Theta$ obeys the diffusion equation on short timescales - short at least compared to the the angular momentum evolution - and the $\Theta$ dynamics can be justifiably neglected, though we retain them for the moment. 
In Eq.~\eqref{Hev}, however, both the Prandtl number and turbulent fluxes are small and can be comparable. They are thus both retained.

We next assume that $F_H<0$ and $F_\Theta>0$. But to make progress, we need to introduce a closure scheme. Noting that the turbulent fluxes depend on the local turbulent strength, which in turn depend on the magnitude of $R$, we set $F_H=F_H(R_\text{eff})$ and $F_\Theta=F_\Theta(R_\text{eff})$, where in the mean field setting and in our adopted units we have $R_\text{eff}= -R_0 (\d\Theta/d X)/(2\d H/\d X)$. Finally we define $$R_0\left(\frac{dF_H}{dR_\text{eff}}\right)_0\equiv F_H'<0, \qquad R_0\left(\frac{dF_\Theta}{dR_\text{eff}}\right)_0\equiv F_\Theta'>0.$$ 

\subsection{Linear stability of homogeneous turbulence}

Consider the state of (quasi-) steady homogeneous COS wave turbulence in a Keplerian disk. The mean fields hence comprise only the background laminar gradients, so that $H=H_0=(1/2)X$ and $\Theta=\Theta_0=-X$. 

We next disturb this background with mean perturbations $H'$ and $\Theta'$. It can be quickly shown that these give rise to a perturbation in the $R$ parameter, which to linear order is
$R'_\text{eff} = -R_0(2\d_X H' + \d_X \Theta')$. We may then write down the coupled linearised equations for the two perturbations:
\begin{align}
\d_T H' &= (\text{Pr}+2F_H')\d_X^2 H' + F_H'\d_X^2 \Theta', \label{Hd}\\    
\d_T \Theta' &= (1+F_\Theta')\d_X^2\Theta'+2F_\Theta'\d_X^2 H'.
\end{align}
If we assume, as mentioned earlier, that the thermal fluctuation will be smeared out rapidly by radiative diffusion, then we can drop the last term on the right in Eq.~\eqref{Hd}. This then yields the diffusion equation with transport coefficient Pr$+2F_H'$. If this is positive then the disturbance decays, but if it is negative then we have antidiffusion and the formation of layers of high and low angular momentum. As a consequence, the criterion for instability is simply that viscous diffusion is sufficiently weak
$\text{Pr} <  -2 F_H'$ (noting that $F_H'<0$) and fails to remedy the sharpening of gradients brought about by the turbulent flux.

Incorporating the thermal dynamics now, we obtain a dispersion relation for normal modes of the type $\propto e^{sT + \text{i} K T}$:
\begin{align}
&s^2+K^2(1+F_\Theta'+2 F_H' + \text{Pr})s \notag \\ 
& \hskip2cm +K^4\left[2F_H'+\text{Pr}(1+F_\Theta')\right]=0,    
\end{align}
and the instability criterion is modified slightly:
$$\text{Pr} < \frac{-2 F_H'}{1+F_\Theta'}.$$
Thus the thermal physics is  somewhat stabilising.
\end{appendix}

\end{document}